\documentclass[3p,times]{elsarticle}
\journal{}
\makeatletter
\def\ps@pprintTitle{%
\let\@oddhead\@empty
\let\@evenhead\@empty
\def\@oddfoot{}%
\let\@evenfoot\@oddfoot}
\makeatother
\usepackage{amsmath,amssymb}
\usepackage{tikz}
\usepackage{listings}
\usepackage{color}

\usepackage{colortbl}
\usepackage{algorithm}
\usepackage{algpseudocode,setspace}
\usepackage{hyperref}
\hypersetup{colorlinks=true,linkcolor=blue,filecolor=magenta,urlcolor=cyan}
\usepackage{xspace}
\newcommand\Rey{\mbox{\textit{Re}}} 

\usepackage{caption}
\usepackage{subcaption}  

\begin{document}
\captionsetup[table]{skip=2pt} 

\begin{frontmatter} 
\title{CaLES: A GPU-accelerated solver for large-eddy simulation \\ of wall-bounded flows\tnoteref{link}} \tnotetext[link]{Source code is openly available under the MIT License at \href{https://github.com/soaringxmc/CaLES}{\nolinkurl{github.com/soaringxmc/CaLES}}.} 

\author[sapienza]{Maochao Xiao} \ead{maochao.xiao@uniroma1.it} 
\author[sapienza]{Alessandro Ceci} 
\author[delft]{Pedro Costa} 
\author[umd]{Johan Larsson} 
\author[sapienza]{Sergio Pirozzoli} \ead{sergio.pirozzoli@uniroma1.it} 
\address[sapienza]{Dipartimento di Ingegneria Meccanica e Aerospaziale, Sapienza Università di Roma, Via Eudossiana 18, 00184 Roma, Italy} 
\address[delft]{Process and Energy Department, TU Delft, Leeghwaterstraat 39, Delft 2628 CB, The Netherlands} 
\address[umd]{University of Maryland, College Park, Maryland 20742, USA} 

\begin{abstract} 
We introduce CaLES, a GPU-accelerated finite-difference solver designed for large-eddy simulations (LES) of incompressible wall-bounded flows in massively parallel environments. Built upon the existing direct numerical simulation (DNS) solver CaNS, CaLES relies on low-storage, third-order Runge-Kutta schemes for temporal discretization, with the option to treat viscous terms via an implicit Crank-Nicolson scheme in one or three directions. A fast direct solver, based on eigenfunction expansions, is used to solve the discretized Poisson/Helmholtz equations. For turbulence modeling, the classical Smagorinsky model with van Driest near-wall damping and the dynamic Smagorinsky model are implemented, along with a logarithmic law wall model. GPU acceleration is achieved through OpenACC directives, following CaNS-2.3.0. Performance assessments were conducted on the Leonardo cluster at CINECA, Italy. Each node is equipped with one Intel Xeon Platinum 8358 CPU (2.60 GHz, 32 cores) and four NVIDIA A100 GPUs (64 GB HBM2e), interconnected via NVLink 3.0 (200 GB/s). The inter-node communication bandwidth is 25 GB/s, supported by a DragonFly+ network architecture with NVIDIA Mellanox InfiniBand HDR. Results indicate that the computational speed on a single GPU is equivalent to approximately 15 CPU nodes, depending on the treatment of viscous terms and the subgrid-scale model, and that the solver efficiently scales across multiple GPUs. The predictive capability of CaLES has been tested using multiple flow cases, including decaying isotropic turbulence, turbulent channel flow, and turbulent duct flow. The high computational efficiency of the solver enables grid convergence studies on extremely fine grids, pinpointing non-monotonic grid convergence for wall-modeled LES.
\end{abstract} 


\end{frontmatter}

\section{Introduction}\label{sec:intro}
Large-eddy simulation (LES) has become an important tool in the design processes of spacecraft, aircraft, automobiles, ships, and other engineering systems, for cases where traditional Reynolds-averaged Navier-Stokes (RANS) models fall short. RANS models often struggle to accurately predict complex flow phenomena, such as flow separation, rotational turbulence, and three-dimensional turbulent boundary layers~\cite{wilcox1998turbulence}. Direct numerical simulation (DNS) can provide accurate results but become computationally prohibitive at high Reynolds numbers due to the high mesh resolutions required. As a result, LES has gained popularity for offering a reasonable balance between computational cost and predictive accuracy, particularly for complex aerodynamic and hydrodynamic applications. In recent years, LES has been increasingly applied to external aerodynamic flows, especially at the edges of flight envelopes, such as high-lift aircraft aerodynamics~\cite{goc2021large} and iced-wing separated flows~\cite{xiao2020numerical}. It has also been used in internal flows, including those within aircraft engines~\cite{arroyo2021towards}. These applications establish LES as a critical component in the industry push toward Certification by Analysis~\cite{mauery2021guide}. Despite significant advancements, both wall models and subgrid-scale (SGS) closures for LES remain areas of active development. For wall modeling, researchers have been exploring robust models that account for more complex flows, such as laminar-turbulent transitions~\cite{larsson2016large} and flow separation~\cite{dupuy2023data} among others. Concurrently, advancements in SGS models have been made, with some of the recent studies emphasizing robust SGS models suited to anisotropic grids~\cite{agrawal2022non}.

It is well established that LES is significantly more computationally demanding than RANS, making the development of faster LES solvers a critical need. An efficient LES solver is not only essential for large-scale production simulations, but also invaluable for the development and testing of subgrid-scale and wall models. Indeed, as data-driven approaches and machine learning techniques are increasingly applied to turbulence modeling, the demand for efficient LES solvers becomes even more pressing. Previous studies have trained SGS models using filtered DNS data~\cite{gamahara2017searching, kang2023neural}. However, relying on DNS data can be problematic, as the SGS tensor in implicitly filtered LES does not perfectly align with the SGS stress terms derived from the filtered Navier-Stokes equations~\cite{bae2022numerical}. This inconsistency is especially significant when grid sizes are significantly larger than the Kolmogorov scale, as the numerical and modeling errors can then be comparable. As a result, SGS models trained on filtered DNS data may perform well in \textit{a priori} tests but fail in \textit{a posteriori} assessments, whereby performance is evaluated in actual LES simulations. Consequently, research has increasingly shifted toward generating training data directly from LES, and turbulence models are optimized for accurate statistical metrics such as mean velocity and wall shear stress~\cite{lozano2023machine}. In the same vein, reinforcement learning has been explored for both SGS and wall modeling~\cite{novati2021automating, bae2022scientific}. Such strategies have been referred to as ``model-consistent'' approaches~\cite{duraisamy2021perspectives}. While these strategies show promise, they often require hundreds or even thousands of LES runs to generate training data or complete one-time training, underscoring the need for fast and efficient solvers. Moreover, the grid convergence properties of wall-modeled LES (WMLES) have become increasingly studied~\cite{zhou2024sensitivity, yang2024grid}, which also requires many LES runs on a series of refined meshes for thorough analysis. Fast LES solvers are therefore desirable to enable grid convergence investigations at high Reynolds numbers.

Table~\ref{tab:solver} gives some popular open-source solvers used in academic research. Whereas all those solvers are capable of performing DNS for canonical flow cases, only URANOS and LESGO currently support LES capabilities. It is well-known that incompressible solvers, such as LESGO, are typically more efficient than compressible solvers like URANOS for simulations of low-Mach-number flows. This efficiency mainly derives from allowing much larger time steps when explicit time-integration schemes are used. However, a key limitation of LESGO is the absence of GPU acceleration, which restricts its scalability and efficiency on modern high-performance computing platforms. Given the high demand for LES in simulating incompressible flows, particularly for machine-learning-based turbulence modeling and grid convergence studies, the development of a GPU-accelerated incompressible LES solver is highly desirable in academia.

\begin{table}[hbt!]
 \centering
 \caption{Open-source DNS/LES solvers for academic research.} \label{tab:solver}
 \small
 \begin{tabular}{ccccc}
   \hline
   Solver & Governing equations     &GPU-supported  &Purposes \\
   STREAmS-2~\cite{bernardini2023streams}  &Compressible NS        &Yes     &DNS \\
   URANOS~\cite{de2024uranos}   &Compressible NS         &Yes     &DNS/LES \\
   AFiD~\cite{zhu2018afid}   &Incompressible NS        &Yes     &DNS \\
   LESGO~\cite{lesgo}   &Incompressible NS        &No         &DNS/LES \\
   CaNS~\cite{costa2021gpu}   &Incompressible NS        &Yes     &DNS \\
   CaLES           &Incompressible NS        &Yes        &DNS/LES \\
   \hline
 \end{tabular}
\end{table}

The present work introduces CaLES, a GPU-accelerated incompressible LES solver specifically designed for wall-bounded flows. CaLES builds on the capabilities of CaNS~\cite{costa2021gpu, costa2018fft}, an open-source DNS solver known for its efficiency in solving the incompressible Navier-Stokes equations. The solver uses a fractional-step method~\cite{kim1985application}. Temporal discretization is carried out using a low-storage, third-order Runge-Kutta scheme~\cite{Wray1990}. Spatial discretization is performed using a second-order finite-difference method on staggered grids~\cite{harlow1965numerical}, which avoids odd-even decoupling phenomena and preserves energy at the discrete level in the inviscid limit~\cite{verstappen2003symmetry}. The solver employs eigenfunction expansions~\cite{schumann1988fast} to efficiently solve the Poisson equation. GPU acceleration is achieved using a combination of CUDA Fortran and OpenACC directives, and performance benchmarks demonstrate that the code performance on 4 NVIDIA Tesla V100 GPUs in a DGX-2 system is roughly equivalent (0.9 times slower to 1.6 times faster) to 2048 cores on state-of-the-art CPU-based supercomputers, and 3.1 to 5.6 times faster when all 16 GPUs in the DGX-2 cluster are used~\cite{costa2021gpu}. CaLES extends CaNS by supporting LES through the inclusion of the classical Smagorinsky model~\cite{smagorinsky1963general} with the van Driest damping function~\cite{van1956turbulent}, and the dynamic Smagorinsky model ~\cite{germano1991dynamic, lilly1992proposed}, along with a logarithmic-law wall model. The solver can simulate various canonical flows in Cartesian single-block domains, including isotropic turbulence, temporally-evolving turbulent boundary layers, channel flows, duct flows, cavity flows, etc. Flexibility and efficiency make it an ideal platform to develop subgrid-scale and wall models, particularly those based on machine learning techniques, and to perform grid convergence studies at high Reynolds numbers.

The remainder of this paper is organized as follows. Section~\ref{sec:method} introduces the governing equations, subgrid-scale models, and numerical methods. Section~\ref{sec:impl} discusses the implementation and performance of the solver. Section~\ref{sec:valid} validates the LES capabilities using decaying isotropic turbulence, turbulent channel flow, and turbulent duct flow. Finally, Section~\ref{sec:conclusion} provides the conclusions of this study.

\section{Methodology}\label{sec:method}
\subsection{Governing equations and subgrid-scale models}\label{sec:equations}
The filtered incompressible Navier-Stokes (NS) equations read as
\begin{equation}\label{eq:conti}
\frac{\partial \overline{u}_i}{\partial x_i} = 0,
\end{equation}
\begin{equation}\label{eq:mom}
\frac{\partial \overline{u}_i}{\partial t} + \frac{\partial \overline{u}_i \overline{u}_j}{\partial x_j} = -\frac{1}{\rho} \frac{\partial \overline{p}}{\partial x_i} + \nu \frac{\partial^2 \overline{u}_i}{\partial x_j \partial x_j} - \frac{\partial \tau_{ij}}{\partial x_j},
\end{equation}
where $\overline{u}_i$ represents the filtered velocity, $\rho$ is the fluid density, $\overline{p}$ is the filtered pressure, and $\nu$ denotes the kinematic viscosity. The term $ \tau_{ij} = \overline{u_i u_j} - \overline{u}_i \overline{u}_j $ is the subgrid-scale stress tensor, which encapsulates the effects of unresolved scales and requires a suitable closure model. The isotropic component of the SGS stress is typically absorbed into the pressure, resulting in a modified pressure field, $\overline{p} \gets \overline{p} + \frac{1}{3}\rho \tau_{kk}$. The remaining deviatoric part is modeled, according to the Boussinesq assumption, as
\begin{equation}\tau_{ij} - \frac{1}{3} \tau_{kk} \delta_{ij} = -2 \nu_t \overline{S}_{ij},
\end{equation}
where $\delta_{ij}$ denotes the Kronecker delta, and $\overline{S}_{ij}$ is the filtered strain-rate tensor. We implemented two representative closure models: the classical Smagorinsky model~\cite{smagorinsky1963general} and its dynamic version~\cite{germano1991dynamic,lilly1992proposed}. The baseline Smagorinsky model reads as~\cite{smagorinsky1963general}
\begin{equation}\label{eq:smag}
\nu_t = \left( C_s \Delta D(y) \right)^2 \overline{S},
\end{equation}
where $C_s$ is a model constant, $\Delta$ is the filter width, $D(y)$ is the near-wall damping function, and  $\overline{S}$ is the rate-of-strain, i.e., $\overline{S} = \sqrt{2 \overline{S}_{ij} \overline{S}_{ij}}$. The van Driest damping function~\cite{van1956turbulent} is commonly used in the presence of no-slip walls, i.e., $D(y) = 1 - \exp\left(-y^*/25\right)$, where ``*'' denotes the wall distance non-dimensionalized by the wall units. A drawback of the standard Smagorinsky model is its inaccurate prediction of the eddy dissipation in laminar and transitional flows, leading to erroneous wall shear stresses and delayed transition to turbulence. Moreover, the optimal value of the model constant $C_s$ depends significantly on the flow features. \citet{lilly1967representation} showed that for isotropic turbulence, with spatial resolution within the inertial subrange, $C_s \approx 0.17$, whereas \citet{deardorff1970numerical} suggested $C_s \approx 0.1$ for wall-bounded turbulent shear flows.
The Smagorinsky model can be improved using a dynamic procedure, whereby the model coefficient is evaluated dynamically by comparing the eddy dissipation at two filter levels. The dynamic Smagorinsky model with Lilly's modification~\cite{germano1991dynamic, lilly1992proposed} is expressed as:
\begin{equation}\label{eq:dsmag}
\nu_t = c_s \Delta^2 \overline{S},
\end{equation}
where
\begin{equation}\label{eq:cs}
c_s = \frac{\langle M_{ij} L_{ij} \rangle}{\langle M_{ij} M_{ij} \rangle},
\end{equation}
\begin{equation}\label{eq:lij}
L_{ij} = \widetilde{\overline{u}_i\overline{u}_j} - \widetilde{\overline u}_i\widetilde{\overline u}_j,
\end{equation}
\begin{equation}\label{eq:mij}
M_{ij} = 2 \Delta^2 \widetilde{\overline S\,\overline{S}_{ij}} - 2 (\alpha \Delta)^2 \widetilde{\overline{S}}\,\widetilde{\overline{S}}_{ij}.
\end{equation}
Here, the bar denotes filtering with filter width $\Delta$, the tilde indicates test filtering with filter width $\alpha \Delta$, and the brackets $\langle\rangle$ represent an averaging operation. The ratio of the two filter widths is commonly set to $\alpha=2.0$. The test-filtered strain rate is computed as
\begin{equation}
\widetilde{\overline{S}}_{ij} = \frac{1}{2} \left( \frac{\partial \widetilde{\overline{u}}_i}{\partial x_j} + \frac{\partial \widetilde{\overline{u}}_j}{\partial x_i} \right).
\end{equation}
The dynamic Smagorinsky model provides reasonable subgrid-scale dissipation and automatically switches off in laminar flows. However, it demands more memory and computational resources than the static version and often requires averaging or clipping to ensure numerical stability~\cite{lilly1992proposed}.

Wall models are required when the near-wall mesh resolution is too coarse to resolve the inner-layer turbulent scales up to the inertial subrange. These models are typically classified into near-wall RANS models~\cite{nikitin2000approach} and wall-stress models~\cite{kawai2012wall}. In classical wall-stress models, the wall-parallel velocity at a specific wall distance is used as input and the wall shear stress is estimated as output, which replaces the no-slip boundary condition. The simplest wall-stress model is the logarithmic wall law:
\begin{equation}\label{eq:wm}
\frac{U_{wm}}{u_{\tau}} = \frac{1}{\kappa} \ln \left( \frac{u_{\tau} h_{wm}}{\nu} \right) + B,
\end{equation}
where $\kappa = 0.41$, $B = 5.2$, $u_\tau$ is the friction velocity, $h_{wm}$ is the wall-modeled layer thickness, and $U_{wm}$ is the wall-parallel velocity magnitude at the top of the wall-modeled layer.

\subsection{Numerical methods}\label{sec:numerics}
The filtered incompressible Navier-Stokes equations are solved using a fractional-step method~\cite{kim1985application}. Time integration is performed with a low-storage, three-step Runge-Kutta scheme~\cite{Wray1990}, while spatial discretization is handled using a second-order finite-difference method on staggered grids~\cite{harlow1965numerical}. The continuity equation~\eqref{eq:conti} and the momentum equation~\eqref{eq:mom} are coupled through a pressure-correction method~\cite{amsden1970smac}. This section provides an overview of the numerical schemes, with a focus on key aspects of the implementation of SGS and wall models. For additional details, refer to the descriptions of CaNS~\cite{costa2018fft}.

At each sub-step, the flow field is updated as
\begin{equation}\label{eq:onestar_exp}
u_i^{*} = u_i^{k} + \Delta t \left[ \alpha_{\kappa + 1} \left( H_i^{k} + \nu L_{jj} u_i^{k} \right) + \beta_{\kappa + 1} \left( H_i^{k-1} + \nu L_{jj} u_i^{k-1} \right) - \gamma_{\kappa + 1} \partial_i p^{k-1/2} \right],
\end{equation}
\begin{equation}\label{eq:poi}
L_{jj} \phi = \frac{\partial_i u_i^{*}}{\gamma_{k+1} \Delta t},
\end{equation}
\begin{equation}\label{eq:correc}
u_i^{k+1} = u_i^{*} - \gamma_{k+1} \Delta t \partial_i \phi,
\end{equation}
\begin{equation}\label{eq:p_update_exp}
p^{k+1/2} = p^{k-1/2} + \phi.
\end{equation}
Here, $k=0$ corresponds to physical time level $n$, and $k+1=3$ corresponds to time level $n+1$. The symbol $u_i^*$ denotes the predicted velocity, and $\phi$ represents the pressure correction. The Runge-Kutta coefficients are $\alpha_{k+1} = (8/15, 5/12, 3/4)$, $\beta_{k+1} = (0, -17/60, -5/12)$, and $\gamma_{k+1} = \alpha_{k+1} + \beta_{k+1}$. In equation~\eqref{eq:onestar_exp}, $H_i$ includes convective and SGS stress terms,
\begin{equation}\label{eq:H_i}
H_i = \frac{\partial u_i u_j}{\partial x_j} + \frac{\partial \tau_{ij}}{\partial x_j},
\end{equation}
and the Laplace operator in the diffusive term is defined as 
\begin{equation}
L_{jj} = \frac{\partial^2}{\partial x_j \partial x_j}.
\end{equation}
For low-Reynolds-number flows or very fine grids, it may be desirable to use implicit temporal discretization for the diffusion terms. With Crank-Nicolson time integration, this results in
\begin{equation}\label{eq:twostar_imp3d}
u_i^{**} = u_i^{k} + \Delta t \left[ \alpha_{\kappa + 1} H_i^{k} + \beta_{\kappa + 1} H_i^{k-1} + \gamma_{\kappa + 1} \left( \nu L_{jj} u_i^{k} - \partial_i p^{k-1/2} \right) \right],
\end{equation}
\begin{equation}\label{eq:onestar_imp3d}
u_i^{*} - \gamma_{\kappa + 1} \frac{\nu \Delta t}{2} L_{jj} \left( u_i^{*} \right) = u_i^{**} - \gamma_{\kappa + 1} \frac{\nu \Delta t}{2} L_{jj} u_i^{k},
\end{equation}
\begin{equation}\label{eq:p_update_imp3d}
p^{k+1/2} = p^{k-1/2} + \phi - \gamma_{k+1} \frac{\nu \Delta t}{2} L_{jj} \phi.
\end{equation}
Equations~\eqref{eq:twostar_imp3d} and \eqref{eq:onestar_imp3d} are intentionally not combined to illustrate that $u_i^{**}$ provides a better approximation of $u_i^{k+1}$ than the sum of the terms on the right-hand side of equation~\eqref{eq:onestar_imp3d}~\cite{costa2021gpu}. Alternatively, implicit treatment of the viscous terms can be performed only in the $y$ direction, resulting in
\begin{equation}\label{eq:twostar_imp1d}
u_{i}^{**} = u_{i}^{k} + \Delta t \left\{ \alpha_{k+1} \left[ H_{i}^{k} + \nu \left( L_{11} + L_{33} \right) u_{i}^{k} \right] + \beta_{k+1} \left[ H_{i}^{k-1} + \nu \left( L_{11} + L_{33} \right) u_{i}^{k-1} \right] + \gamma_{k+1} \left( \nu L_{22} u_{i}^{k} - \partial_{i} p^{k-1/2} \right) \right\},
\end{equation}
\begin{equation}\label{eq:onestar_imp1d}
\left( 1 - \gamma_{\kappa + 1} \frac{\nu \Delta t}{2} L_{22} \right) u_i^{*} = u_i^{**} - \gamma_{\kappa + 1} \frac{\nu \Delta t}{2} L_{22} u_i^{k},
\end{equation}
\begin{equation}\label{eq:p_update_imp1d}
p^{k+1/2} = p^{k-1/2} + \phi - \gamma_{k+1} \frac{\nu \Delta t}{2} L_{22} \phi .\end{equation}
In the solver, $z$ is designated as the coordinate direction along which non-uniform grid spacing can be applied, typically corresponding to the wall-normal direction in a plane channel. However, to maintain consistency with the conventions in the turbulence community, we refer to the wall-normal direction as $y$ throughout this paper. The Poisson equation~\eqref{eq:poi}, once discretized, is solved using an eigenfunction expansion method~\cite{schumann1988fast}, which allows for an efficient solution using the Thomas algorithm along the $y$ direction. When implicit treatment of the viscous diffusive terms is applied, the resulting three modified Helmholtz equations in equation~\eqref{eq:onestar_imp3d} are solved with the same direct solver used for the Poisson equation.

The SGS stress terms are always handled explicitly, regardless of the time integration scheme used for the viscous terms. This allows the SGS stress terms to be grouped with the convective terms, as shown in equation~\eqref{eq:H_i}. The SGS model is evaluated at cell centers, where both the SGS viscosity and the strain-rate tensor are stored. The diagonal components of the strain-rate tensor are directly computed at the cell centers, while the non-diagonal components are first calculated at the cell edge midpoints and then averaged to the cell centers. In the dynamic procedure, two-dimensional (2D) or three-dimensional (3D) box filters can be applied. The implementation of the filters assumes uniform grid spacing for simplicity. The 3D box filter cannot be used in the first off-wall layer, hence a 2D box filter is applied through linear extrapolation of the wall-parallel velocity to the ghost points, followed by the application of a 3D filter. In this layer, we set $\alpha = 4^{1/3}$ in equation~\eqref{eq:mij}, which mathematically corresponds to a 2D box filter. Our WMLES tests on channel flows indicate that using this value in this first layer yields more accurate results than $\alpha = 8^{1/3}$. The averaging operation in equation~\eqref{eq:cs} is performed in the homogeneous directions, and the averaged coefficient is clipped to zero when it is negative. The velocity components are averaged to the cell centers before applying the test filter to evaluate $L_{ij}$ in equation~\eqref{eq:lij}.

\begin{figure}[htbp]
   \centering
   \includegraphics[width=0.7\linewidth]{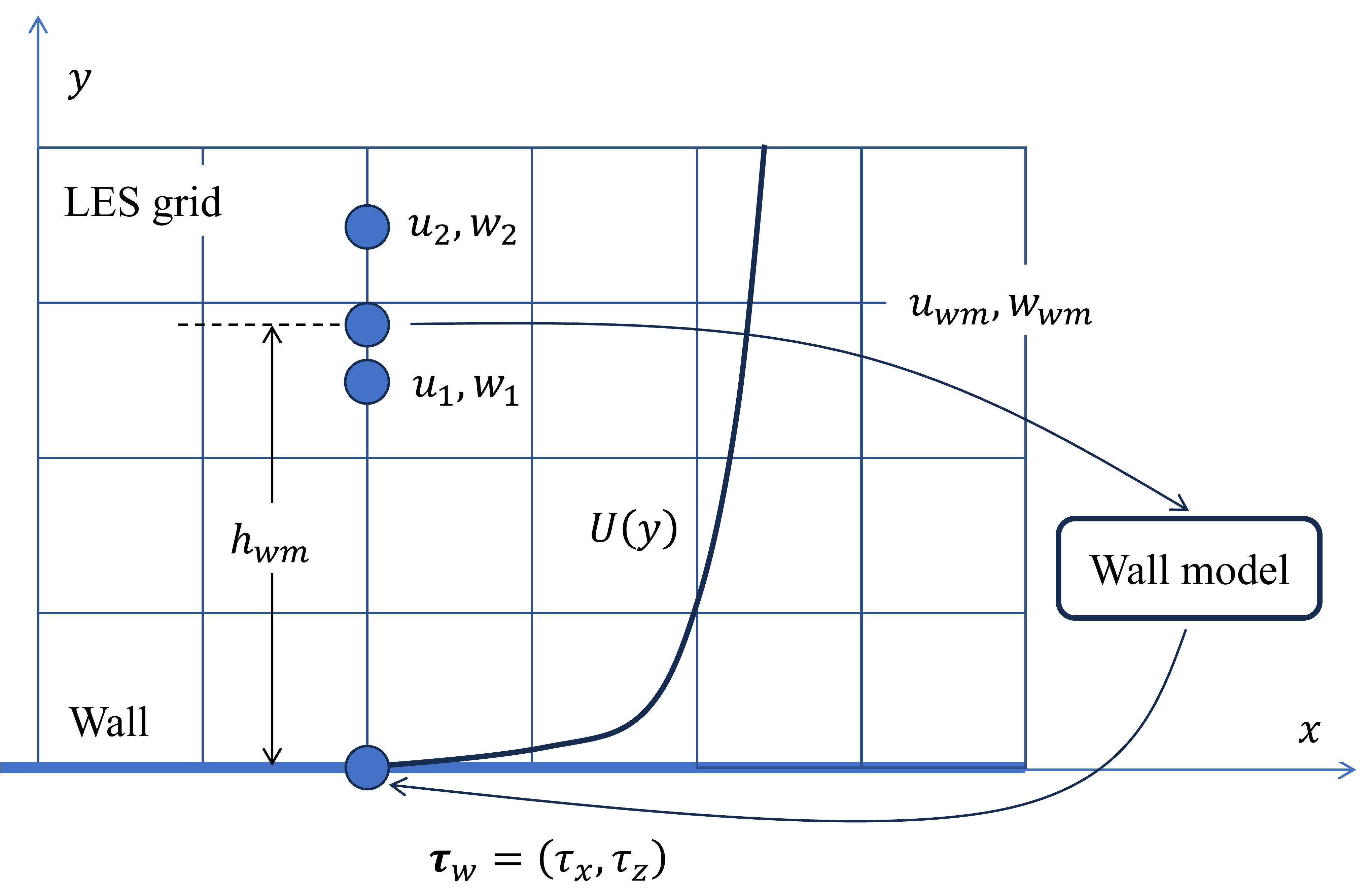}
   \caption{Schematic of the wall-model implementation.}
   \label{fig:wall_model}
\end{figure}

For the wall model, the Newton–Raphson iterative method is used to determine the wall shear stress from equation~\eqref{eq:wm}, typically requiring 3 to 7 iterations to achieve convergence in wall shear stress within a relative tolerance of $0.01\%$. Figure~\ref{fig:wall_model} illustrates how the wall model is coupled to the LES solution. The input velocity at the top of the wall-modeled layer is the magnitude of the instantaneous wall-parallel velocity, which is evaluated as $U_{wm}=(u_{wm}^2 + w_{wm}^2)^{1/2}$, where $u_{wm}$ and $w_{wm}$ are the $x$- and $z$-direction velocities, respectively. The two components are obtained via linear interpolation in the wall-normal direction between locations 1 and 2. The resulting wall shear stress is then used as the boundary condition for the wall-parallel velocity vector. The impermeability condition is enforced for the wall-normal velocity component. When a staggered grid is used, the two wall-parallel velocity components (e.g., $u$ and $w$) are stored at different cell face centers. Consequently, when the wall model is used to enforce the boundary condition for $u$, as illustrated in figure~\ref{fig:wall_model}, interpolation of $w$ to the location of $u$ is required to evaluate the wall-parallel velocity vector. Similarly, for $w$, interpolation of $u$ to the location of $w$ is performed. This approach minimizes the number of interpolations, which is desirable when computing either SGS viscosity or wall models on staggered grids. When utilizing a wall model, the wall-normal derivatives of the wall-parallel velocity components are determined from first-order one-sided finite-difference scheme in the first off-wall layer, thus avoiding crossing the under-resolved layer between the wall and the first off-wall wall-parallel velocity location~\cite{kawai2012wall}. In CaLES, one-sided finite differencing is achieved through linear extrapolation of the wall-parallel velocity to the ghost points, followed by second-order central differencing. This procedure is crucial for accurate evaluation of the strain-rate tensor required to evaluate the SGS viscosity. The viscous terms in the filtered Navier-Stokes equations are evaluated as usual in the first off-wall layer, as the wall shear stress is directly provided by the wall model, and the layer between the first and second off-wall locations of the wall-parallel velocity can be regarded as properly resolved. When the viscous terms are handled implicitly in all three directions, the wall model can only be applied in the $y$-direction, as homogeneous boundary conditions are required in the other two directions where Fourier transforms are applied. However, this limitation may be irrelevant, as explicit time integration is commonly used for wall-modeled LES due to the large thickness of the first off-wall layer of cells. This restriction does not apply when the viscous terms are handled implicitly only in the $y$-direction.

\section{Implementation and performance}\label{sec:impl}

\subsection{Overall implementation strategy}
CaLES is developed using CaNS-2.3.0 as its baseline solver. Algorithm~\ref{alg:overall_solution} outlines the overall solution procedure for explicit time integration. When the viscous terms are handled implicitly, $u_i^*$ is computed using equations~\eqref{eq:twostar_imp3d} and \eqref{eq:onestar_imp3d}, and $p^{k+1/2}$ is updated from equation~\eqref{eq:p_update_imp3d}. When the viscous terms are handled implicitly only in the $y$-direction, $u_i^*$ is determined by equations~\eqref{eq:twostar_imp1d} and \eqref{eq:onestar_imp1d}, while $p^{k+1/2}$ is updated from equation~\eqref{eq:p_update_imp1d}. To ensure consistency with boundary conditions, ghost cells are updated immediately after any variable is updated. Algorithm~\ref{alg:bc} details the procedure for applying boundary conditions. The calculation of wall shear stress using the wall model must be performed after all other boundary conditions have been applied, as the wall model computation relies on the updated ghost-point values at boundary points.

\begin{algorithm}
\caption{Overall solution procedure.}
\label{alg:overall_solution}
\begin{algorithmic}[1]
   \State Initialize velocity $u_i$ and pressure $p$.
   \State Compute eddy viscosity $\nu_t$.
   \State Set iteration counter $n = 0$.
   \While{$n \le n_{\text{max}}$}
       \State Increment iteration counter: $n \gets n + 1$.
       \State Determine time step $\Delta t$.
       \For{$k = 0$ to $2$}
           \State Compute intermediate velocity $u_i^*$ using equation~\eqref{eq:onestar_exp};
           \State Compute the right-hand side of the Poisson equation and solve for $\phi$ in equation~\eqref{eq:poi};
           \State Update $u_i^{k+1}$ using the correction procedure of equation~\eqref{eq:correc};
           \State Update $p^{k+1/2}$ using equation~\eqref{eq:p_update_exp};
           \State Compute $\nu_t^{k+1}$.
       \EndFor
   \EndWhile
   \State Terminate simulation.
\end{algorithmic}
\end{algorithm}

\begin{algorithm}
\caption{Boundary condition treatment.}
\label{alg:bc}
\begin{algorithmic}[1]
   \State Perform ghost-cell exchange between blocks.
   \State Update all boundary conditions except wall-model boundary conditions.
   \State Compute wall shear stress using equation~\eqref{eq:wm}.
   \State Update wall-model boundary conditions.
\end{algorithmic}
\end{algorithm}

The GPU porting was implemented using OpenACC directives. A key factor to achieving high acceleration is to minimize data transfer between GPU and CPU memory. In the solver, large arrays are transferred only during the initialization or the first time step using unstructured data lifetimes. From the second iteration onward, only scalars or small arrays are transferred between CPU and GPU memory, except when flow-field output is required. According to the OpenACC Programming and Best Practices Guide~\cite{openacc_guide_2021}, parallel regions are defined using either the ``kernels'' construct or the``parallel'' construct. Specifically, the ``kernels'' construct allows the compiler to automatically exploit parallelism in a region of code, while the ``parallel'' construct, often used in conjunction with the ``loop'' construct and the ``collapse'' clause, is applied to accelerate key loops for optimal performance. We use the ``kernels'' construct for simple array assignments, and the ``parallel'' construct to speed up loops for the computation of advective, viscous and SGS stress terms, as well as for time advancement. Additionally, the ``async'' clause is applied to the kernels, parallel, update, and data directives (both structured and unstructured), enabling the CPU to continue with other tasks while the accelerator performs operations, without waiting for their completion. The FFTs required by the Poisson/Helmholtz solver are carried out using the cuFFT library from the CUDA Toolkit. For additional details, refer to~\cite{costa2021gpu}. 

In line with CaNS-2.3.0, parallelization of the code is achieved through MPI, with each rank allocated to one GPU. The structured grid block is partitioned into subdomains using a 2D pencil-like decomposition. The pencil axis is recommended to be aligned with the $x$-direction for optimal efficiency, except when viscous terms are handled implicitly in the $y$-direction, in which case alignment with the $y$-axis is preferable. The cuDecomp library~\cite{romero2022distributed} manages the transpose operations required for FFT-based transforms and ghost-cell exchanges. The library not only optimizes the pencil domain decomposition layout, but also finds the most efficient communication backends for transposes and ghost-cell exchanges. This process involves runtime testing of different grid decomposition layouts and communication backends to identify the best-performing combination. Notably, transpose operations and ghost-cell exchanges can utilize different communication backends. Supported communication methods include CUDA-aware MPI point-to-point, MPI all-to-all, NVIDIA Collective Communication Library (NCCL), and NVIDIA Shared Memory (NVSHMEM), with different staging strategies. Given that the optimal setup is highly system-dependent, the hardware-adaptive decomposition provided by cuDecomp is crucial for efficient resource utilization~\cite{romero2022distributed}.



\subsection{Performance analysis}
\begin{figure}[htbp]
   \centering
   \begin{subfigure}{0.49\textwidth}
       \centering
       \includegraphics[width=\linewidth]{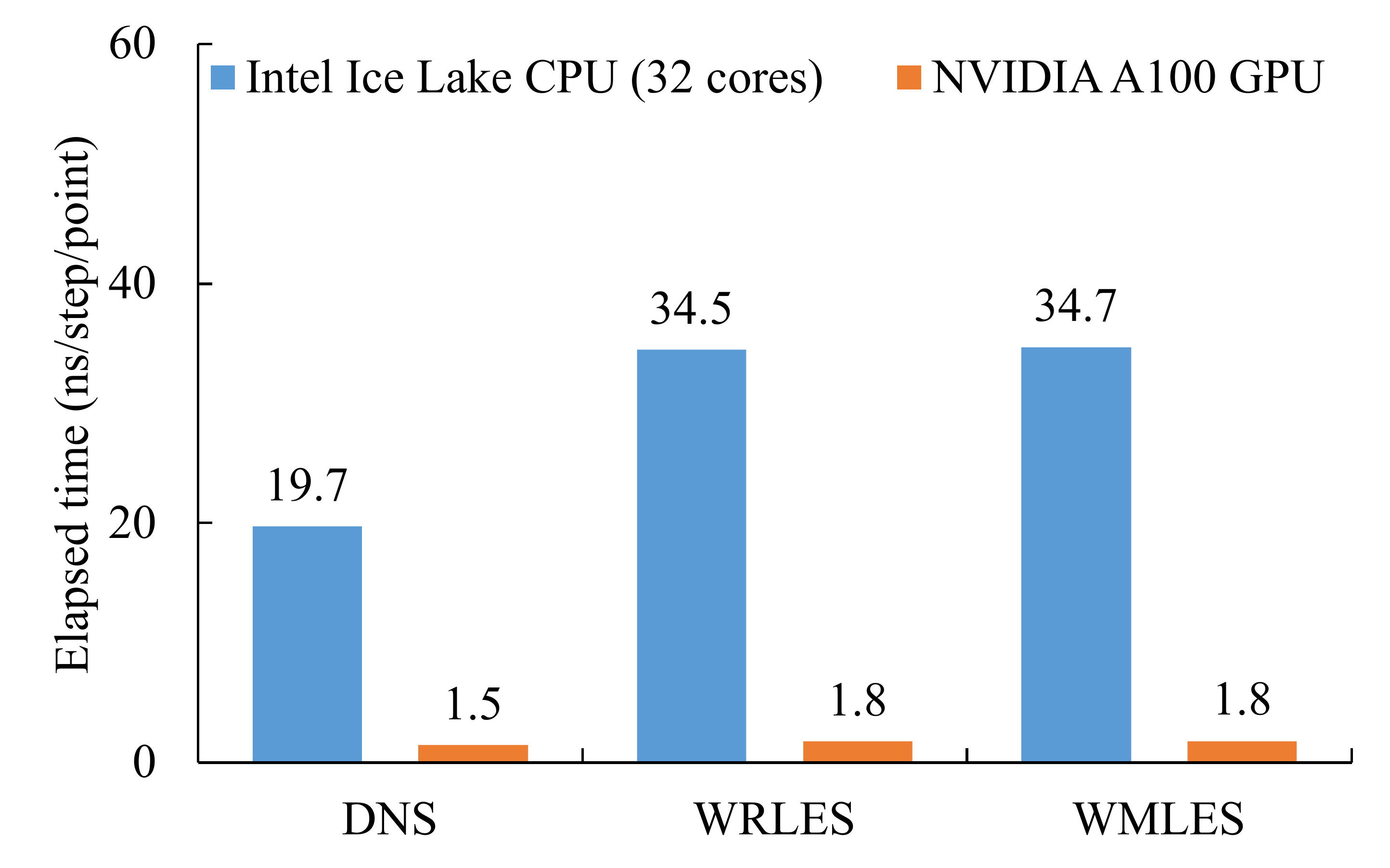}
       \label{fig:walltime_exp}
   \end{subfigure}
   \hfill
   \begin{subfigure}{0.49\textwidth}
       \centering
       \includegraphics[width=\linewidth]{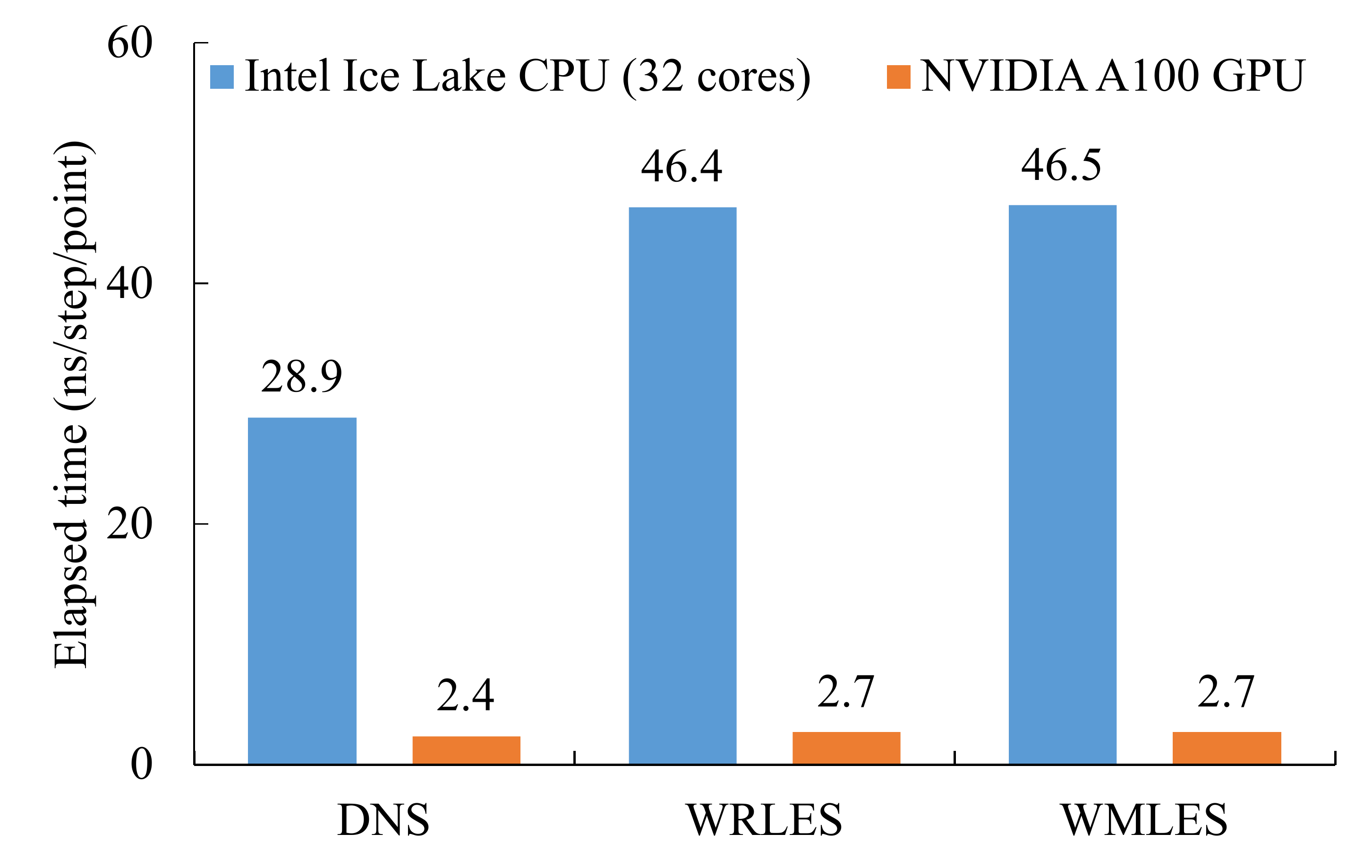}
       \label{fig:sec_time_exp}
   \end{subfigure}
   \caption{Elapsed wall-clock time per time step per grid point for different methods when the viscous terms are handled explicitly (a), or implicitly in the $y$ direction (b).}
   \label{fig:wall_time}
\end{figure}

Assessment of the code performance was conducted on the Booster partition of the Leonardo cluster at CINECA, Italy. Each node of the cluster is equipped with an Intel Xeon Platinum 8358 CPU (2.60 GHz, 32 cores) and four NVIDIA A100 GPUs (64 GB HBM2e). The intra-node communication bandwidth is 200 GB/s, supported by NVLink 3.0. The inter-node bandwidth is 25 GB/s, facilitated by the DragonFly+ network architecture using NVIDIA Mellanox InfiniBand HDR, giving each GPU an effective communication rate of approximately 6.25 GB/s. The performance comparison is made between a single GPU card and a CPU node. The test case under consideration is flow in a plane channel with two walls in the $y$ direction, and periodic boundary conditions in the other two directions. The grid size is $(N_x,N_y,N_z)=(512, 384, 1440)$, which is approximately the maximum grid size that can fit into GPU memory when the Smagorinsky model with the van Driest damping function is activated and the viscous terms are handled implicitly along the $y$ direction. It is noteworthy that, during runs on a single GPU, data sharing is also performed when handling periodic boundary conditions in the two decomposed directions, with the pencil-axis as the non-decomposed direction. Figure~\ref{fig:wall_time} compares the wall-clock time for different methods when the viscous terms are handled explicitly or implicitly in the $y$ direction. The inclusion of the wall model results in negligible computational overhead. However, applying the static Smagorinsky model with the van Driest damping function increases the computational cost by approximately 0.3~ns per step per grid point. When the viscous terms are handled explicitly, the speed-up factors relative to the CPU node are 13$\times$ for DNS and 19$\times$ for wall-resolved LES (WRLES). If the viscous terms are handled implicitly in the $y$ direction, the speed-up factors are 12$\times$ for DNS and 17$\times$ for WRLES. The greater speed-up of WRLES is due to the increased computational intensity introduced by the subgrid-scale model.

\begin{figure}[htbp]
   \centering
   \begin{subfigure}{0.49\textwidth}
   \centering
   \includegraphics[width=\linewidth]{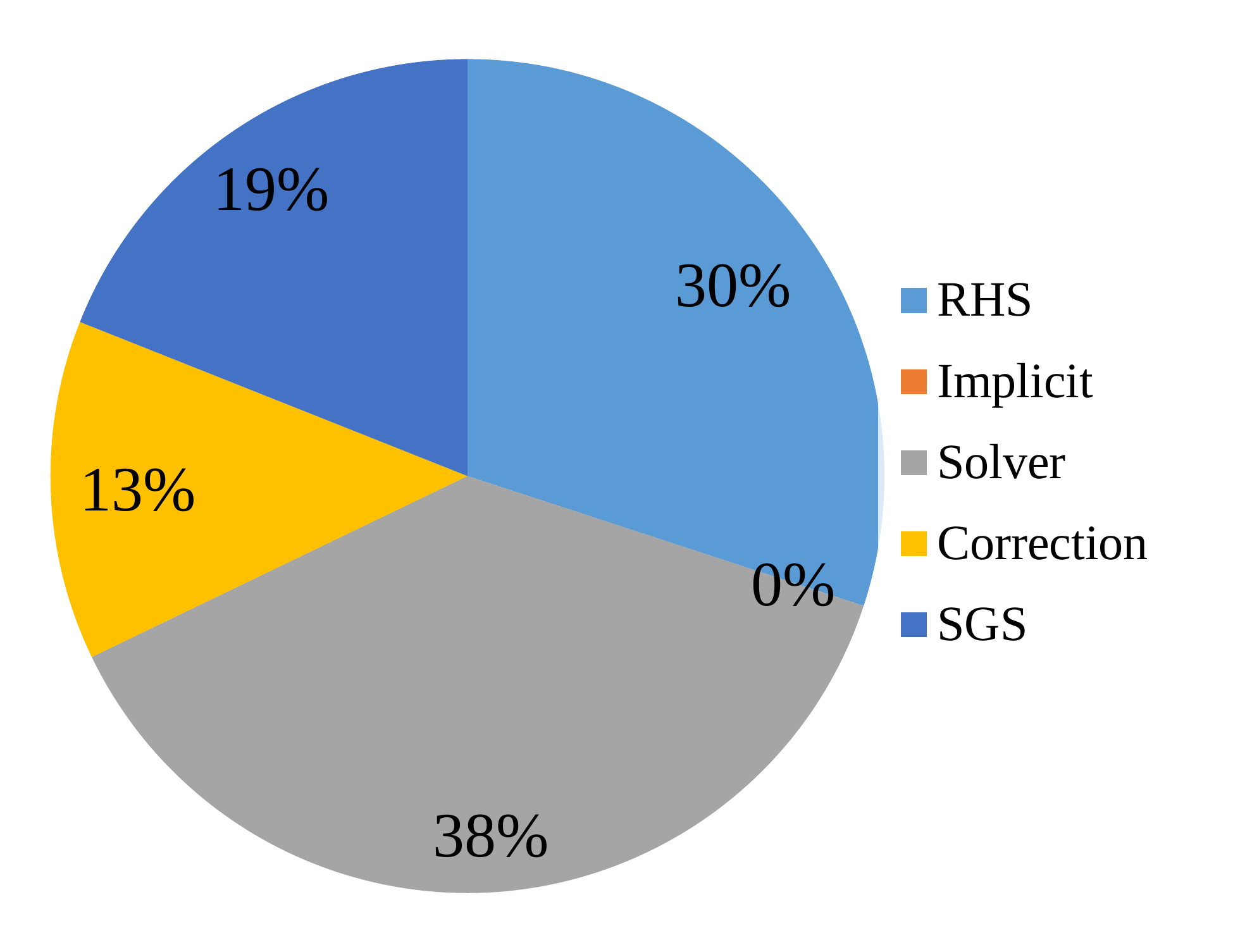}
   \caption{}
   \label{fig:walltime_imp1d}
   \end{subfigure}
   \hfill
   \begin{subfigure}{0.49\textwidth}
   \centering
   \includegraphics[width=\linewidth]{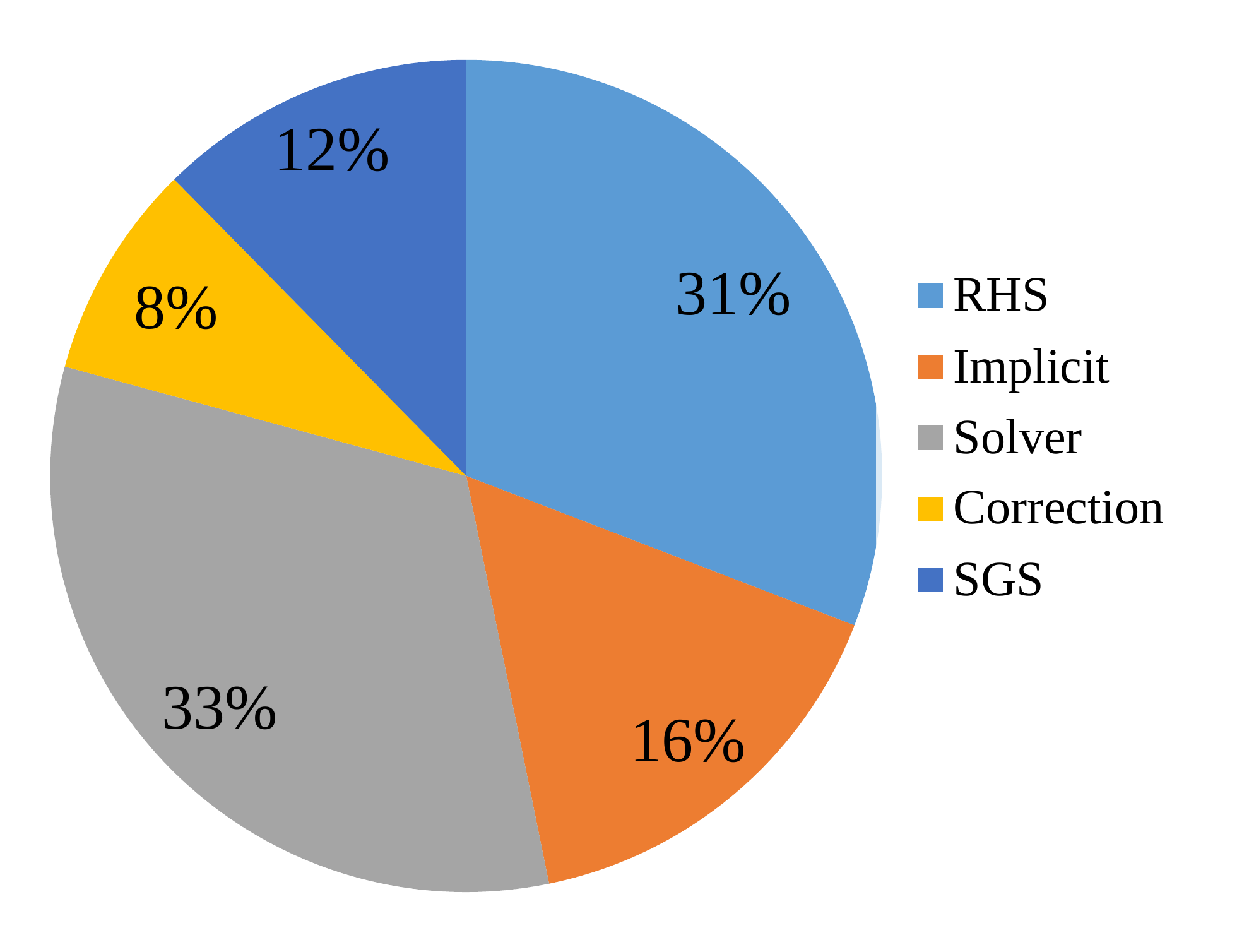}
   \caption{}
   \label{fig:sec_time_imp1d}
   \end{subfigure}
   \caption{Comparison of wall-clock time breakdown for different computational components, when the viscous terms are handled explicitly (a), or implicitly in the $y$ direction (b). ``RHS'' denotes the time spent calculating the right-hand side of the momentum equation, given the eddy viscosity; ``Implicit'' indicates the time required to solve the Helmholtz equations when the viscous terms are handled implicitly in the $y$ direction; ``Solver'' corresponds to the time spent solving the Poisson equation; ``Correction'' denotes the time allocated for the correction procedure; and ``SGS'' represents the time required to evaluate the eddy viscosity using the classical Smagorinsky model with the van Driest damping function.}
   \label{fig:sec_time}
\end{figure}

Figure~\ref{fig:sec_time} reports the wall-clock time breakdown for different computational components. The most time is consumed by the Poisson solver and the computation of the right-hand side of the momentum equation. The calculation of eddy viscosity using the Smagorinsky model with the van Driest damping function is the third most time-consuming part. When the viscous terms are handled implicitly in the $y$ direction, the Helmholtz solver is the third most time-consuming component, followed by the eddy viscosity calculation. Table~\ref{tab:memory} presents the estimated memory footprint per grid point for different methods. The incorporation of the Smagorinsky model with the van Driest damping function increases memory usage per grid point by 32 bytes, or approximately $20\%$ of the DNS memory footprint. In contrast, the dynamic Smagorinsky model requires an additional 216 bytes, or about $150\%$ of the DNS memory footprint. Note that memory requirements for WMLES simulations exactly match those of the corresponding WRLES cases and are therefore omitted from the table.

\begin{table}[htbp]
 \centering
 \caption{Estimated memory footprint per grid point for different methods. ``SM'' denotes the Smagorinsky model with the van Driest damping function, and ``DSM'' is the dynamic Smagorinsky model.} \label{tab:memory}
 \small
 \begin{tabular}{cccc}
   \hline
   Method         & DNS & WRLES (SM) & WRLES (DSM) \\
   Explicit       & 136 bytes & 168 bytes        & 352 bytes \\
   Implicit-$y$   & 160 bytes & 192 bytes        & 376 bytes \\
   \hline
 \end{tabular}
\end{table}

\begin{figure}[htbp]
   \centering
   \begin{subfigure}{0.50\textwidth}
       \centering
       \includegraphics[width=\linewidth]{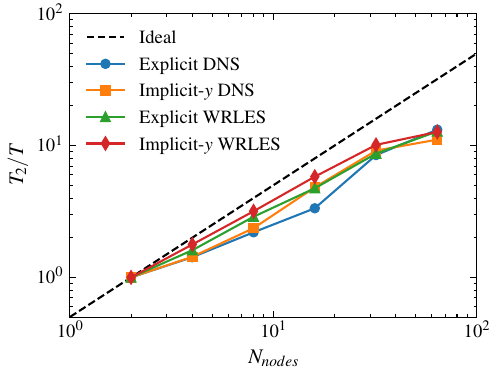}
       \caption{}
       \label{fig:sub1}
   \end{subfigure}
   \hfill
   \begin{subfigure}{0.48\textwidth}
       \centering
       \includegraphics[width=\linewidth]{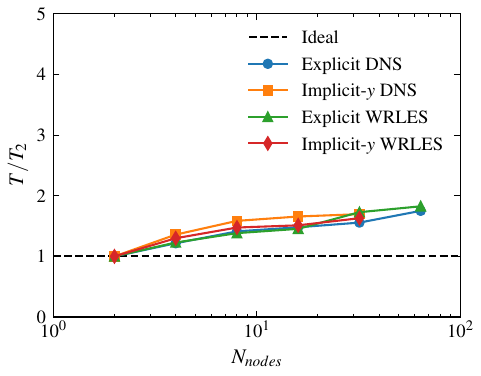}
       \caption{}
       \label{fig:sub2}
   \end{subfigure}
   \caption{Strong (a) and weak (b) code scaling performance. Here, $N_{nodes}$ denotes the number of nodes, $T_2$ represents the wall-clock time for $N_{nodes}=2$, and $T$ is the wall-clock time for a given $N_{nodes}$.}
   \label{fig:scaling}
\end{figure}

We now proceed to analyze the strong and weak scalability performance of the solver. For the strong scaling experiments, a fixed grid size of $(N_x, N_y, N_z) = (512, 384, 1440 \times 4 \times 2)$ is used. This grid is generated by refining the previous single GPU grid by a factor of 8 along the spanwise $z$ direction, resulting in a size close to the maximum allowable grid on two nodes when the viscous terms are handled implicitly in the $y$ direction with the Smagorinsky model and the van Driest damping function activated. For the weak scaling tests, the grid size was set to $(512, 384, 1440\times 4 \times N_{nodes})$. We begin the scaling tests from two nodes instead of one, to avoid performance degradation caused by transitioning from fast intra-node communication to slower inter-node communication. The wall-clock time for two nodes, $T_2$, is used as the reference for reporting the code scaling performance. The domain decomposition is aligned with the $x$-axis when the viscous terms are handled explicitly, and with the $y$-axis when the viscous terms are handled implicitly in the $y$ direction. Figure~\ref{fig:scaling} presents the results for both strong and weak scaling. In GPU-resident, distributed-memory simulations of turbulent flows, weak scaling is the most critical performance metric. Maximizing GPU occupancy is always desirable, making it essential that the code maintains high performance for a fixed problem size per computational subdomain or MPI task, and consequently per GPU. Figure~\ref{fig:scaling}(b) demonstrates that the weak scaling performance across the different computational configurations has small variations. The computational time increases by approximately $80\%$ as the number of nodes increases from 2 to 64. When the viscous terms are handled implicitly in the $y$ direction, there are no data points for $N_{nodes} = 64$ due to GPU memory limitations. The fully implicit scheme is not analyzed here due to its low computational efficiency when Fourier transforms are applied to solve the momentum equations. In such cases, more efficient techniques, such as the alternating-direction-implicit (ADI) scheme~\cite{kim1985application}, may be employed due to its $O(N)$ computational complexity for a single dimension and its additional options for parallel implementation. By contrast, FFT methods have $O(N \log N)$ complexity and typically require extensive data transpositions. However, it should be noted that ADI can still be computationally expensive in a distributed-memory setting, as it requires either transposing the domain or solving a sequence of three tri-diagonal systems in all three directions.


\section{Validation}\label{sec:valid}
The solver is validated using three representative cases: homogeneous decaying isotropic turbulence (DIT), turbulent channel, and turbulent duct flow. The DIT case is employed to validate the SGS models, while the channel case at $\Rey_b = 20,000$ is to validate the wall-resolved LES capability. The channel flow at $\Rey_b = 250,000$ and the duct flow are instead employed to validate the wall-modeled LES capability. In these simulations, the viscous terms are handled explicitly, except for the wall-resolved LES of the channel flow, where they are handled implicitly in the $y$ direction. For all simulations, the time step is dynamically adjusted to its maximum allowable value for numerical stability, multiplied by a safety factor of $0.95$.

\subsection{Decaying isotropic turbulence}
We first validate the LES capability using freely decaying isotropic turbulence. The physical experiment was performed by~\citet{comte1971simple}, with decaying turbulence generated behind a mesh with size $M = 5.08$~cm, and freestream velocity $U_0 = 10$~m/s. The Taylor microscale Reynolds number ($\Rey_\lambda = u_{rms} \lambda / \nu$, where $u_{rms}$ is the root-mean-square of a fluctuating velocity component, $\lambda$ is the Taylor microscale, and $\nu$ is the kinematic viscosity) is $71.6$ at $tU_0/M = 42$, decreasing to $60.6$ at $tU_0/M = 171$. In a reference frame moving with the average flow velocity, the problem is modeled as freely decaying isotropic turbulence. We simulate this by considering the flow inside a cubic domain with periodic boundary conditions, where the box edge length is $9 \times 2\pi$~cm ($\approx 11M$). Two grid resolutions are employed, with 32 and 64 cells in each direction, respectively. The corresponding grid spacings are $\Delta/\eta = 60$ and $30$, where $\eta$ is the Kolmogorov length scale at $tU_0/M = 42$. The computations are initialized with a synthetic turbulent field whose energy spectrum matches the filtered experimental spectrum at the initial time $t U_0/M = 42$. The filtering is done either with a spectral cutoff filter or a physical box filter. When the spectral cutoff filter is applied, the initial field is directly generated on the LES grids using an open-source tool~\cite{saad2017scalable}. For the physical box filter, a field is first generated on a grid of $256^3$, then it is filtered onto the LES grids. 

\begin{figure}[htbp]
   \centering
   \begin{subfigure}{0.49\textwidth}
       \centering
       \includegraphics[width=\linewidth]{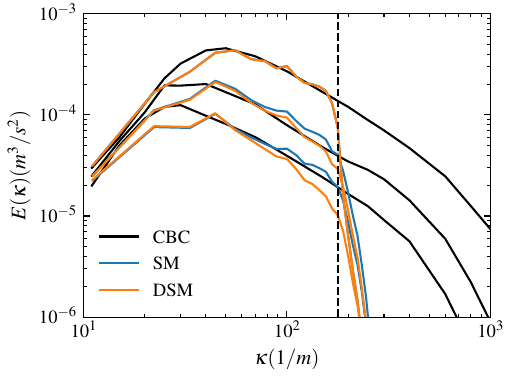}
       \caption{}
       \label{fig:sub1}
   \end{subfigure}
   \hfill
   \begin{subfigure}{0.49\textwidth}
       \centering
       \includegraphics[width=\linewidth]{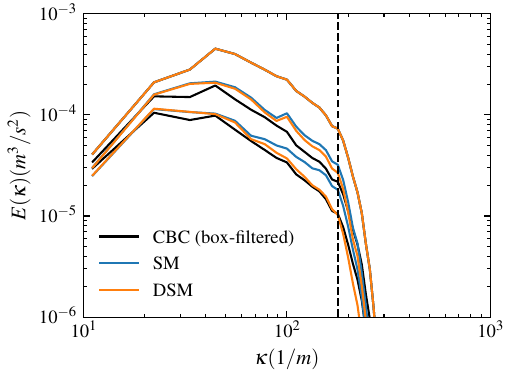}
       \caption{}
       \label{fig:sub2}
   \end{subfigure}    
   \vfill
   \begin{subfigure}{0.49\textwidth}
   \centering
   \includegraphics[width=\linewidth]{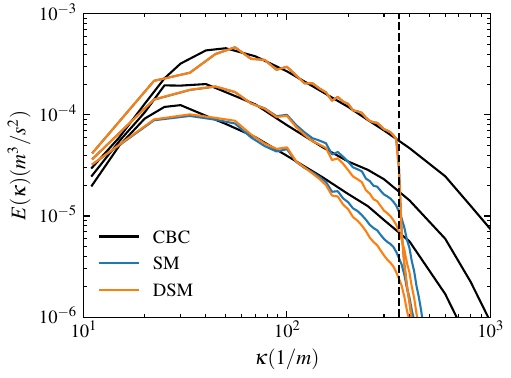}
   \caption{}
   \label{fig:sub2}
   \end{subfigure}
   \hfill
   \begin{subfigure}{0.49\textwidth}
   \centering
   \includegraphics[width=\linewidth]{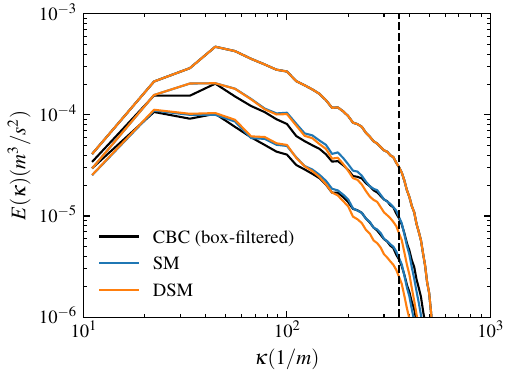}
   \caption{}
   \label{fig:sub2}
   \end{subfigure}
   \caption{Decaying isotropic turbulence: computed velocity spectra with $32^3$ (a,b) and $64^3$ (c,d) grid points. The experimental spectra~\cite{comte1971simple} are shown unfiltered (a, c) and box-filtered (b, d). In each plot, the results for three time instants $tU_0/M = 42$, $98$, and $171$ are displayed from top to bottom. The vertical dashed lines denote the Nyquist limits.}
   \label{fig:dit}
\end{figure}

Figure~\ref{fig:dit} compares the computed energy spectra with experimental results at three time instants, namely $t U_0/M = 42$, $98$ and $171$. The classical Smagorinsky model (SM) with $C_s = 0.18$ accurately predicts the results on both meshes, regardless of the filter used. The only exception is the coarse grid when box filter is applied, in which case the model does not yield sufficient subgrid-scale dissipation. Our tests show that a larger model constant, $C_s = 0.22$, yields good agreement with the filtered experimental spectra (not shown in figure~\ref{fig:dit}). The dependence of the model constant on the grid resolution is a common drawback of static SGS models~\citep{cocle2009scale, meneveau1997dynamic}.
In contrast, the dynamic Smagorinsky model (DSM) does not rely on model constants and reasonably agrees with experimental data. Notably, the computed velocity spectra exhibit near-perfect agreement with experimental results when the box filter is applied, which aligns with the implementation of DSM, where the test filter is a box filter. Larger deviations are generally observed at the scales close to the Nyquist limits, likely due to the numerical errors associated with finite-differencing at high wavenumbers~\citep{nicoud2011using}.

\clearpage
\subsection{Wall-resolved turbulent plane channel flow}

Fully developed channel flow is simulated in a domain of $(L_x, L_y, L_z) = (12.8h, 2.0h, 4.8h)$, with $h$ the channel half-height. No-slip boundary conditions are imposed in the $y$ direction and periodic boundary conditions are applied to the $x$ and $z$ directions. The bulk velocity is maintained constant in time using a time-varying, spatially uniform body force. The bulk Reynolds number is defined as $\Rey_b = 2u_bh/\nu$, with $u_b$ the bulk velocity and $\nu$ the kinematic viscosity. The friction Reynolds number is defined as $\Rey_{\tau} = u_{\tau} h / \nu$, where $u_{\tau}=(\tau_w/\rho)^{1/2}$ is the friction velocity, and the friction coefficient is defined as $C_f = \tau_w / (\rho u_b^2 / 2)$. The bulk Reynolds number is $20,000$ and reference DNS data~\citep{lee2015direct} have $\Rey_{\tau,DNS} = 543.5$ and $C_{f,DNS} = 0.00591$. Table~\ref{tab:WRLES} reports the computational parameters for the wall-resolved LES cases, where $N_x$, $N_y$, and $N_z$ denote the number of grid points in each direction, and $\Delta x$, $\Delta y$, and $\Delta z$ represent the corresponding grid spacings. Specifically, $\Delta y_c$ denotes the grid spacing at the channel centerline, and $\Delta y_w$ is the height of the first off-wall layer. The superscript ``+'' indicates normalization by DNS wall units, whereas superscript ``*'' used below indicates normalization by the wall units of the simulation. The symbols ``SM'' and ``DSM'' denote use of the classical Smagorinsky model with the van Driest damping function and of the dynamic Smagorinsky model, respectively. \#ETT denotes the time-averaging interval, expressed in terms of the eddy turnover time $h/u_{\tau,DNS}$. Four sets of grids are used to assess grid convergence, with $\Delta x^+$ decreasing from approximately $40$ to $10$ while maintaining an aspect ratio of $AR = \Delta x/\Delta z = 1.8$. At the channel centerline, the wall-normal grid spacing is approximately equal to the spanwise spacing. Table~\ref{tab:WRLES} also includes the skin friction coefficient, with its standard uncertainty estimated using a modified batch means method~\cite{russo2017fast}. Its relative error is determined as
\begin{equation}\label{eq:eps_f}
   \epsilon_f = \frac{C_f - C_{f,DNS}}{C_{f,DNS}}.
\end{equation}
The results show that the SM model becomes increasingly accurate as the grid is refined, although it shows approximately $10\%$ error on the coarsest grid. In contrast, the DSM model consistently exhibits errors of less than approximately $3\%$ across all the grid resolutions.

\begin{table}[htbp]
 \centering
 \caption{Computational parameters for WRLES of channel flow.} \label{tab:WRLES}
 \small
 \begin{tabular}{ccccccccccc}
   \hline
   Mesh ($N_x \times N_y \times N_z$) & $\Delta x^+$ & $\Delta z^+$ & $\Delta y_c^+$ & $\Delta y_w^+$ & $AR$ & SGS & $\Rey_\tau$ & $C_f$   & $\epsilon_f$  & \#ETT \\
   $192  \times 128  \times 128 $ &   36.2 &   20.4 &   21.5 &   0.59 & 1.8 & SM     & 568.4     & $0.00646 \pm 0.14\%$ &   9.38\% & 32.6 \\
   $288  \times 192  \times 192 $ &   24.2 &   13.6 &   14.3 &   0.39 & 1.8 & SM     & 567.3     & $0.00644 \pm 0.17\%$ &   8.95\% & 32.6 \\
   $384  \times 256  \times 256 $ &   18.1 &   10.2 &   10.8 &   0.29 & 1.8 & SM     & 556.8     & $0.00620 \pm 0.16\%$ &   4.95\% & 32.6 \\
   $576  \times 384  \times 384 $ &   12.1 &    6.8 &    7.2 &   0.19 & 1.8 & SM     & 544.8     & $0.00594 \pm 0.10\%$ &   0.50\% & 32.6 \\
   $192  \times 128  \times 128 $ &   36.2 &   20.4 &   21.5 &   0.59 & 1.8 & DSM    & 537.3     & $0.00577 \pm 0.13\%$ &  -2.25\% & 32.6 \\
   $288  \times 192  \times 192 $ &   24.2 &   13.6 &   14.3 &   0.39 & 1.8 & DSM    & 541.6     & $0.00587 \pm 0.13\%$ &  -0.69\% & 32.6 \\
   $384  \times 256  \times 256 $ &   18.1 &   10.2 &   10.8 &   0.29 & 1.8 & DSM    & 539.4     & $0.00582 \pm 0.16\%$ &  -1.49\% & 32.6 \\
   $576  \times 384  \times 384 $ &   12.1 &    6.8 &    7.2 &   0.19 & 1.8 & DSM    & 534.7     & $0.00572 \pm 0.17\%$ &  -3.22\% & 32.6 \\
   \hline
 \end{tabular}
\end{table}

\begin{figure}[htbp]
   \centering
   \begin{subfigure}{0.56\textwidth}
   \centering
   \includegraphics[width=\linewidth]{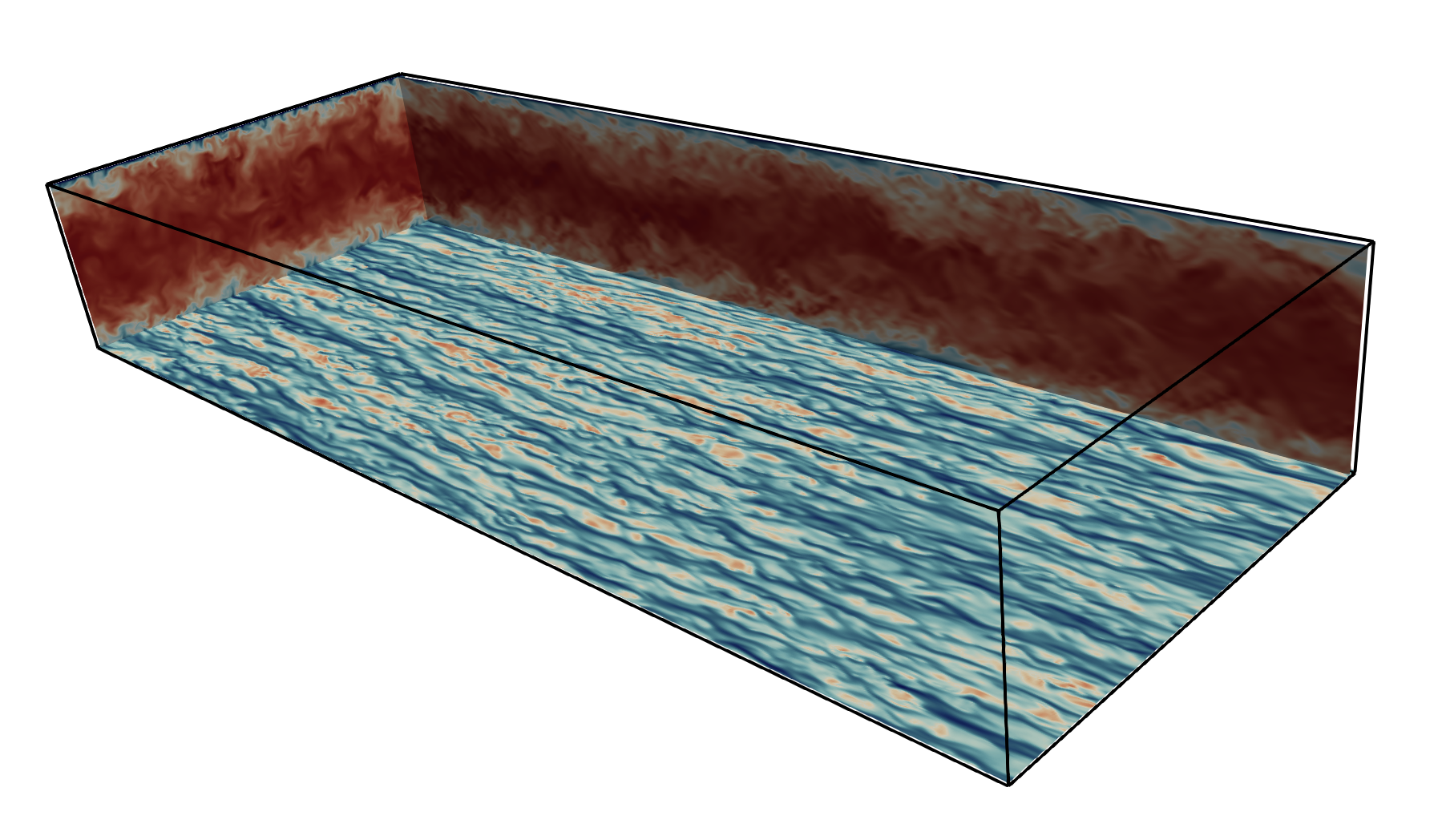}
   \caption{}
   \label{fig:sub2}
   \end{subfigure}
   \hfill
   \begin{subfigure}{0.43\textwidth}
   \centering
   \includegraphics[width=\linewidth]{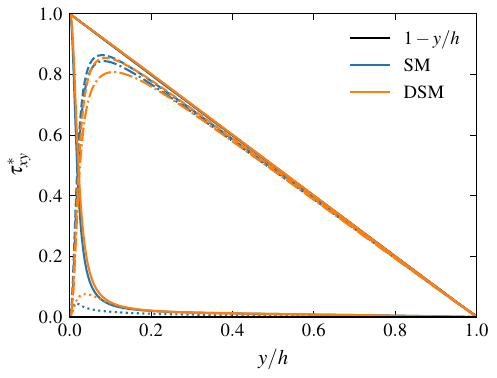}
   \caption{}
   \label{fig:sub2}
   \end{subfigure}
   \caption{Turbulent channel flow: visualization of streamwise velocity obtained with WRLES using the SM model on a mesh with $\Delta_z^+ = 6.8$ (a) and profiles of mean total (solid), resolved and modeled turbulent (dashed), resolved turbulent (dash-dotted), modeled turbulent (dotted), and viscous (solid) shear stress obtained with WRLES using the SM and DSM models on a grid with $\Delta_z^+ = 20.4$ (b). In (a), the wall-parallel plane located at $y^+ \approx 15$. In (b), normalization is based on the wall units of each simulation.}
   \label{fig:channel_wrles}
\end{figure}

Figure~\ref{fig:channel_wrles}(a) shows a visualization of the flow computed on the finest grid, where the near-wall low- and high-speed streaks are clearly observed. To verify the LES implementation, figure~\ref{fig:channel_wrles}(b) presents the profiles of the various contributions to the total shear stress for the grid with $\Delta z^+ = 20.4$. Achievement of a linear distribution of the total shear stress provides evidence of general reliability of the LES implementation. Overall, the DSM model generates higher levels of modeled stress along with a lower peak of the resolved shear stress, as compared to the SM model with the van Driest damping function, because the DSM yields higher eddy viscosity levels.

\begin{figure}[htbp]
   \centering
   \begin{subfigure}{0.49\textwidth}
       \centering
       \includegraphics[width=\linewidth]{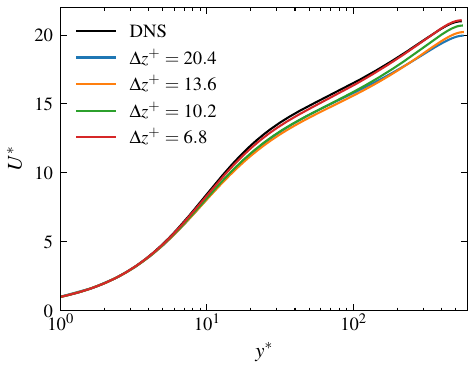}
       \caption{}
       \label{fig:sub1}
   \end{subfigure}
   \hfill
   \begin{subfigure}{0.49\textwidth}
       \centering
       \includegraphics[width=\linewidth]{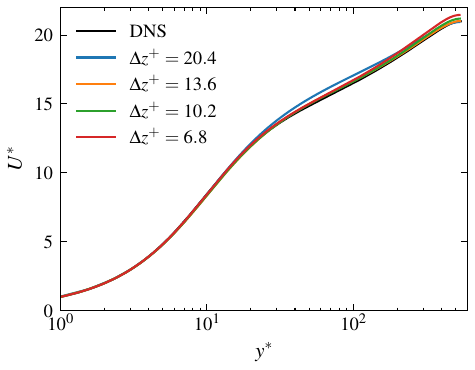}
       \caption{}
       \label{fig:sub2}
   \end{subfigure}    
   \caption{Turbulent channel flow: profiles of mean streamwise velocity obtained with WRLES using the SM (a) and DSM (b) models. Normalization is based on wall units of each simulation. The DNS data is from~\cite{lee2015direct}.}
   \label{fig:wrles-u}
\end{figure}

\begin{figure}[htbp]
   \centering
   \begin{subfigure}{0.49\textwidth}
   \centering
   \includegraphics[width=\linewidth]{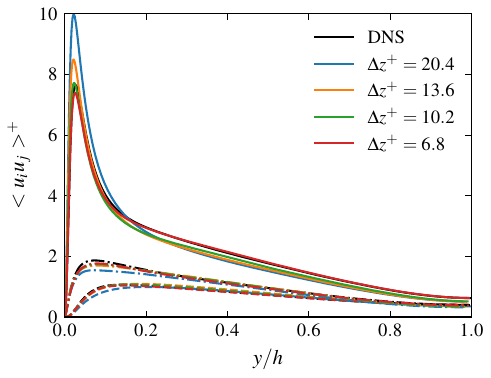}
   \caption{}
   \label{fig:sub2}
   \end{subfigure}
   \hfill
   \begin{subfigure}{0.49\textwidth}
   \centering
   \includegraphics[width=\linewidth]{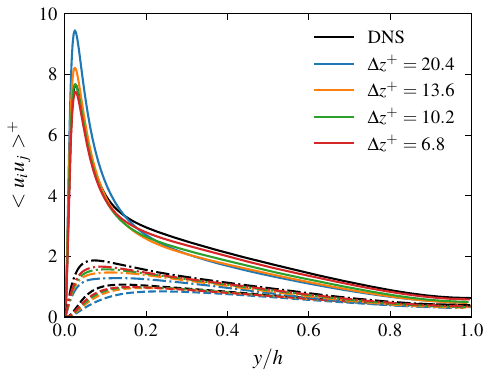}
   \caption{}
   \label{fig:sub2}
   \end{subfigure}
   \vfill
   \begin{subfigure}{0.49\textwidth}
   \centering
   \includegraphics[width=\linewidth]{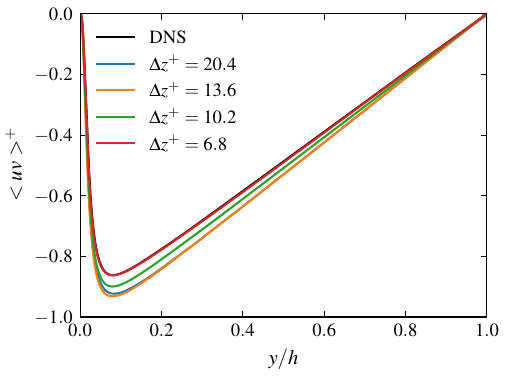}
   \caption{}
   \label{fig:sub2}
   \end{subfigure}
   \hfill
   \begin{subfigure}{0.49\textwidth}
   \centering
   \includegraphics[width=\linewidth]{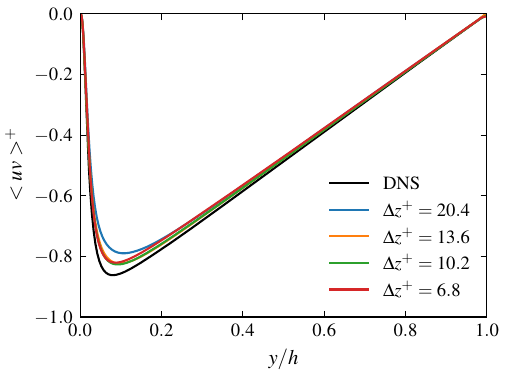}
   \caption{}
   \label{fig:sub2}
   \end{subfigure}
   \caption{Turbulent channel flow: profiles of resolved turbulent normal stress (a,b) and shear stress (c,d) obtained with WRLES using the SM (a,c) and DSM (b,d) models. Normalization is based on the DNS wall units. The DNS data is from~\cite{lee2015direct}. Line codes: $\langle uu\rangle$ (solid), $\langle vv\rangle$ (dashed), $\langle ww\rangle$ (dash-dotted).}
   \label{fig:wrles-rey}
\end{figure}

Figure~\ref{fig:wrles-u} presents the mean streamwise velocity profiles obtained with the SM and DSM models. The velocity profiles normalized by the wall units of each simulation exhibit sensitivity to grid refinement, particularly with the SM model. However, although not shown, when normalized by the wall units of DNS, the velocity profiles show little dependence on grid refinement, with all four grid resolutions closely matching the DNS data. Consequently, the grid sensitivity of the velocity profiles, when normalized by the wall units of each simulation, primarily stems from the grid dependence of the skin friction, particularly when the SM model is applied (see table~\ref{tab:WRLES}). Figure~\ref{fig:wrles-rey} shows the turbulent normal and shear stresses, normalized by the DNS wall units, disregarding the modeled stress. When the SM model is applied, the resolved shear stress is over-predicted, whereas it is under-predicted with the DSM model. This behavior aligns with the corresponding over- and under-estimation of the friction coefficient by the two models. As for the normal stress, the peak of the streamwise component is over-estimated, as also observed in other studies~\cite{vreman2004eddy}. Overall, grid refinement leads to convergence of all the profiles towards the DNS data.

\clearpage
\subsection{Wall-modeled turbulent plane channel flow}

\begin{figure}[htbp]
   \centering
   \begin{subfigure}{0.56\textwidth}
   \centering
   \includegraphics[width=\linewidth]{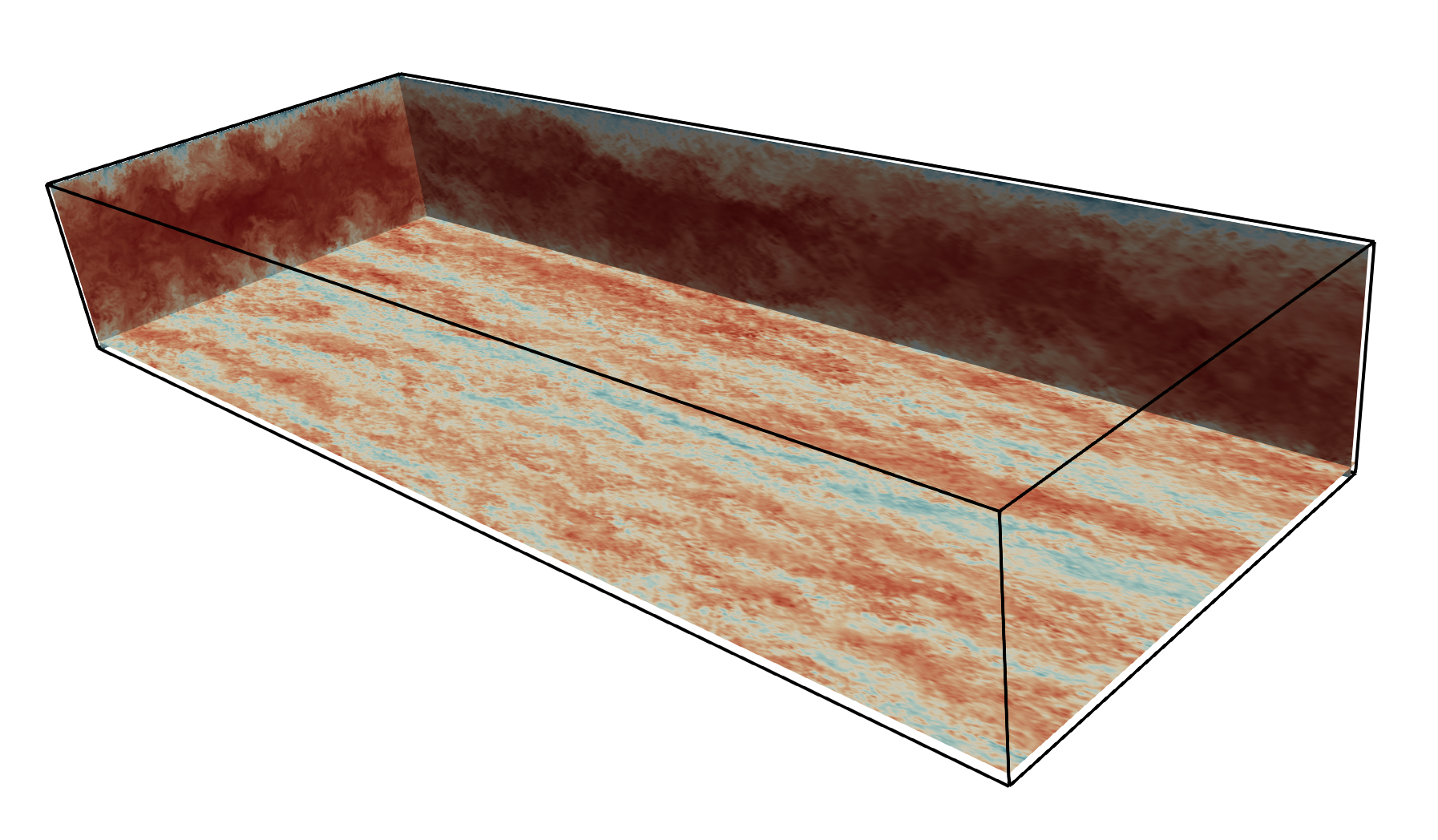}
   \caption{}
   \label{fig:sub2}
   \end{subfigure}
   \hfill
   \begin{subfigure}{0.43\textwidth}
   \centering
   \includegraphics[width=\linewidth]{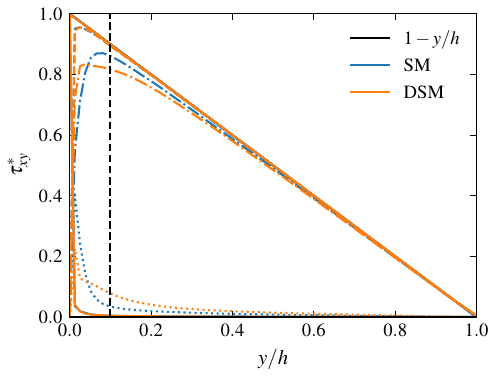}
   \caption{}
   \label{fig:sub2}
   \end{subfigure}
   \caption{Turbulent channel flow: visualization of streamwise velocity obtained with WMLES using the SM model on a mesh with $\Delta_z/h = 0.012$ (a) and profiles of mean total (solid), resolved and modeled turbulent (dashed), resolved turbulent (dash-dotted), modeled turbulent (dotted) and viscous (solid) shear stress computed with WMLES using the SM and DSM models, for the grid with $\Delta_z/h = 0.050$ and $AR=1.0$ (b). In (a), the wall-parallel plane is located at $y/h = 0.1$. In (b), the dashed line denotes $y/h=0.1$. Normalization is based on the wall units of each simulation.}
   \label{fig:channel_wmles}
\end{figure}

The computational setup is identical to that of the previous WRLES case, except that the bulk Reynolds number is $250,000$ and that wall-model boundary conditions are imposed on the walls. The reference data~\citep{lee2015direct} have $\Rey_{\tau, \mathrm{DNS}} = 5185.9$ and $C_{f, \mathrm{DNS}} = 0.00344$. Table~\ref{tab:WMLES} provides the computational parameters for the WMLES cases. The parameter definitions are the same as in table~\ref{tab:WRLES}, but the cell spacings are given in terms of the half-channel height. We use aspect ratios of $AR = 1.0$ and $AR = 2.0$, with thirteen grids generated through uniform refinement in all three directions for the grid convergence study. The finest grid has approximately $\Delta z^+ = 10$, which qualifies it as a WRLES case. For all the cases, the wall-modeled layer thickness is $h_{wm}=0.1h$. Figure~\ref{fig:channel_wmles}(a) visualizes the turbulent channel flow obtained with WMLES using the SM model on a mesh with $\Delta_z/h = 0.012$. The low- and high-speed streaks are clearly visible on the plane at $y/h = 0.1$. Figure~\ref{fig:channel_wmles}(b) presents the profiles of various contributions to the total shear stress for the grid with $\Delta_z/h = 0.050$. The linear distribution of total shear stress is predicted accurately, demonstrating the correct implementation of WMLES. Overall, DSM yields higher levels of modeled stress, with a lower peak in the resolved shear stress, compared to the SM model. This aligns with the WRLES results (figure~\ref{fig:channel_wrles}(b)).

\begin{table}[htbp]
 \centering
 \caption{Computational parameters for WMLES of turbulent channel flow.}\label{tab:WMLES}
 \small
 \begin{tabular}{ccccccccccc}
   \hline
   Mesh ($N_x \times N_y \times N_z$) & $\Delta x/h$ & $\Delta z/h$ & $\Delta y_c/h$ & $\Delta y_w/h$ & $AR$ & SGS & $\Rey_\tau$ & $C_f$   & $\epsilon_f$      & \#ETT \\
$128  \times 32   \times 48  $ & 0.100  & 0.100  & 0.100  & 0.0252 & 1.0 & SM     & 5165.0    & $0.00341 \pm 0.10\%$ &  -0.81\% & 20.7 \\
$192  \times 48   \times 72  $ & 0.067  & 0.067  & 0.067  & 0.0167 & 1.0 & SM     & 5155.4    & $0.00340 \pm 0.16\%$ &  -1.17\% & 20.7 \\
$256  \times 64   \times 96  $ & 0.050  & 0.050  & 0.050  & 0.0125 & 1.0 & SM     & 5178.9    & $0.00343 \pm 0.17\%$ &  -0.27\% & 20.7 \\
$384  \times 96   \times 144 $ & 0.033  & 0.033  & 0.033  & 0.0083 & 1.0 & SM     & 5147.3    & $0.00339 \pm 0.19\%$ &  -1.48\% & 20.7 \\
$512  \times 128  \times 192 $ & 0.025  & 0.025  & 0.025  & 0.0063 & 1.0 & SM     & 5115.7    & $0.00335 \pm 0.06\%$ &  -2.69\% & 20.7 \\
$640  \times 160  \times 240 $ & 0.020  & 0.020  & 0.020  & 0.0050 & 1.0 & SM     & 5101.7    & $0.00333 \pm 0.27\%$ &  -3.22\% & 20.7 \\
$768  \times 192  \times 288 $ & 0.017  & 0.017  & 0.017  & 0.0042 & 1.0 & SM     & 5097.0    & $0.00333 \pm 0.22\%$ &  -3.40\% & 20.7 \\
$896  \times 224  \times 336 $ & 0.014  & 0.014  & 0.014  & 0.0036 & 1.0 & SM     & 5109.9    & $0.00334 \pm 0.23\%$ &  -2.91\% & 20.7 \\
$1024 \times 256  \times 384 $ & 0.013  & 0.012  & 0.012  & 0.0031 & 1.0 & SM     & 5121.6    & $0.00336 \pm 0.24\%$ &  -2.46\% & 20.7 \\
$1536 \times 256  \times 576 $ & 0.008  & 0.008  & 0.012  & 0.0031 & 1.0 & SM     & 5181.7    & $0.00344 \pm 0.45\%$ &  -0.16\% & 20.7 \\
$2048 \times 512  \times 768 $ & 0.006  & 0.006  & 0.006  & 0.0016 & 1.0 & SM     & 5248.3    & $0.00353 \pm 0.64\%$ &   2.42\% & 20.7 \\
$4096 \times 1024 \times 1536$ & 0.003  & 0.003  & 0.003  & 0.0008 & 1.0 & SM     & 5249.3    & $0.00353 \pm 0.60\%$ &   2.46\% & 20.7 \\
$6144 \times 1536 \times 2304$ & 0.002  & 0.002  & 0.002  & 0.0005 & 1.0 & SM     & 5248.4    & $0.00353 \pm 0.56\%$ &   2.43\% & 20.7 \\
$128  \times 32   \times 48  $ & 0.100  & 0.100  & 0.100  & 0.0252 & 1.0 & DSM    & 5229.5    & $0.00350 \pm 0.27\%$ &   1.69\% & 20.7 \\
$192  \times 48   \times 72  $ & 0.067  & 0.067  & 0.067  & 0.0167 & 1.0 & DSM    & 5235.3    & $0.00351 \pm 0.26\%$ &   1.91\% & 20.7 \\
$256  \times 64   \times 96  $ & 0.050  & 0.050  & 0.050  & 0.0125 & 1.0 & DSM    & 5243.1    & $0.00352 \pm 0.08\%$ &   2.22\% & 20.7 \\
$384  \times 96   \times 144 $ & 0.033  & 0.033  & 0.033  & 0.0083 & 1.0 & DSM    & 5217.6    & $0.00348 \pm 0.20\%$ &   1.23\% & 20.7 \\
$512  \times 128  \times 192 $ & 0.025  & 0.025  & 0.025  & 0.0063 & 1.0 & DSM    & 5181.1    & $0.00344 \pm 0.09\%$ &  -0.19\% & 20.7 \\
$640  \times 160  \times 240 $ & 0.020  & 0.020  & 0.020  & 0.0050 & 1.0 & DSM    & 5163.1    & $0.00341 \pm 0.15\%$ &  -0.88\% & 20.7 \\
$768  \times 192  \times 288 $ & 0.017  & 0.017  & 0.017  & 0.0042 & 1.0 & DSM    & 5151.3    & $0.00340 \pm 0.31\%$ &  -1.33\% & 20.7 \\
$896  \times 224  \times 336 $ & 0.014  & 0.014  & 0.014  & 0.0036 & 1.0 & DSM    & 5150.5    & $0.00340 \pm 0.16\%$ &  -1.36\% & 20.7 \\
$1024 \times 256  \times 384 $ & 0.013  & 0.012  & 0.012  & 0.0031 & 1.0 & DSM    & 5161.8    & $0.00341 \pm 0.29\%$ &  -0.93\% & 20.7 \\
$1536 \times 256  \times 576 $ & 0.008  & 0.008  & 0.012  & 0.0031 & 1.0 & DSM    & 5179.5    & $0.00343 \pm 0.15\%$ &  -0.25\% & 20.7 \\
$2048 \times 512  \times 768 $ & 0.006  & 0.006  & 0.006  & 0.0016 & 1.0 & DSM    & 5207.5    & $0.00347 \pm 0.73\%$ &   0.83\% & 20.7 \\
$4096 \times 1024 \times 1536$ & 0.003  & 0.003  & 0.003  & 0.0008 & 1.0 & DSM    & 5236.4    & $0.00351 \pm 0.25\%$ &   1.96\% & 20.7 \\
$6144 \times 1536 \times 2304$ & 0.002  & 0.002  & 0.002  & 0.0005 & 1.0 & DSM    & 5197.2    & $0.00346 \pm 0.43\%$ &   0.44\% & 20.7 \\
$64   \times 32   \times 48  $ & 0.200  & 0.100  & 0.100  & 0.0252 & 2.0 & SM     & 5227.3    & $0.00350 \pm 0.22\%$ &   1.60\% & 20.7 \\
$96   \times 48   \times 72  $ & 0.133  & 0.067  & 0.067  & 0.0167 & 2.0 & SM     & 5335.4    & $0.00364 \pm 0.09\%$ &   5.85\% & 20.7 \\
$128  \times 64   \times 96  $ & 0.100  & 0.050  & 0.050  & 0.0125 & 2.0 & SM     & 5346.2    & $0.00366 \pm 0.09\%$ &   6.28\% & 20.7 \\
$192  \times 96   \times 144 $ & 0.067  & 0.033  & 0.033  & 0.0083 & 2.0 & SM     & 5325.2    & $0.00363 \pm 0.12\%$ &   5.44\% & 20.7 \\
$256  \times 128  \times 192 $ & 0.050  & 0.025  & 0.025  & 0.0063 & 2.0 & SM     & 5264.4    & $0.00355 \pm 0.09\%$ &   3.05\% & 20.7 \\
$320  \times 160  \times 240 $ & 0.040  & 0.020  & 0.020  & 0.0050 & 2.0 & SM     & 5208.7    & $0.00347 \pm 0.06\%$ &   0.88\% & 20.7 \\
$384  \times 192  \times 288 $ & 0.033  & 0.017  & 0.017  & 0.0042 & 2.0 & SM     & 5153.5    & $0.00340 \pm 0.16\%$ &  -1.24\% & 20.7 \\
$448  \times 224  \times 336 $ & 0.029  & 0.014  & 0.014  & 0.0036 & 2.0 & SM     & 5113.3    & $0.00335 \pm 0.18\%$ &  -2.78\% & 20.7 \\
$512  \times 256  \times 384 $ & 0.025  & 0.012  & 0.012  & 0.0031 & 2.0 & SM     & 5075.8    & $0.00330 \pm 0.18\%$ &  -4.20\% & 20.7 \\
$768  \times 384  \times 576 $ & 0.017  & 0.008  & 0.008  & 0.0021 & 2.0 & SM     & 5057.8    & $0.00327 \pm 0.37\%$ &  -4.88\% & 20.7 \\
$1024 \times 512  \times 768 $ & 0.013  & 0.006  & 0.006  & 0.0016 & 2.0 & SM     & 5165.5    & $0.00342 \pm 2.07\%$ &  -0.79\% & 20.7 \\
$2048 \times 1024 \times 1536$ & 0.006  & 0.003  & 0.003  & 0.0008 & 2.0 & SM     & 5196.9    & $0.00346 \pm 1.12\%$ &   0.42\% & 20.7 \\
$3072 \times 1536 \times 2304$ & 0.004  & 0.002  & 0.002  & 0.0005 & 2.0 & SM     & 5192.6    & $0.00345 \pm 0.53\%$ &   0.26\% & 20.7 \\
$64   \times 32   \times 48  $ & 0.200  & 0.100  & 0.100  & 0.0252 & 2.0 & DSM    & 5360.1    & $0.00368 \pm 0.12\%$ &   6.83\% & 20.7 \\
$96   \times 48   \times 72  $ & 0.133  & 0.067  & 0.067  & 0.0167 & 2.0 & DSM    & 5356.5    & $0.00367 \pm 0.08\%$ &   6.69\% & 20.7 \\
$128  \times 64   \times 96  $ & 0.100  & 0.050  & 0.050  & 0.0125 & 2.0 & DSM    & 5355.5    & $0.00367 \pm 0.07\%$ &   6.65\% & 20.7 \\
$192  \times 96   \times 144 $ & 0.067  & 0.033  & 0.033  & 0.0083 & 2.0 & DSM    & 5338.3    & $0.00365 \pm 0.08\%$ &   5.96\% & 20.7 \\
$256  \times 128  \times 192 $ & 0.050  & 0.025  & 0.025  & 0.0063 & 2.0 & DSM    & 5294.1    & $0.00359 \pm 0.09\%$ &   4.22\% & 20.7 \\
$320  \times 160  \times 240 $ & 0.040  & 0.020  & 0.020  & 0.0050 & 2.0 & DSM    & 5244.4    & $0.00352 \pm 0.04\%$ &   2.27\% & 20.7 \\
$384  \times 192  \times 288 $ & 0.033  & 0.017  & 0.017  & 0.0042 & 2.0 & DSM    & 5192.8    & $0.00345 \pm 0.10\%$ &   0.27\% & 20.7 \\
$448  \times 224  \times 336 $ & 0.029  & 0.014  & 0.014  & 0.0036 & 2.0 & DSM    & 5142.5    & $0.00339 \pm 0.13\%$ &  -1.67\% & 20.7 \\
$512  \times 256  \times 384 $ & 0.025  & 0.012  & 0.012  & 0.0031 & 2.0 & DSM    & 5111.4    & $0.00334 \pm 0.21\%$ &  -2.85\% & 20.7 \\
$768  \times 384  \times 576 $ & 0.017  & 0.008  & 0.008  & 0.0021 & 2.0 & DSM    & 5066.6    & $0.00329 \pm 0.69\%$ &  -4.55\% & 20.7 \\
$1024 \times 512  \times 768 $ & 0.013  & 0.006  & 0.006  & 0.0016 & 2.0 & DSM    & 5141.1    & $0.00338 \pm 0.90\%$ &  -1.72\% & 20.7 \\
$2048 \times 1024 \times 1536$ & 0.006  & 0.003  & 0.003  & 0.0008 & 2.0 & DSM    & 5205.0    & $0.00347 \pm 0.72\%$ &   0.74\% & 20.7 \\
$3072 \times 1536 \times 2304$ & 0.004  & 0.002  & 0.002  & 0.0005 & 2.0 & DSM    & 5219.5    & $0.00349 \pm 0.53\%$ &   1.30\% & 20.7 \\
   \hline
 \end{tabular}
\end{table}

\begin{figure}[htbp]
   \centering
   \begin{subfigure}{0.49\textwidth}
       \centering
       \includegraphics[width=\linewidth]{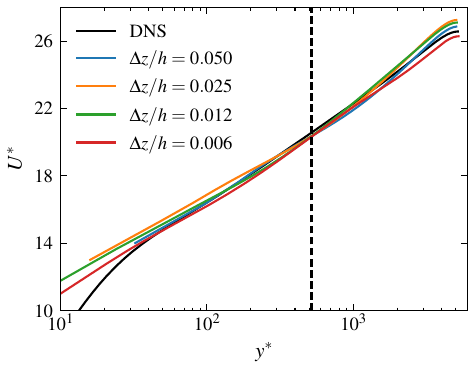}
       \caption{}
       \label{fig:sub1}
   \end{subfigure}
   \hfill
   \begin{subfigure}{0.49\textwidth}
       \centering
       \includegraphics[width=\linewidth]{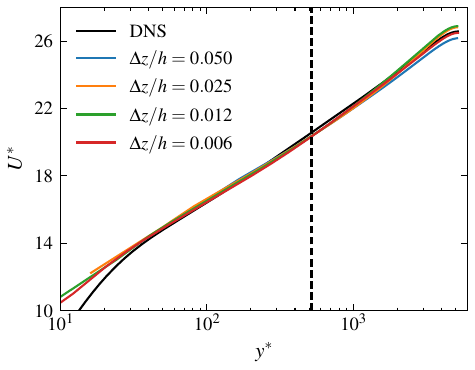}
       \caption{}
       \label{fig:sub2}
   \end{subfigure}
   \vfill
   \begin{subfigure}{0.49\textwidth}
       \centering
       \includegraphics[width=\linewidth]{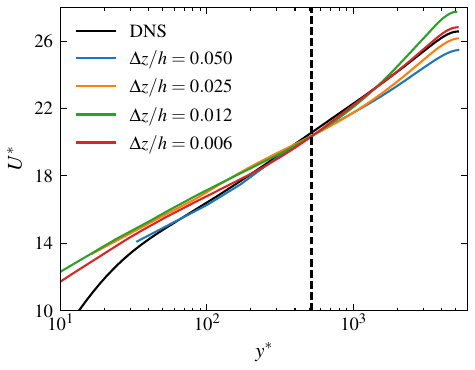}
       \caption{}
       \label{fig:sub1}
   \end{subfigure}
   \hfill
   \begin{subfigure}{0.49\textwidth}
       \centering
       \includegraphics[width=\linewidth]{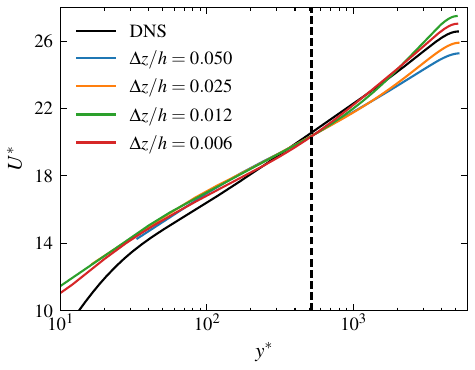}
       \caption{}
       \label{fig:sub2}
   \end{subfigure}    
   \caption{Turbulent channel flow: profiles of mean streamwise velocity obtained with WMLES using the SM (a,c) and DSM (b,d) models on the grids with $AR=1.0$ (a,b) and $AR=2.0$ (c,d). The dashed line denotes $y/h=0.1$. Normalization is based on the wall units of each simulation. The DNS data is from~\cite{lee2015direct}.}
   \label{fig:wmles-u}
\end{figure}

\begin{figure}[htbp]
   \centering
   \begin{subfigure}{0.49\textwidth}
   \centering
   \includegraphics[width=\linewidth]{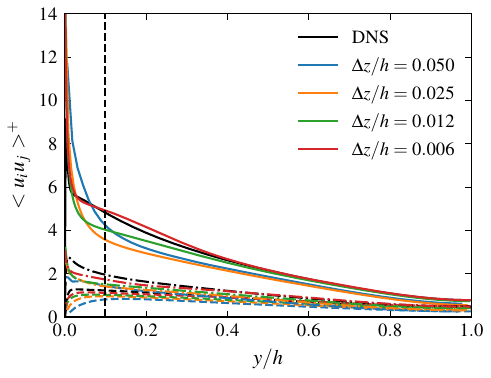}
   \caption{}
   \label{fig:sub2}
   \end{subfigure}
   \hfill
   \begin{subfigure}{0.49\textwidth}
   \centering
   \includegraphics[width=\linewidth]{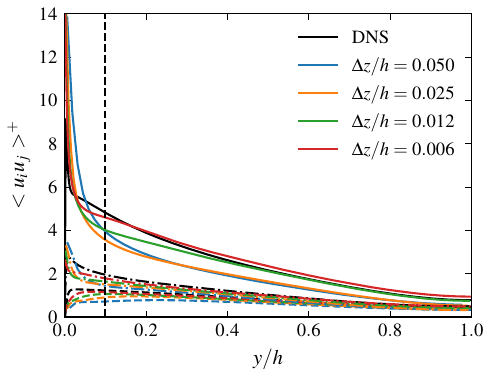}
   \caption{}
   \label{fig:sub2}
   \end{subfigure}
   \vfill
   \begin{subfigure}{0.49\textwidth}
   \centering
   \includegraphics[width=\linewidth]{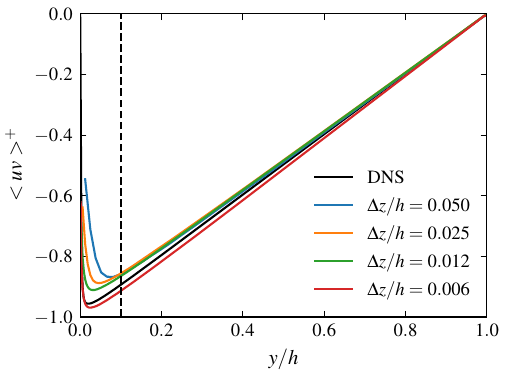}
   \caption{}
   \label{fig:sub2}
   \end{subfigure}
   \hfill
   \begin{subfigure}{0.49\textwidth}
   \centering
   \includegraphics[width=\linewidth]{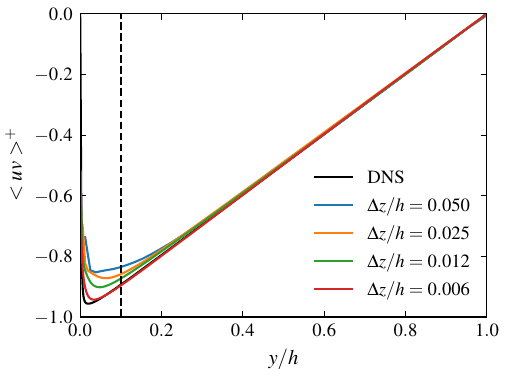}
   \caption{}
   \label{fig:sub2}
   \end{subfigure}
   \caption{Turbulent channel flow: profiles of resolved turbulent normal stress (a,b) and shear stress (c,d) obtained with WMLES using the SM (a,c) and DSM (b,d) models on the grids with $AR=1.0$. The dashed line denotes $y/h=0.1$. Normalization is based on the wall units of the DNS (a,b) and the wall units of each simulation (c,d). Normalization is based on the DNS wall units. The DNS data is from~\cite{lee2015direct}. Line codes: $\langle uu\rangle$ (solid), $\langle vv\rangle$ (dashed), $\langle ww\rangle$ (dash-dotted).}
   \label{fig:wmles-ar1-rey}
\end{figure}

Figure~\ref{fig:wmles-u} presents the computed velocity results on the grids with $AR=1.0$ and $AR=2.0$ and figure~\ref{fig:wmles-ar1-rey} shows the resolved turbulent normal and shear stresses for $AR=1.0$. Profiles on four different grids are displayed to demonstrate grid convergence. For $AR=1.0$, the profiles agree well with the DNS data, although the resolved turbulent stress profiles show noticeable variations with grid refinement. Notably, the streamwise velocity fluctuations are over-predicted, and the resolved turbulent shear stress generally increases as the grid resolution improves. The LES-computed velocity profile in the wall-modeled layer should not be taken seriously, as it is expected to be replaced by the wall model, i.e., the logarithmic law in equation~\eqref{eq:wm}. In contrast, the computed velocity profiles for $AR=2.0$ become more sensitive to grid refinement, although the results on the fine grids exhibit good accuracy.

\begin{figure}[htbp]
   \centering
   \begin{subfigure}{0.49\textwidth}
       \centering
       \includegraphics[width=\linewidth]{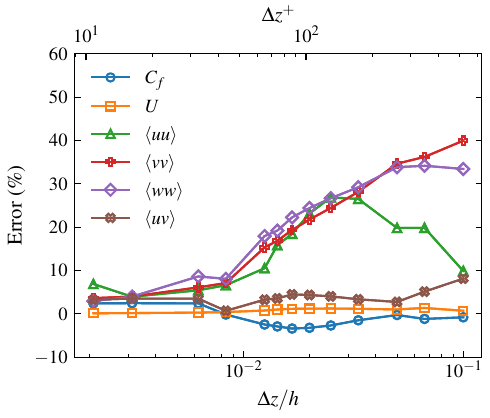}
       \caption{}
       \label{fig:sub1}
   \end{subfigure}
   \hfill
   \begin{subfigure}{0.49\textwidth}
       \centering
       \includegraphics[width=\linewidth]{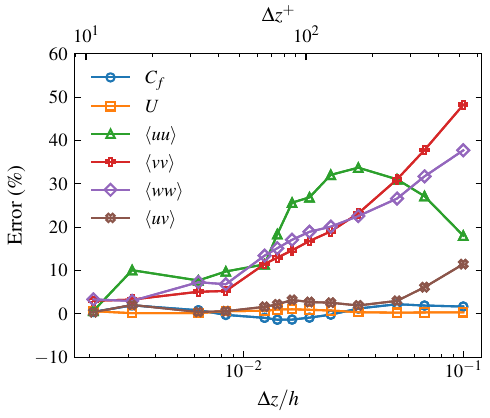}
       \caption{}
       \label{fig:sub2}
   \end{subfigure} 
   \vfill
   \begin{subfigure}{0.49\textwidth}
   \centering
   \includegraphics[width=\linewidth]{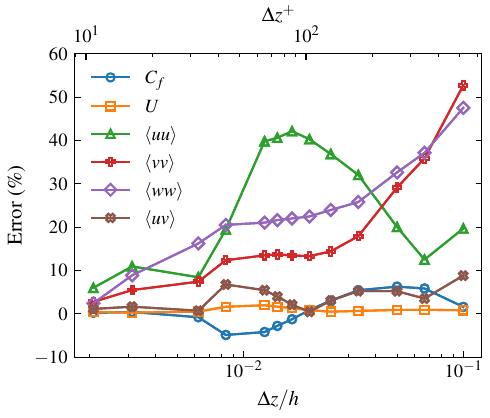}
   \caption{}
   \label{fig:sub2}
   \end{subfigure}
   \hfill
   \begin{subfigure}{0.49\textwidth}
   \centering
   \includegraphics[width=\linewidth]{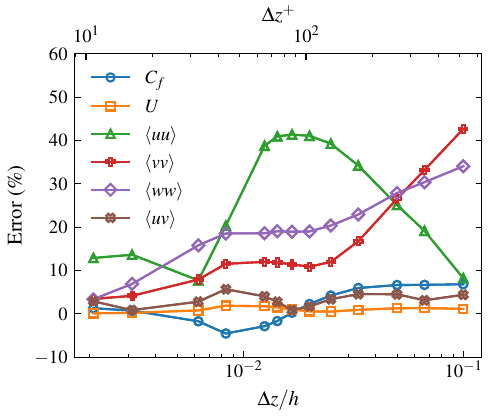}
   \caption{}
   \label{fig:sub2}
   \end{subfigure}
   \caption{Turbulent channel flow: prediction error as a function of $\Delta z/h$ or $\Delta z^+$ for various statistical properties, as obtained from WMLES with SM (a,c) and DSM (b,d) models for grids with $AR=1.0$ (a,b) and $AR=2.0$ (c,d).}
   \label{fig:convergence}
\end{figure}

\clearpage
For grid convergence study, figure~\ref{fig:convergence} plots the variations of prediction errors with $\Delta z/h$ for different quantities, including $C_f$, $U$, $\langle uu \rangle$, $\langle vv \rangle$, $\langle ww \rangle$, and $\langle uv \rangle$. For $C_f$, the relative error is defined in the same way as in equation~\eqref{eq:eps_f}, and for the other quantities, the relative error is defined as
\begin{equation}\label{eq:error} 
\epsilon_\phi = \frac{\left[\int_{y/h=0.1}^{y/h=1.0} \left(\phi-\phi_{ref}\right)^2 d\left(y/h\right)\right]^{1/2}}{\left|\int_{y/h=0.1}^{y/h=1.0} \phi_{ref} d\left(y/h\right)\right|},
\end{equation}
where $\phi$ denotes outer-scaled quantities. Notably, the integration is performed from $y=0.1h$ to $y=h$, excluding the wall-modeled layer, since this layer is contaminated by significant numerical errors and is expected to be replaced by the velocity profile yielded from the wall model. Figure~\ref{fig:convergence} pinpoints non-monotonic grid convergence, for both the SGS models on grids with the two aspect ratios, as also noted by \citet{meyers2007plane} for WRLES. In our results, the streamwise component of the turbulent normal stress exhibits the most marked non-monotonic convergence. Non-monotonic convergence is also observed in the skin friction and turbulent shear stress. On the finest grid with $AR=1.0$, the error in wall friction is $2.43\%$ for the SM model and $0.44\%$ for the DSM model. The errors for the finest grid with $AR=2.0$ are $0.26\%$ and $1.30\%$, respectively. Notably, the sign of the error on the finest grids is always positive. We have also conducted WMLES for cases with $\Rey_b=20,000$. Although the results are not shown here, the signs of the errors for very fine grids are also positive. This is consistent with the fact that mean wall shear stress obtained with WMLES is biased toward larger values when the instantaneous velocity is used as input for the wall model. Additionally, we note that grids with $AR = 1.0$ exhibit better grid convergence compared to those with $AR = 2.0$, as variations with grid refinement become smaller in the errors of wall shear stress and mean velocity profile, and therefore the non-monotonic grid convergence behavior becomes less pronounced.

\clearpage
\subsection{Wall-modeled turbulent square duct flow}

\begin{figure}[htbp]
   \centering
   \begin{subfigure}{0.45\textwidth}
   \centering
   \includegraphics[width=\linewidth]{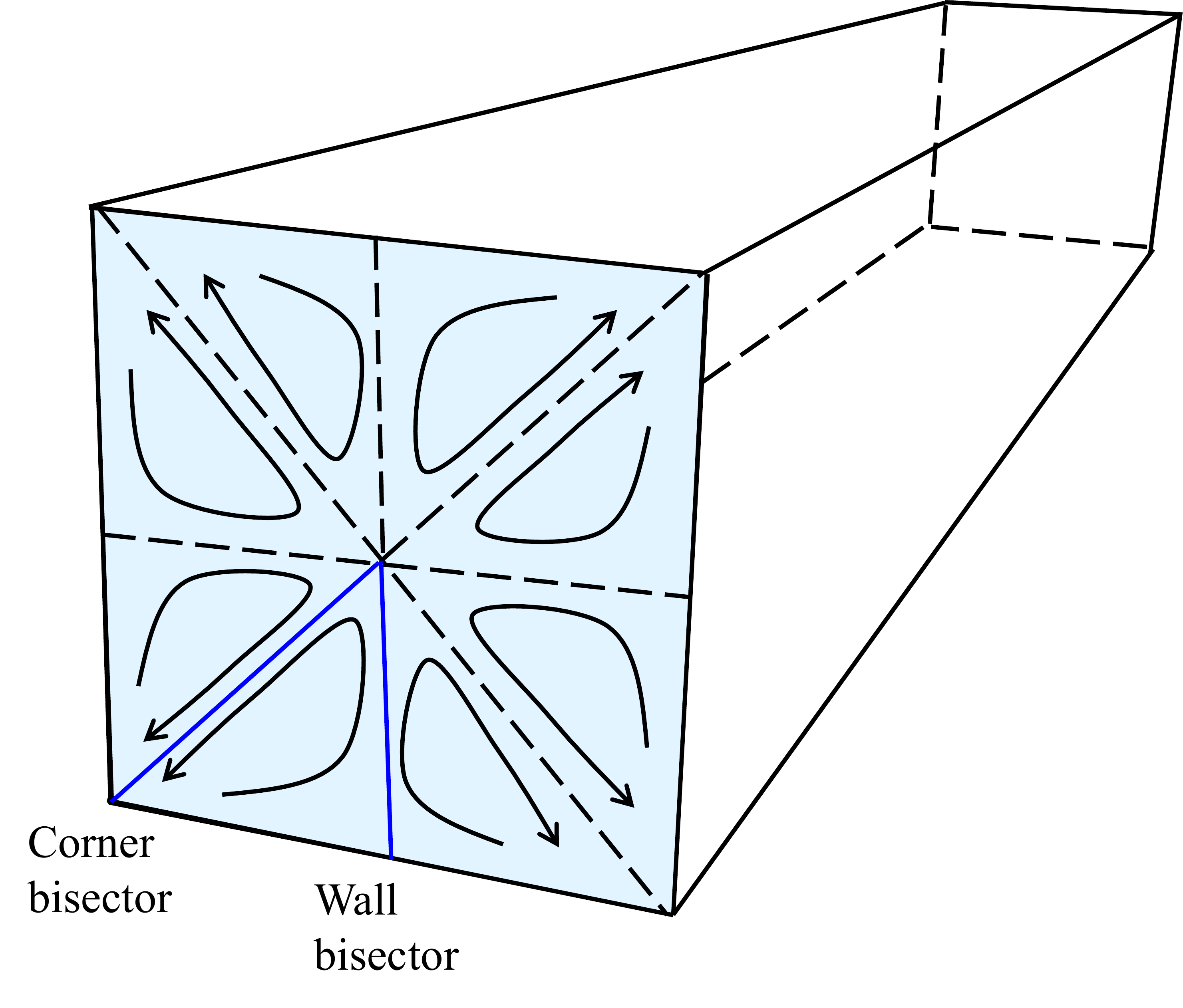}
   \caption{}
   \label{fig:sub2}
   \end{subfigure}
   \hfill
   \begin{subfigure}{0.45\textwidth}
   \centering
   \includegraphics[width=\linewidth]{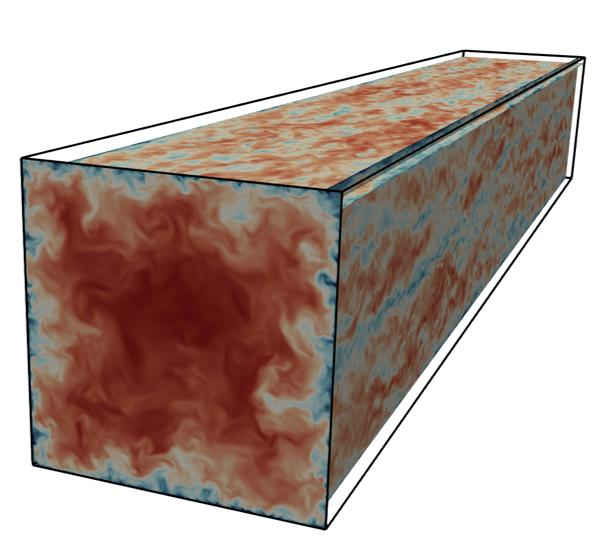}
   \caption{}
   \label{fig:sub2}
   \end{subfigure}
   \caption{Computational setup for the square duct flow (a) and visualization of turbulent flow obtained with WMLES using the SM model on a mesh with $\Delta_z/h = 0.006$ (b). The planes display contours of the streamwise velocity, with the wall-parallel plane located at $y/h = 0.1$.}
   \label{fig:duct}
\end{figure}

Fully developed turbulent duct flow is simulated using WMLES, following the setup in figure~\ref{fig:duct}(a). The computational domain is $(L_x, L_y, L_z) = (12.8h, 2.0h, 2.0h)$, where $h$ is half of the duct side length. Wall-modeled boundary conditions are applied in the $y$ and $z$ directions, whereas periodic boundary conditions are enforced in the streamwise $x$ direction. The simulation maintains the bulk velocity constant by applying a time-varying, spatially uniform body force. The bulk Reynolds number ($\Rey_b = 2u_bh/\nu$) is 40,000. DNS at the same Reynolds number was conducted in~\cite{pirozzoli2018turbulence}, which yielded $\Rey_{\tau,DNS} = 1055$, $C_{f,DNS} = 0.00557$. In our simulations, the wall-modeled layer thickness is set to $h_{wm}=0.1h$. Table~\ref{tab:duct} provides the computational parameters, where $\Rey_\tau = u_\tau h/\nu$ represents the perimeter-averaged friction Reynolds number, and $C_f = \tau_w / (\rho u_b^2 / 2)$ is the perimeter-averaged skin friction coefficient. Six meshes with uniform grid spacings in all three directions are used for the grid convergence study, with the finest having $\Delta_z^+ = 6.3$. Figure~\ref{fig:duct}(b) visualizes the turbulent flow on the finest grid obtained with WMLES using the SM model with the van Driest damping function. This visualization clearly captures the low- and high-speed streaks, as well as the coherent structures responsible for the secondary flow. Table~\ref{tab:duct} lists the relative error in the skin friction coefficient, with absolute values within approximately $3\%$ for all the cases. This supports general applicability of the logarithmic law as a wall model to predict friction in flows with moderate geometrical complexity. Notably, the convergence behavior of the skin friction coefficient is non-monotonic for both SGS models.

\begin{table}[hbtp]
 \centering
 \caption{Parameters for the WMLES of turbulent square duct flow.}\label{tab:duct}
 \small
 \begin{tabular}{cccccccccc}
   \hline
   Mesh ($N_x \times N_y \times N_z$) & $\Delta x/h$ & $\Delta y/h$ & $\Delta z/h$ & $AR$ & SGS & $\Rey_\tau$ & $C_f$   & $\epsilon_f$      & \#ETT \\
   $128  \times 80   \times 80  $ & 0.100  & 0.025  & 0.025  & 4.0 & SM     & 1065.9    & $0.00568 \pm 0.08\%$ &   2.08\% & 52.8 \\
   $256  \times 80   \times 80  $ & 0.050  & 0.025  & 0.025  & 2.0 & SM     & 1044.5    & $0.00546 \pm 0.09\%$ &  -1.97\% & 52.8 \\
   $512  \times 80   \times 80  $ & 0.025  & 0.025  & 0.025  & 1.0 & SM     & 1038.6    & $0.00539 \pm 0.11\%$ &  -3.08\% & 52.8 \\
   $768  \times 120  \times 120 $ & 0.017  & 0.017  & 0.017  & 1.0 & SM     & 1041.4    & $0.00542 \pm 0.17\%$ &  -2.55\% & 52.8 \\
   $1024 \times 160  \times 160 $ & 0.013  & 0.013  & 0.013  & 1.0 & SM     & 1040.9    & $0.00542 \pm 0.11\%$ &  -2.66\% & 52.8 \\
   $2048 \times 320  \times 320 $ & 0.006  & 0.006  & 0.006  & 1.0 & SM     & 1044.3    & $0.00545 \pm 0.17\%$ &  -2.03\% & 52.8 \\
   $128  \times 80   \times 80  $ & 0.100  & 0.025  & 0.025  & 4.0 & DSM    & 1066.3    & $0.00569 \pm 0.10\%$ &   2.16\% & 52.8 \\
   $256  \times 80   \times 80  $ & 0.050  & 0.025  & 0.025  & 2.0 & DSM    & 1051.0    & $0.00552 \pm 0.10\%$ &  -0.76\% & 52.8 \\
   $512  \times 80   \times 80  $ & 0.025  & 0.025  & 0.025  & 1.0 & DSM    & 1046.3    & $0.00547 \pm 0.10\%$ &  -1.65\% & 52.8 \\
   $768  \times 120  \times 120 $ & 0.017  & 0.017  & 0.017  & 1.0 & DSM    & 1047.0    & $0.00548 \pm 0.11\%$ &  -1.51\% & 52.8 \\
   $1024 \times 160  \times 160 $ & 0.013  & 0.013  & 0.013  & 1.0 & DSM    & 1045.7    & $0.00547 \pm 0.09\%$ &  -1.76\% & 52.8 \\
   $2048 \times 320  \times 320 $ & 0.006  & 0.006  & 0.006  & 1.0 & DSM    & 1044.9    & $0.00546 \pm 0.13\%$ &  -1.91\% & 52.8 \\
   \hline
 \end{tabular}
\end{table}

\begin{figure}[H]
   \centering
   \begin{subfigure}{0.19\textwidth}
       \centering
       \includegraphics[width=\linewidth]{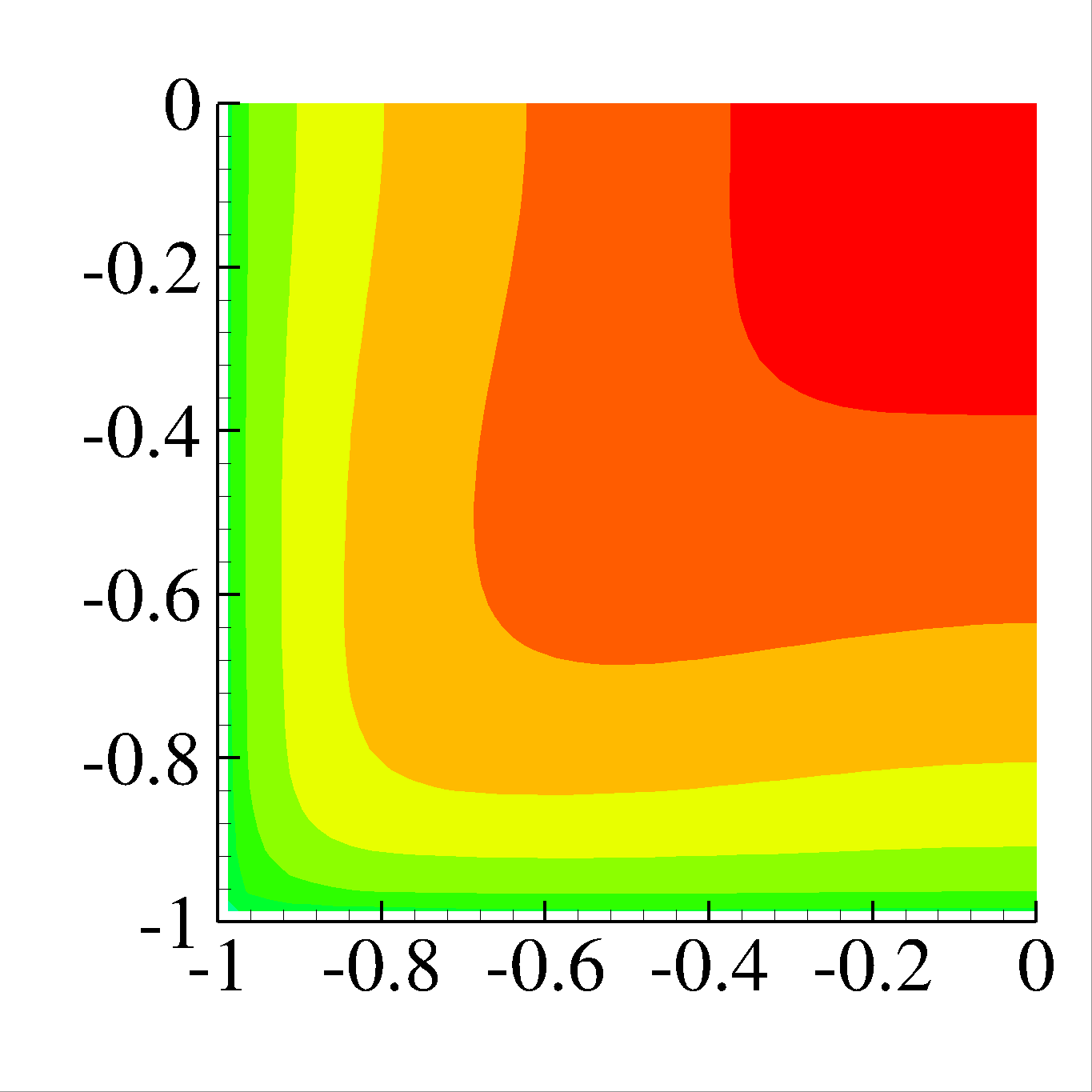}
       \label{fig:sub2}
   \end{subfigure}
   \hspace{-0.5em}
   \begin{subfigure}{0.19\textwidth}
       \centering
       \includegraphics[width=\linewidth]{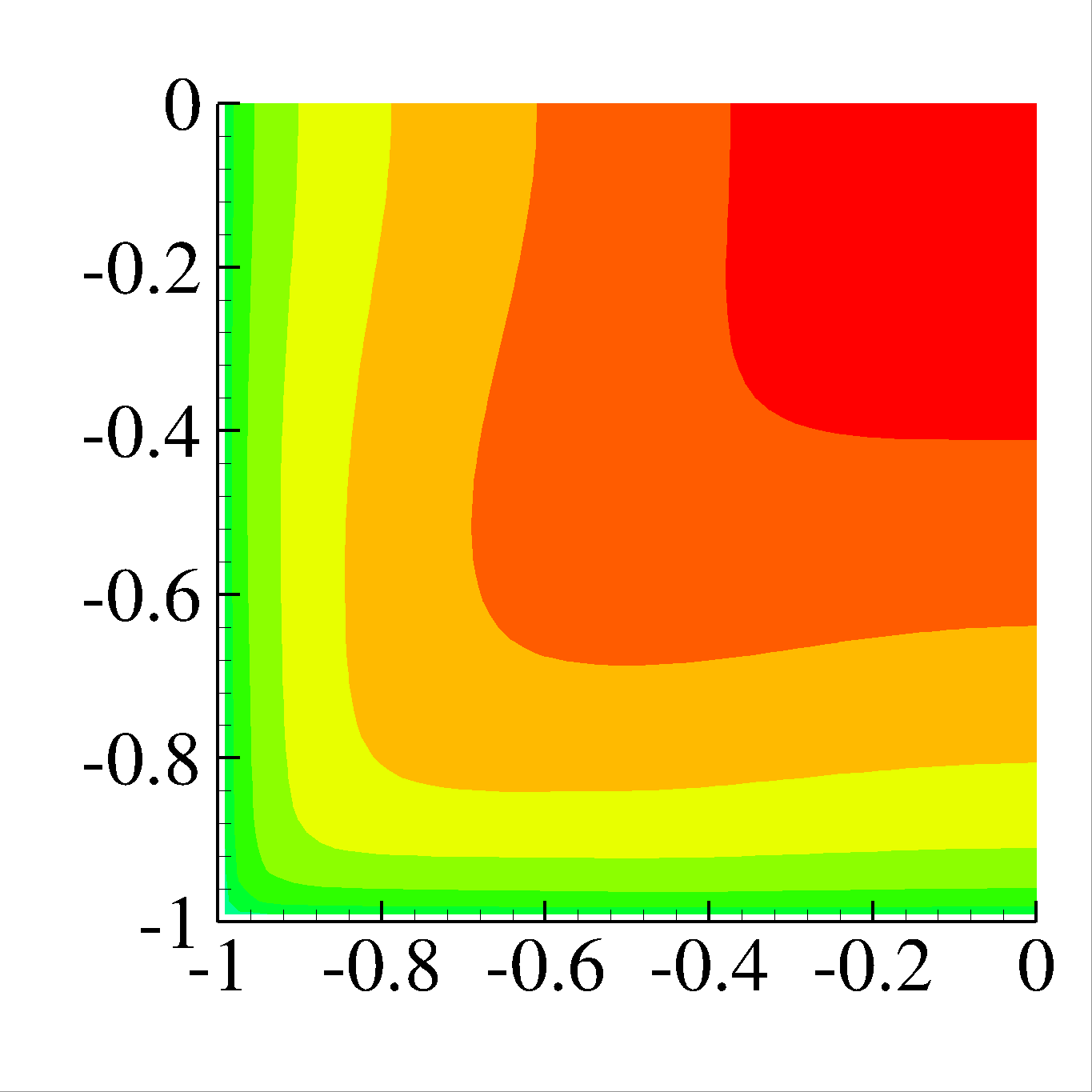}
       \label{fig:sub2}
   \end{subfigure}
   \hspace{-0.5em}
   \begin{subfigure}{0.19\textwidth}
       \centering
       \includegraphics[width=\linewidth]{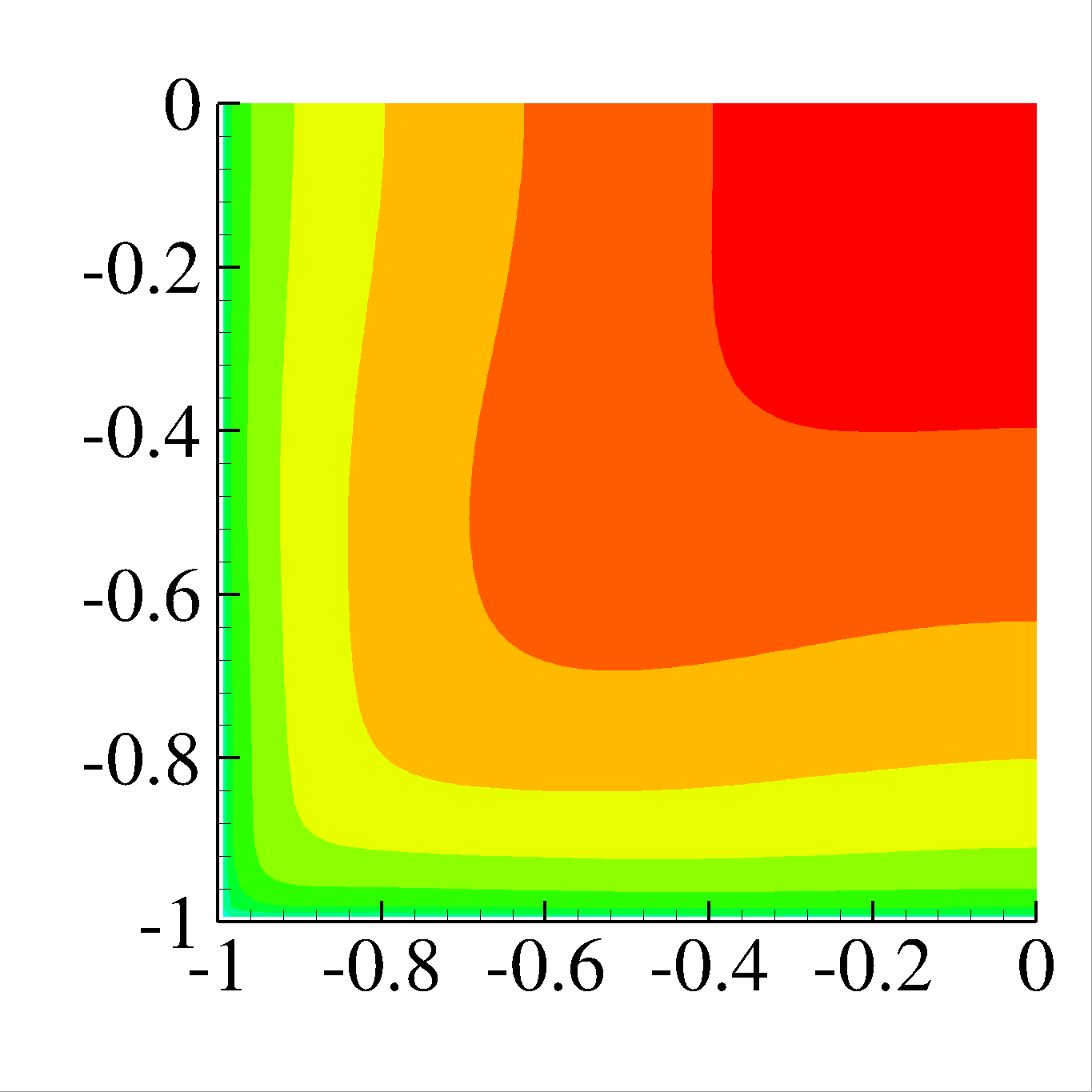}
       \label{fig:sub2}
   \end{subfigure}   
   \hspace{-0.5em}
   \begin{subfigure}{0.19\textwidth}
       \centering
       \includegraphics[width=\linewidth]{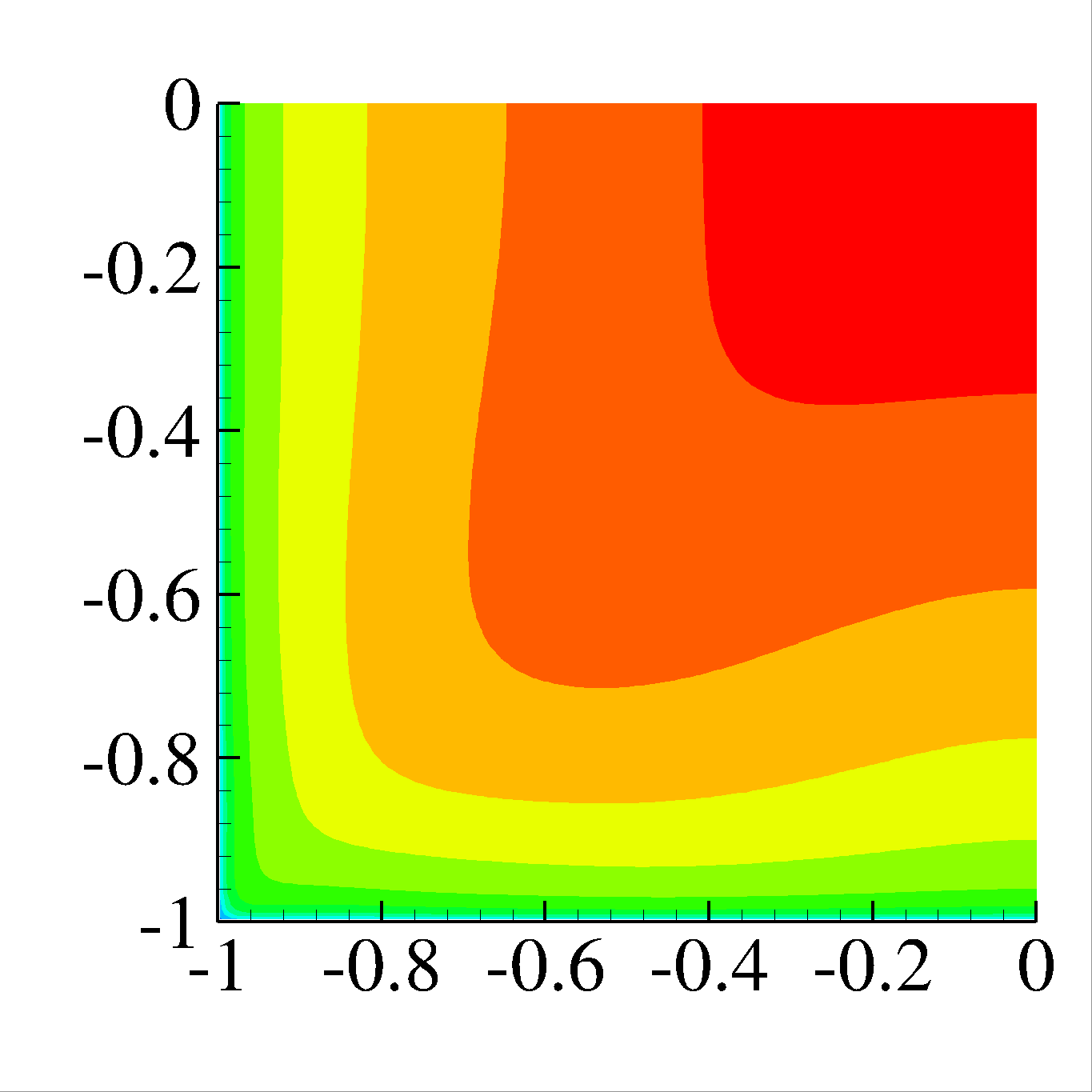}
       \label{fig:sub2}
   \end{subfigure}
   \hspace{-0.5em}
   \begin{subfigure}{0.24\textwidth}
       \centering
       \includegraphics[width=\linewidth]{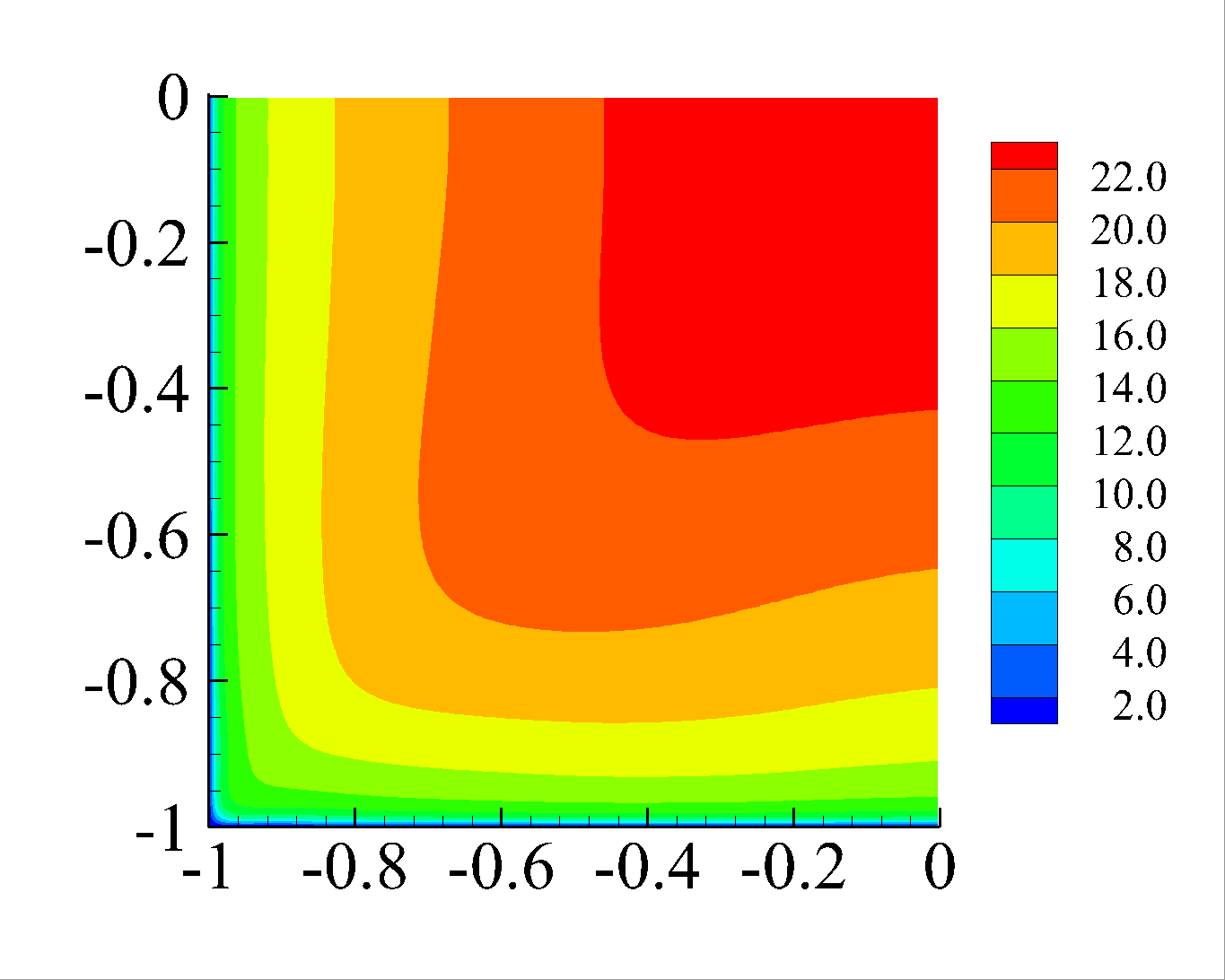}
       \label{fig:sub1}
   \end{subfigure}
   \vspace{-3.0em}
   
   \begin{subfigure}{0.19\textwidth}
   \centering
   \includegraphics[width=\linewidth]{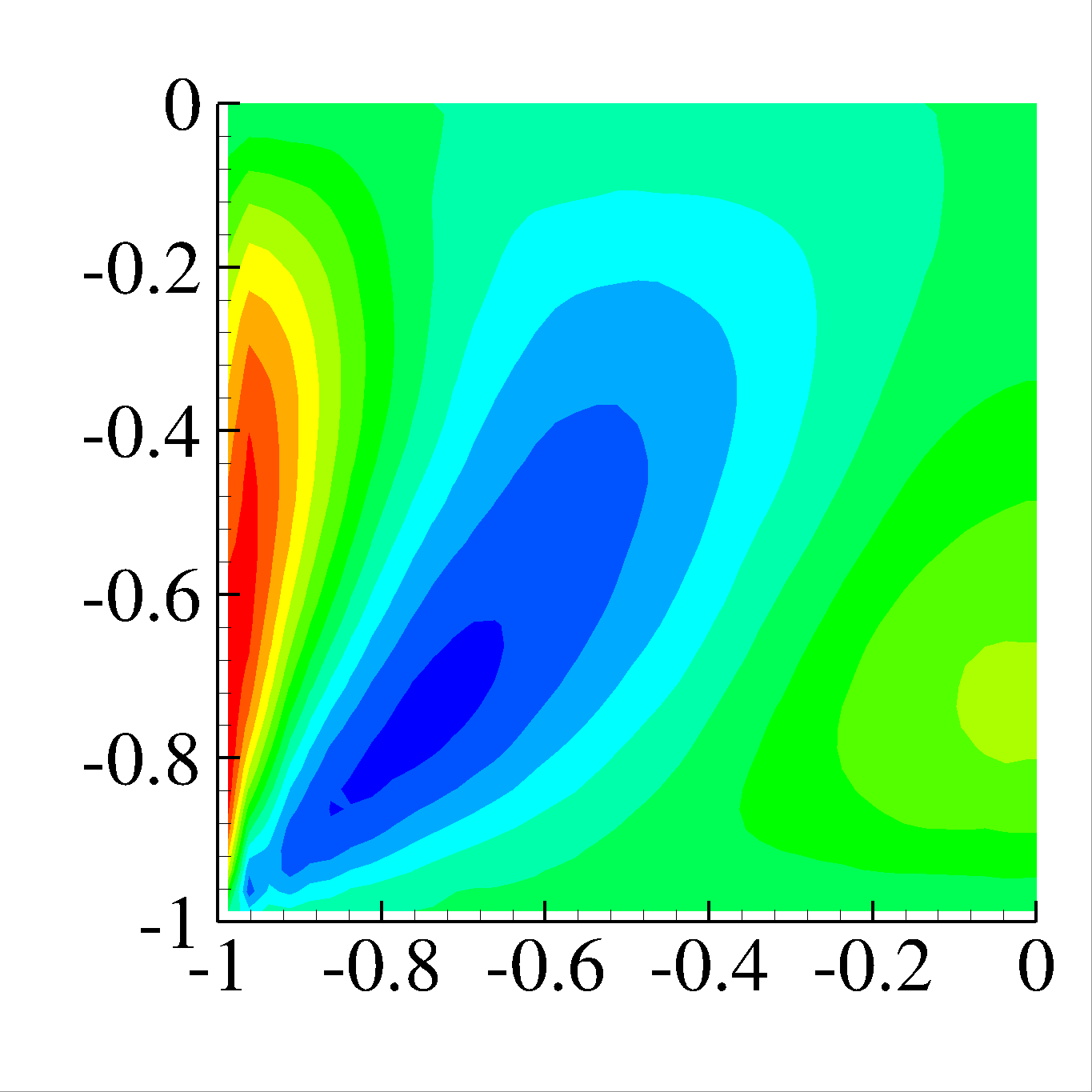}
   \label{fig:sub2}
   \end{subfigure}
   \hspace{-0.5em}
   \begin{subfigure}{0.19\textwidth}
   \centering
   \includegraphics[width=\linewidth]{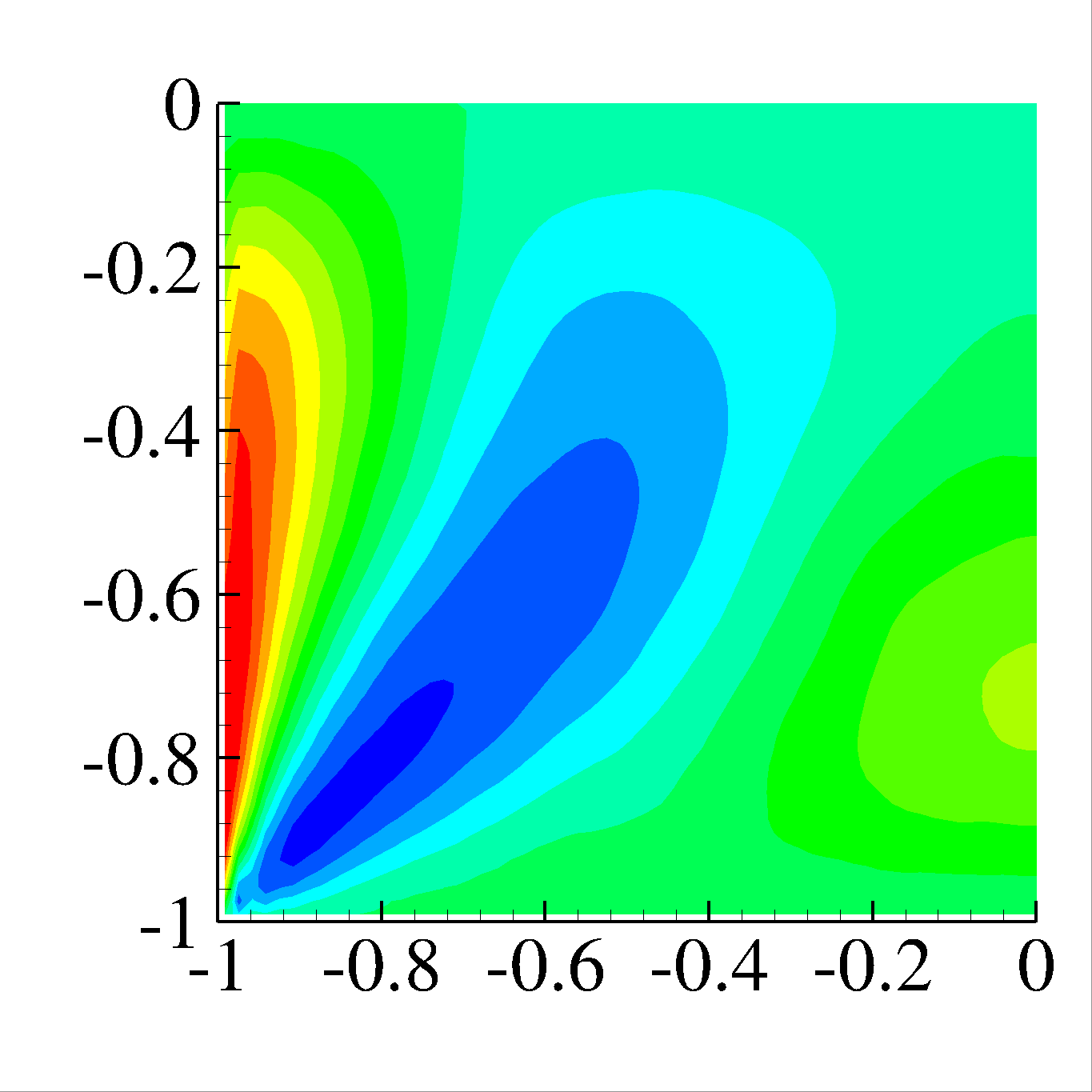}
   \label{fig:sub2}
   \end{subfigure}
   \hspace{-0.5em}
   \begin{subfigure}{0.19\textwidth}
   \centering
   \includegraphics[width=\linewidth]{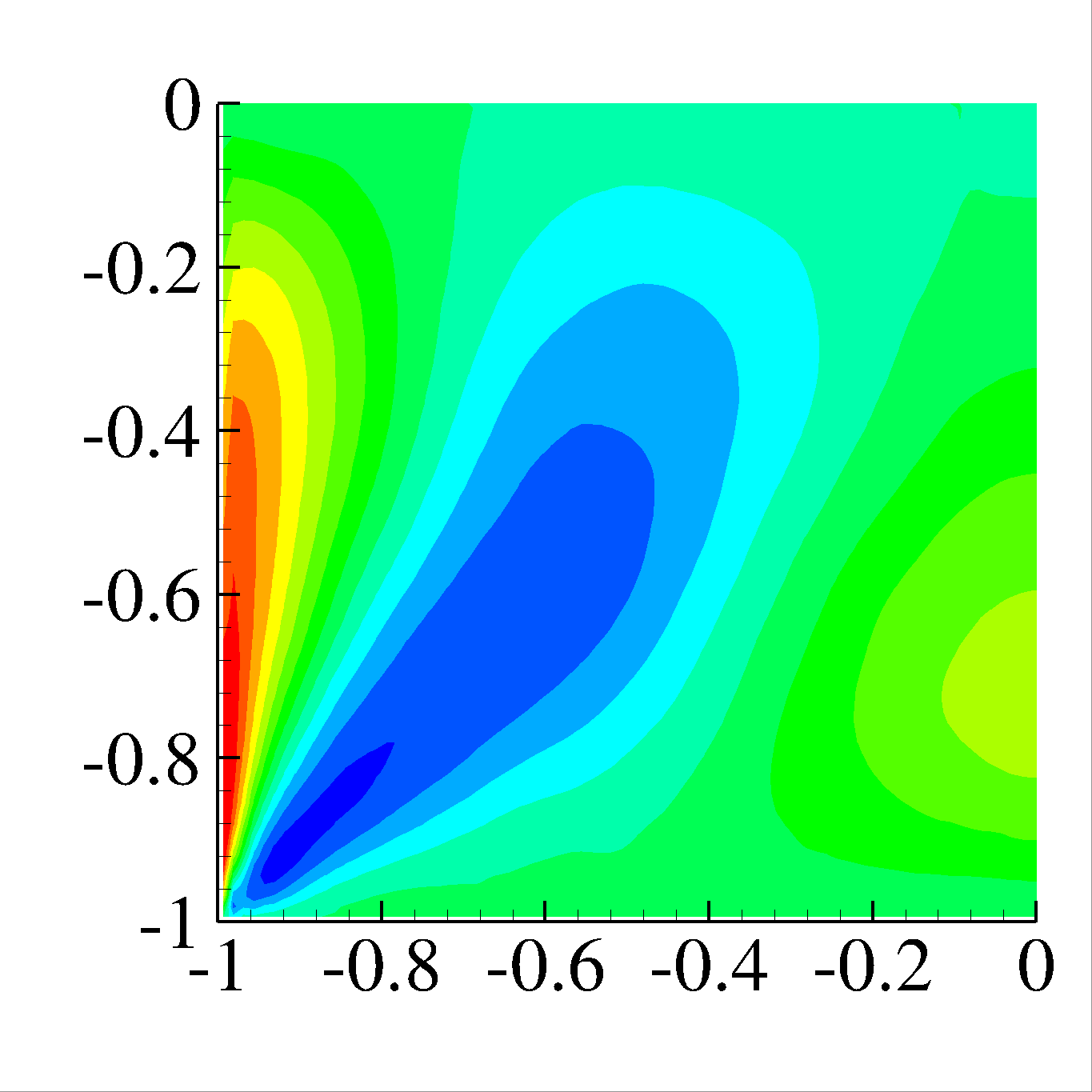}
   \label{fig:sub2}
   \end{subfigure}
   \hspace{-0.5em}
   \begin{subfigure}{0.19\textwidth}
   \centering
   \includegraphics[width=\linewidth]{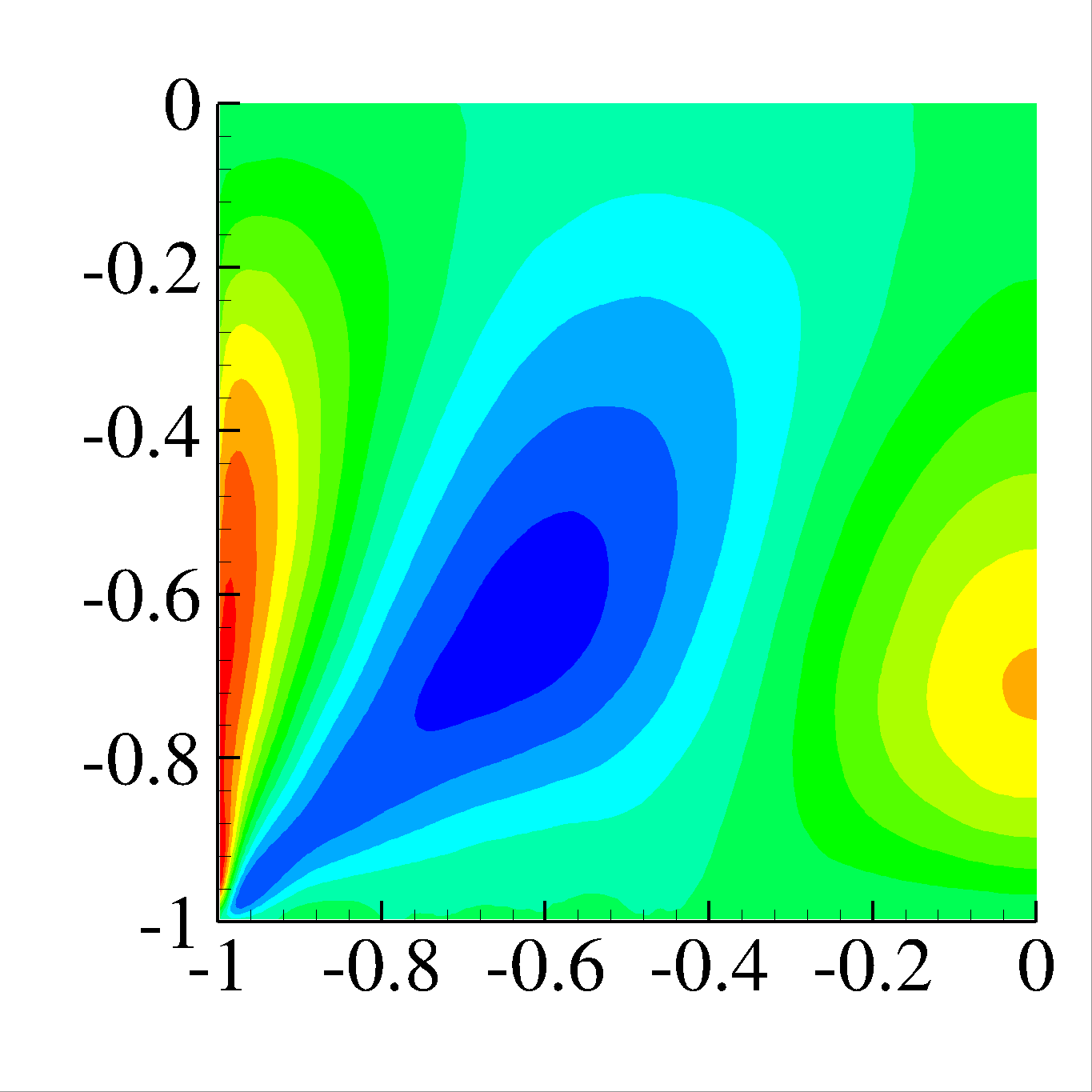}
   \label{fig:sub2}
   \end{subfigure}
   \hspace{-0.5em}
   \begin{subfigure}{0.24\textwidth}
   \centering
       \includegraphics[width=\linewidth]{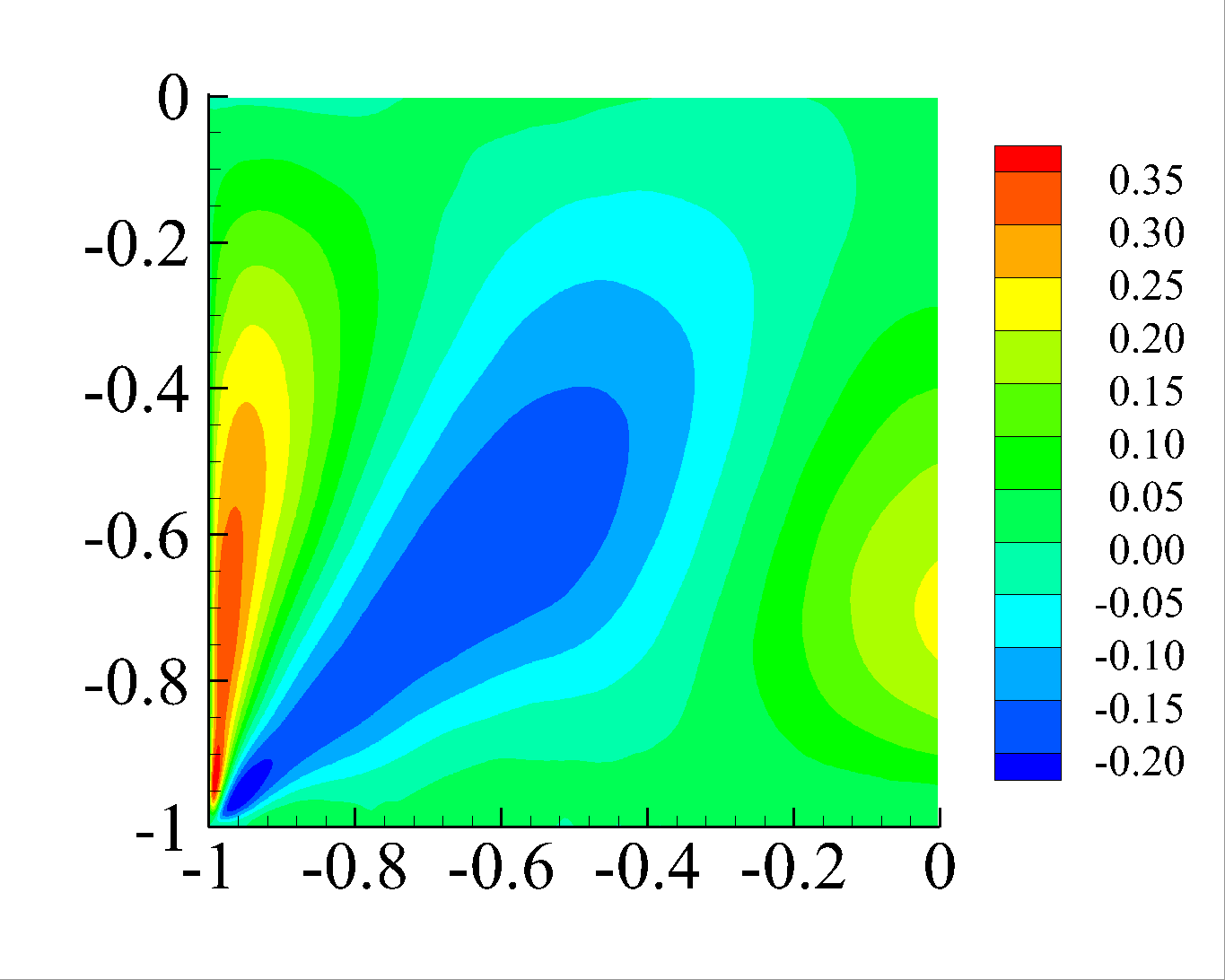}
   \label{fig:sub2}
   \end{subfigure}
   \vspace{-3.0em}
   
   \begin{subfigure}{0.19\textwidth}
   \centering
   \includegraphics[width=\linewidth]{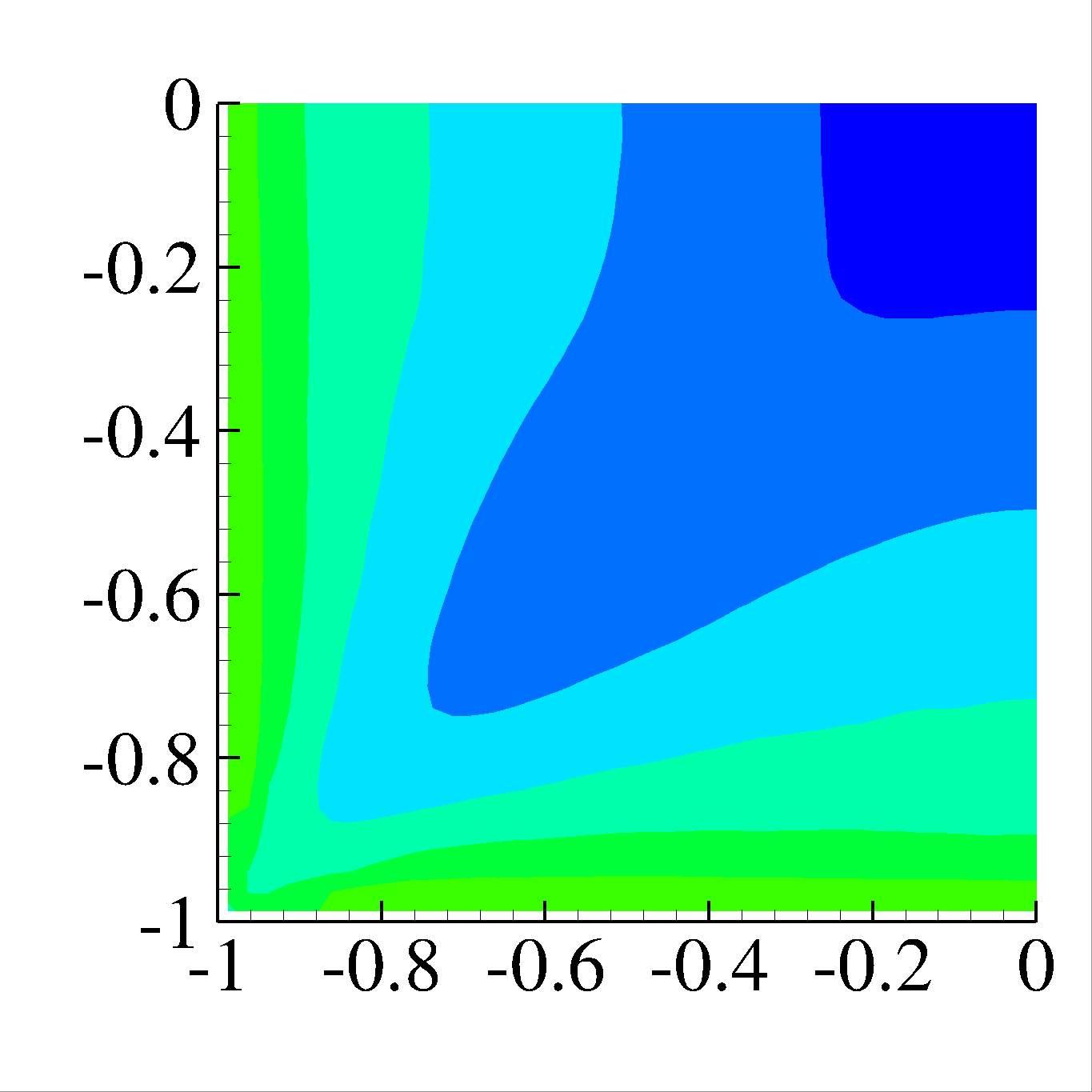}
   \label{fig:sub2}
   \end{subfigure}
   \hspace{-0.5em}
   \begin{subfigure}{0.19\textwidth}
   \centering
   \includegraphics[width=\linewidth]{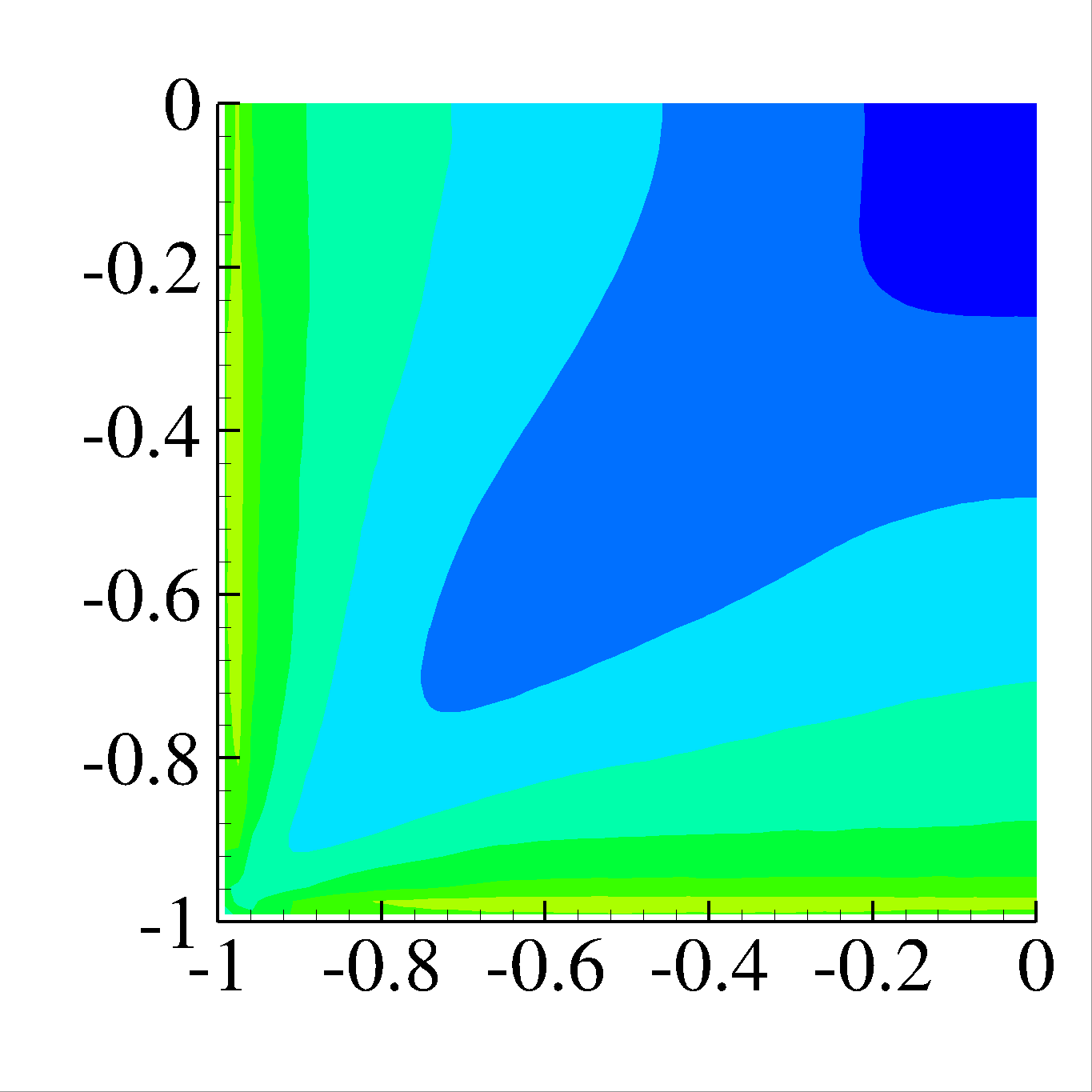}
   \label{fig:sub2}
   \end{subfigure}
   \hspace{-0.5em}
   \begin{subfigure}{0.19\textwidth}
   \centering
   \includegraphics[width=\linewidth]{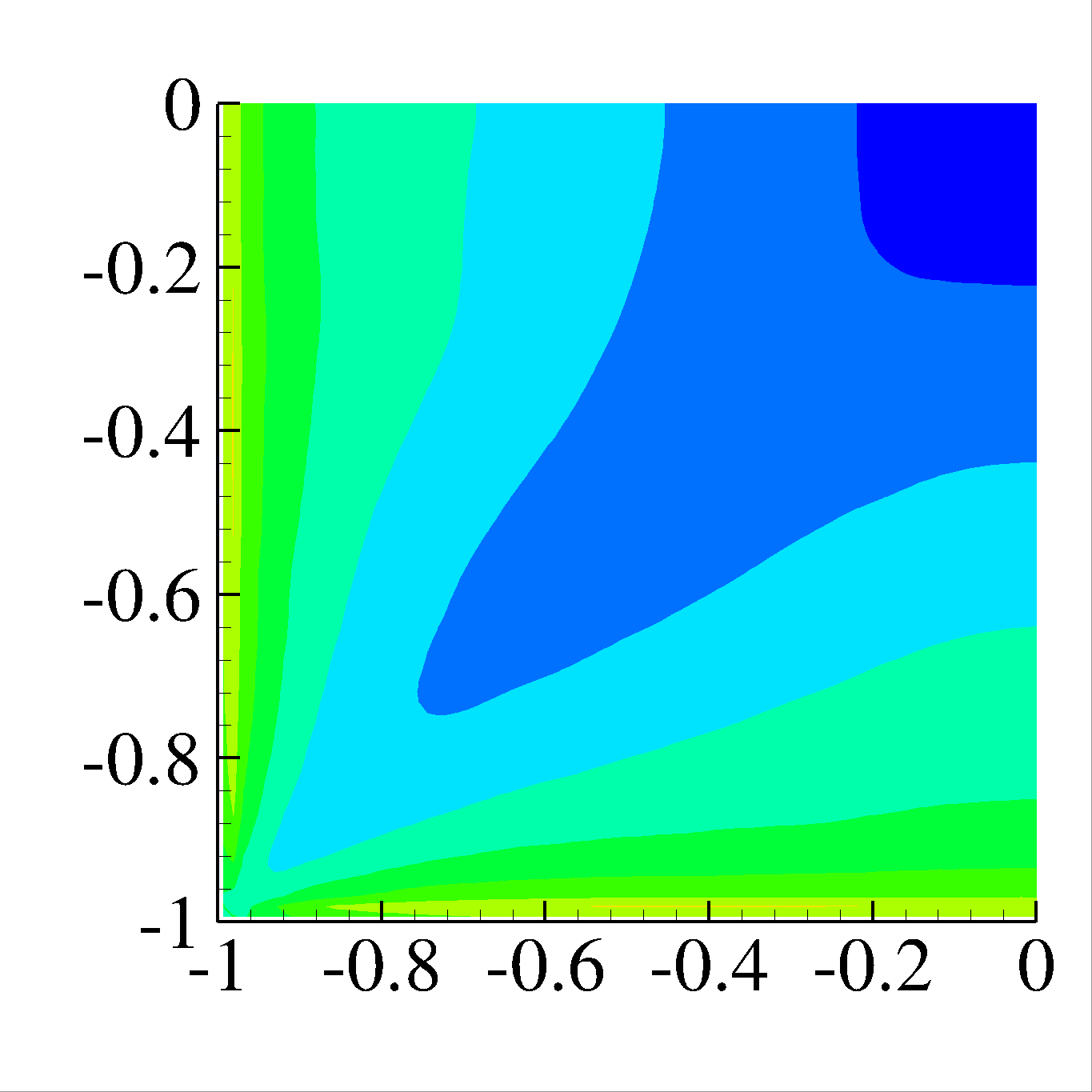}
   \label{fig:sub2}
   \end{subfigure}
   \hspace{-0.5em}
   \begin{subfigure}{0.19\textwidth}
   \centering
   \includegraphics[width=\linewidth]{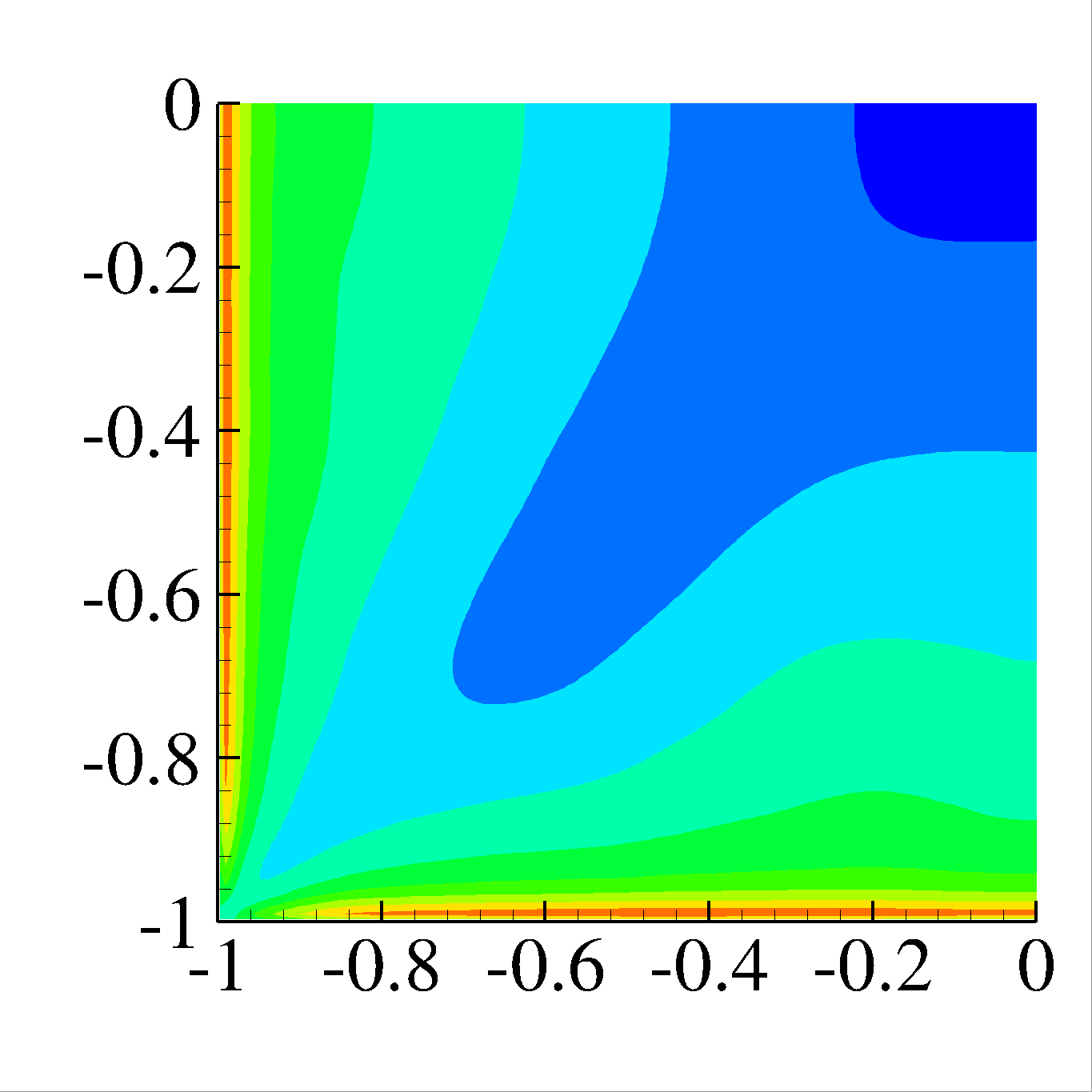}
   \label{fig:sub2}
   \end{subfigure}
   \hspace{-0.5em}
   \begin{subfigure}{0.24\textwidth}
   \centering
   \includegraphics[width=\linewidth]{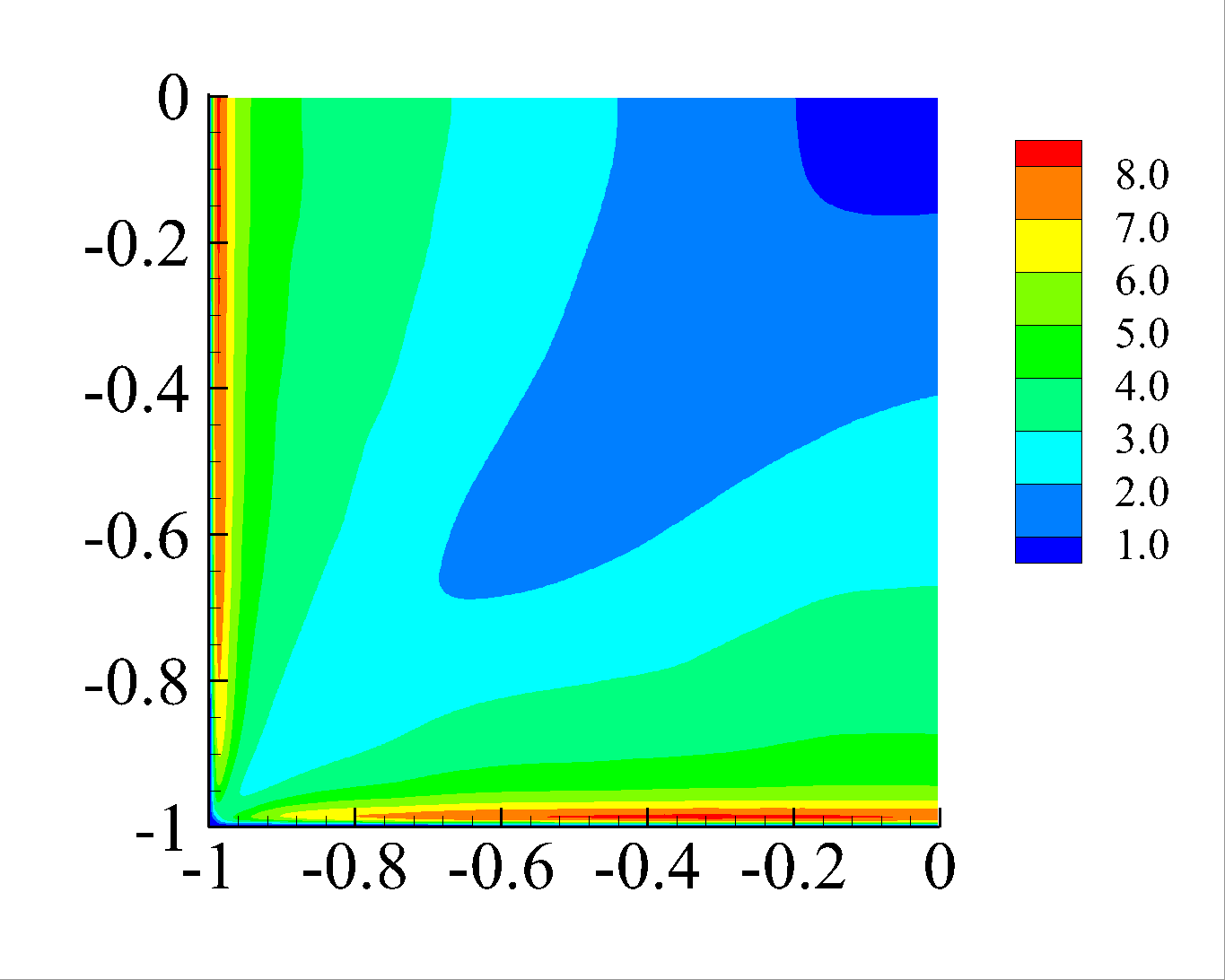}
   \label{fig:sub2}
   \end{subfigure}
   \vspace{-3.0em}
   
   \begin{subfigure}{0.19\textwidth}
   \centering
   \includegraphics[width=\linewidth]{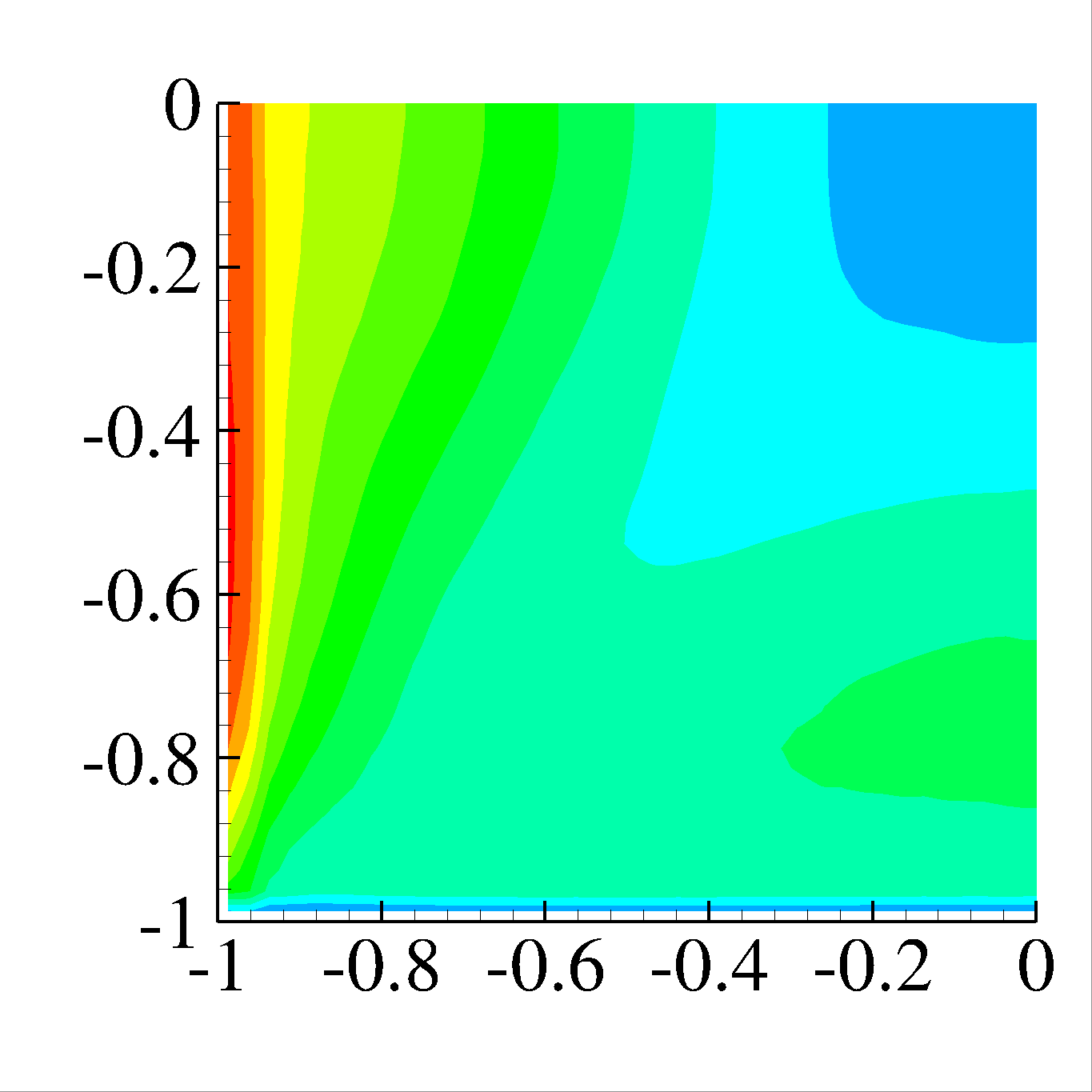}
   \label{fig:sub2}
   \end{subfigure}
   \hspace{-0.5em}
   \begin{subfigure}{0.19\textwidth}
   \centering
   \includegraphics[width=\linewidth]{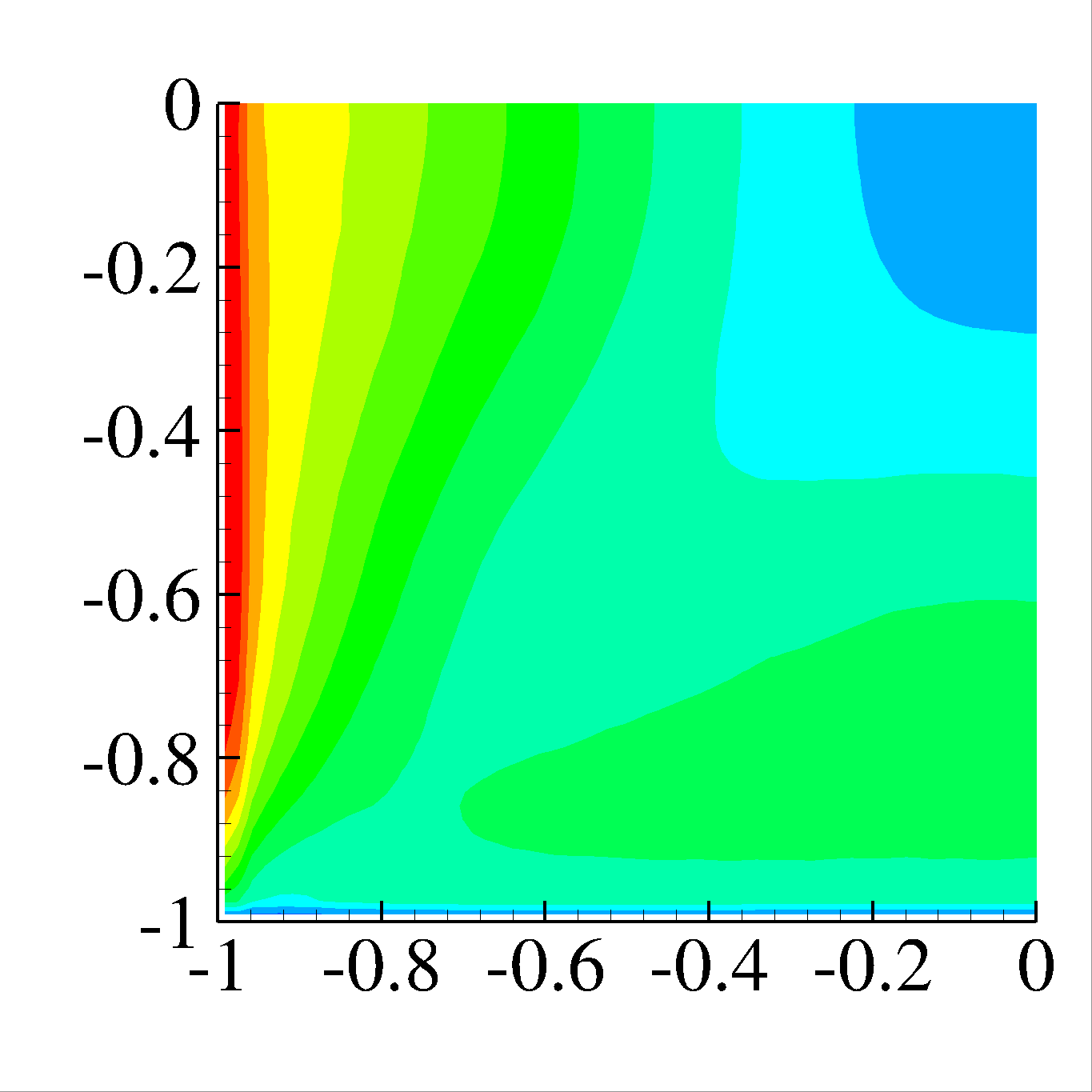}
   \label{fig:sub2}
   \end{subfigure}
   \hspace{-0.5em}
   \begin{subfigure}{0.19\textwidth}
   \centering
   \includegraphics[width=\linewidth]{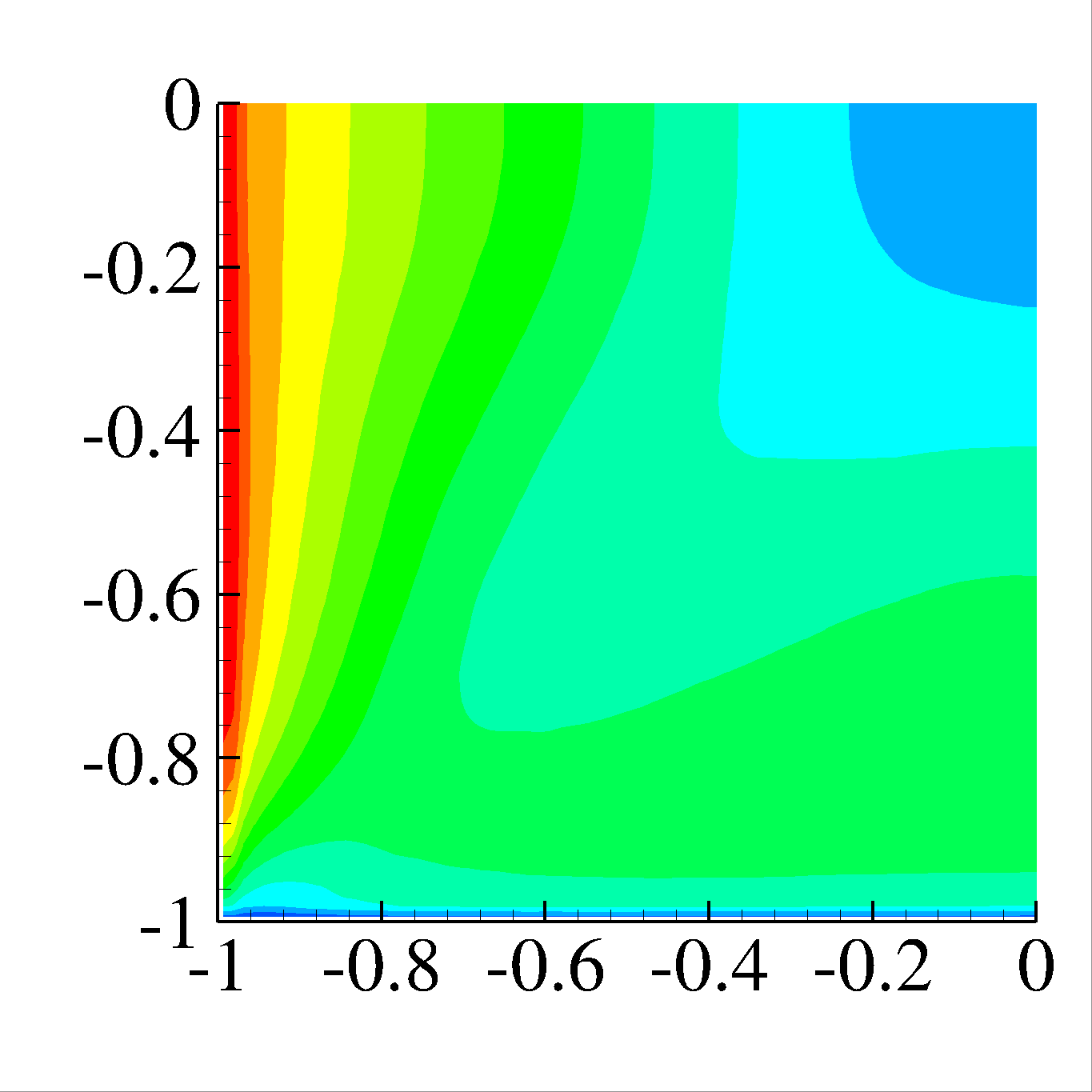}
   \label{fig:sub2}
   \end{subfigure}
   \hspace{-0.5em}
   \begin{subfigure}{0.19\textwidth}
   \centering
   \includegraphics[width=\linewidth]{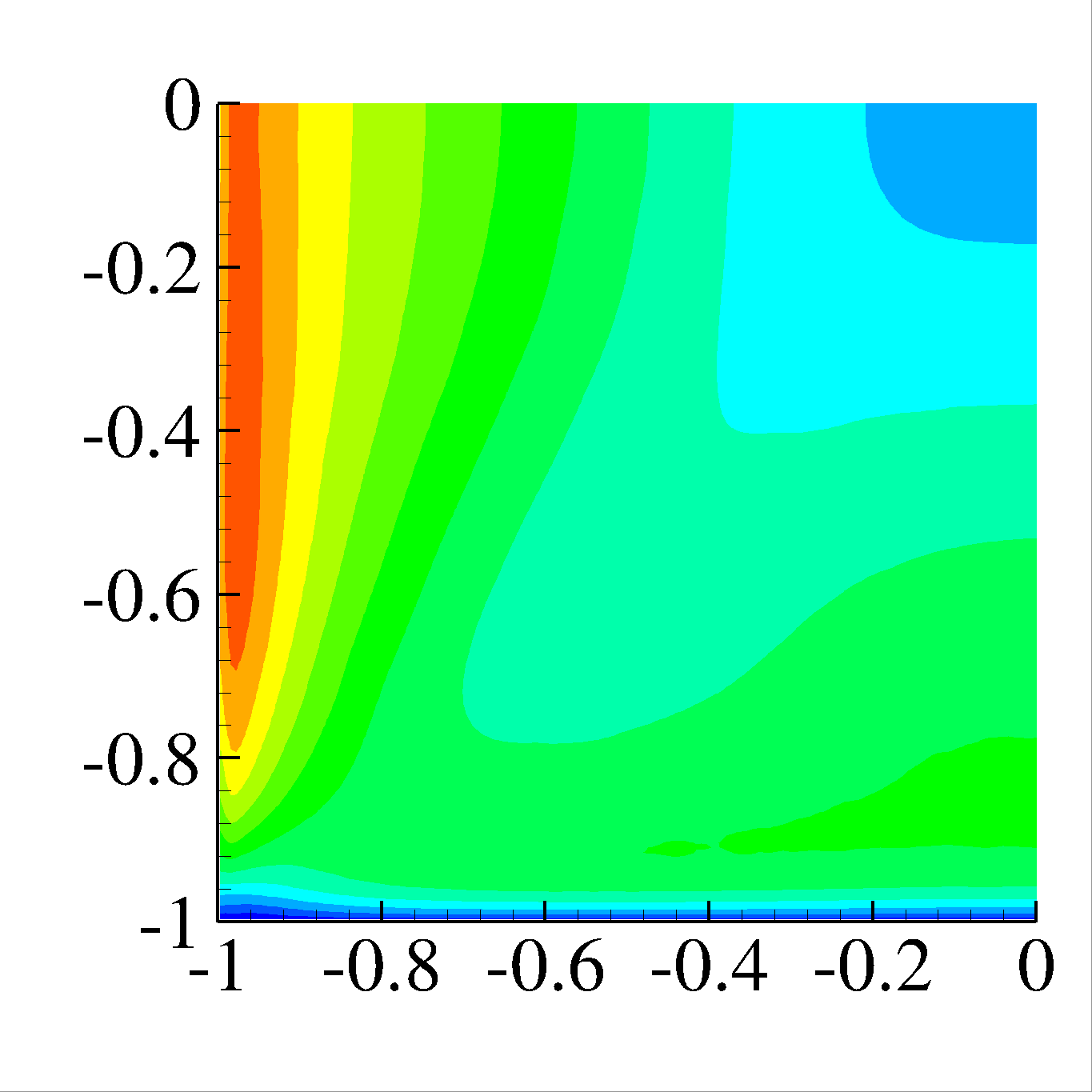}
   \label{fig:sub2}
   \end{subfigure}
   \hspace{-0.5em}
   \begin{subfigure}{0.24\textwidth}
   \centering
   \includegraphics[width=\linewidth]{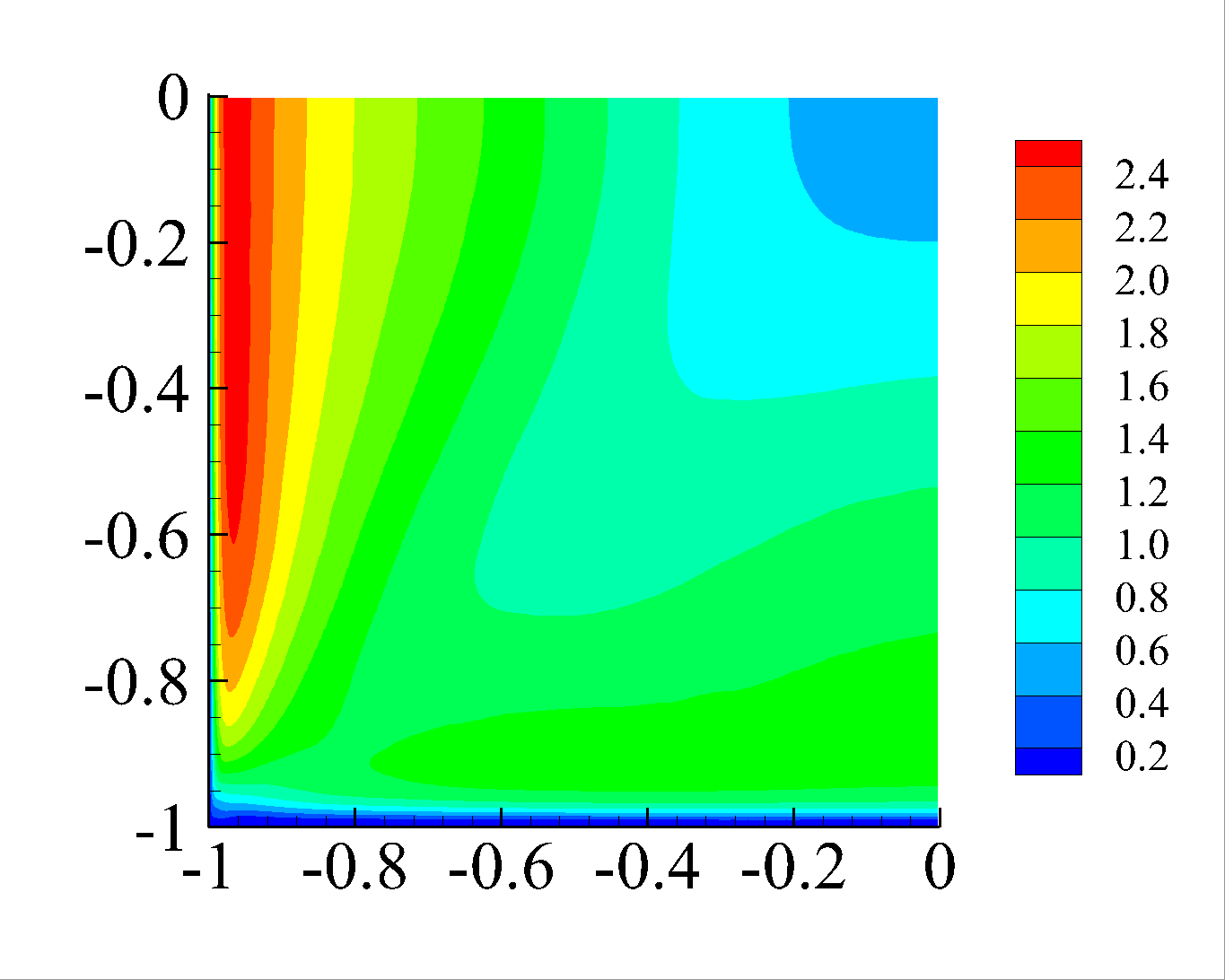}
   \label{fig:sub2}
   \end{subfigure}
   \vspace{-3.0em}
   
   \begin{subfigure}{0.19\textwidth}
   \centering
   \includegraphics[width=\linewidth]{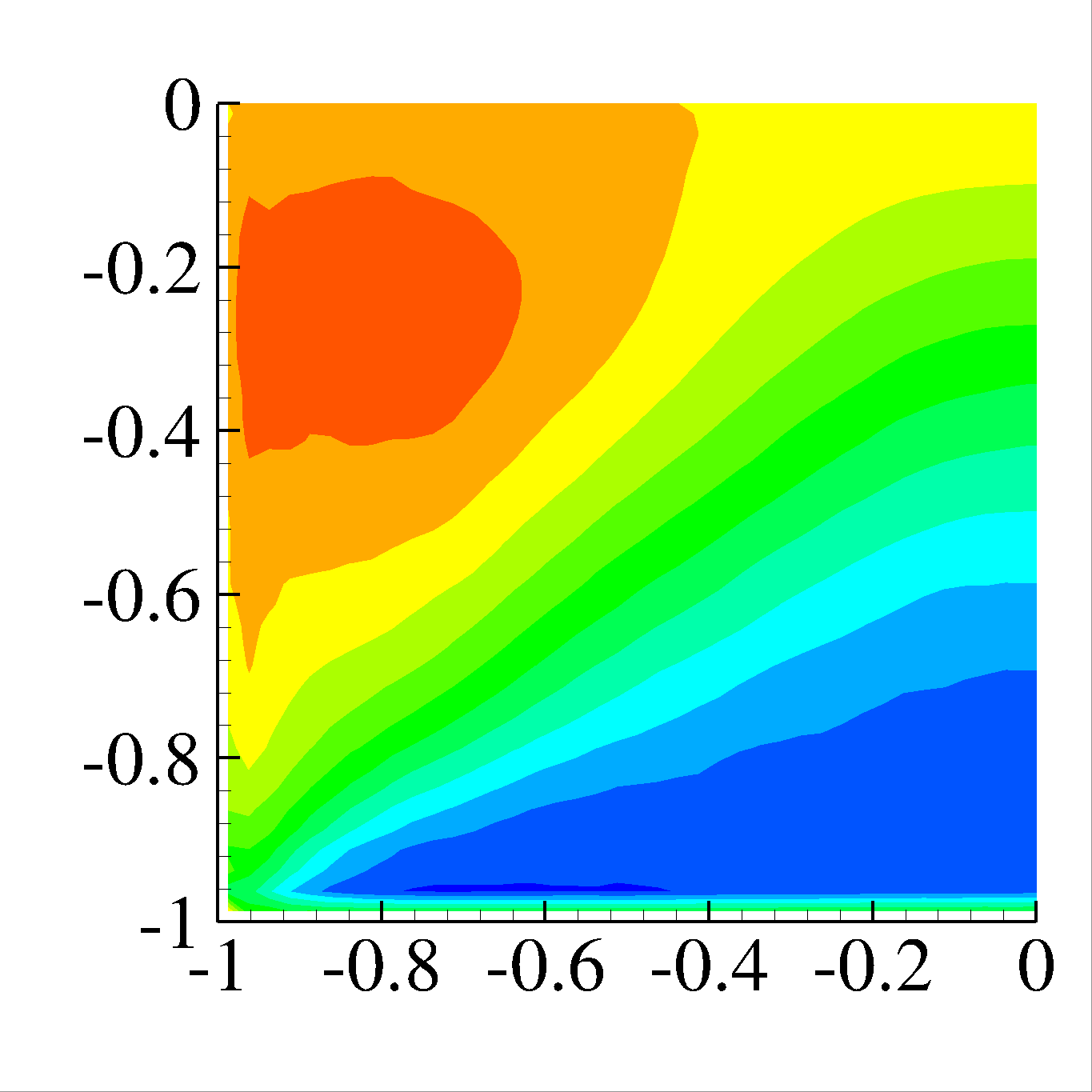}
   \label{fig:sub2}
   \end{subfigure}
   \hspace{-0.5em}
   \begin{subfigure}{0.19\textwidth}
   \centering
   \includegraphics[width=\linewidth]{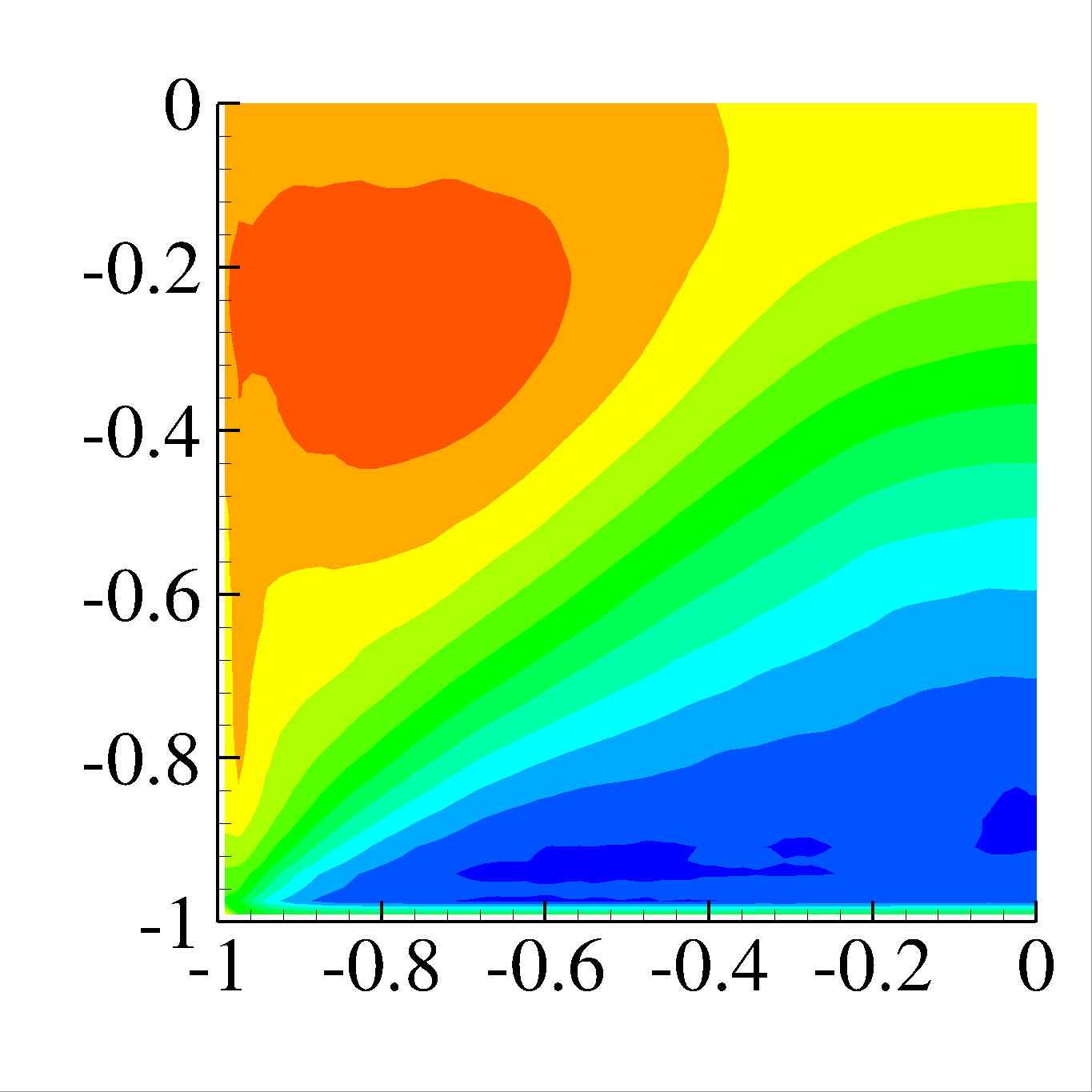}
   \label{fig:sub2}
   \end{subfigure}
   \hspace{-0.5em}
   \begin{subfigure}{0.19\textwidth}
   \centering
   \includegraphics[width=\linewidth]{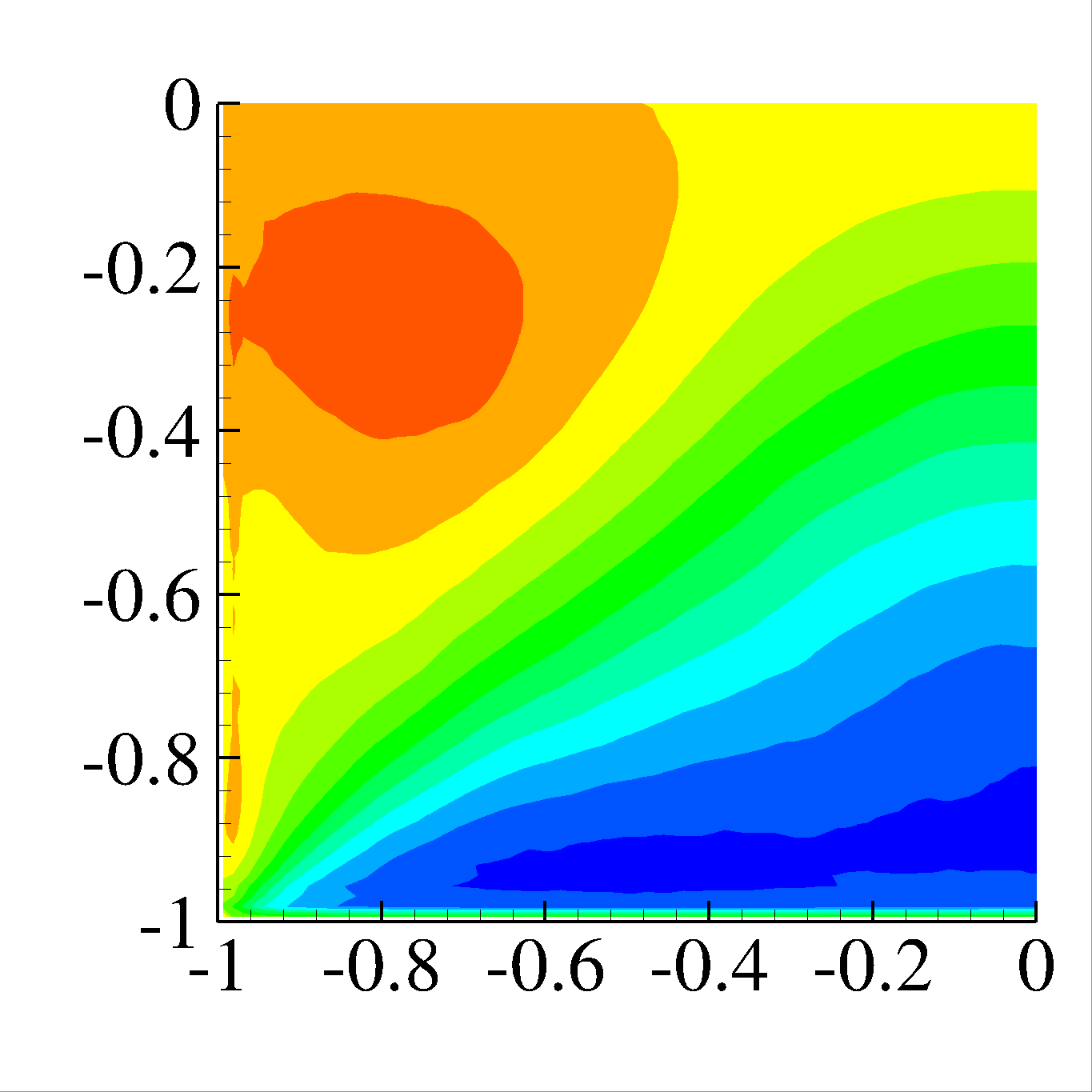}
   \label{fig:sub2}
   \end{subfigure}
   \hspace{-0.5em}
   \begin{subfigure}{0.19\textwidth}
   \centering
   \includegraphics[width=\linewidth]{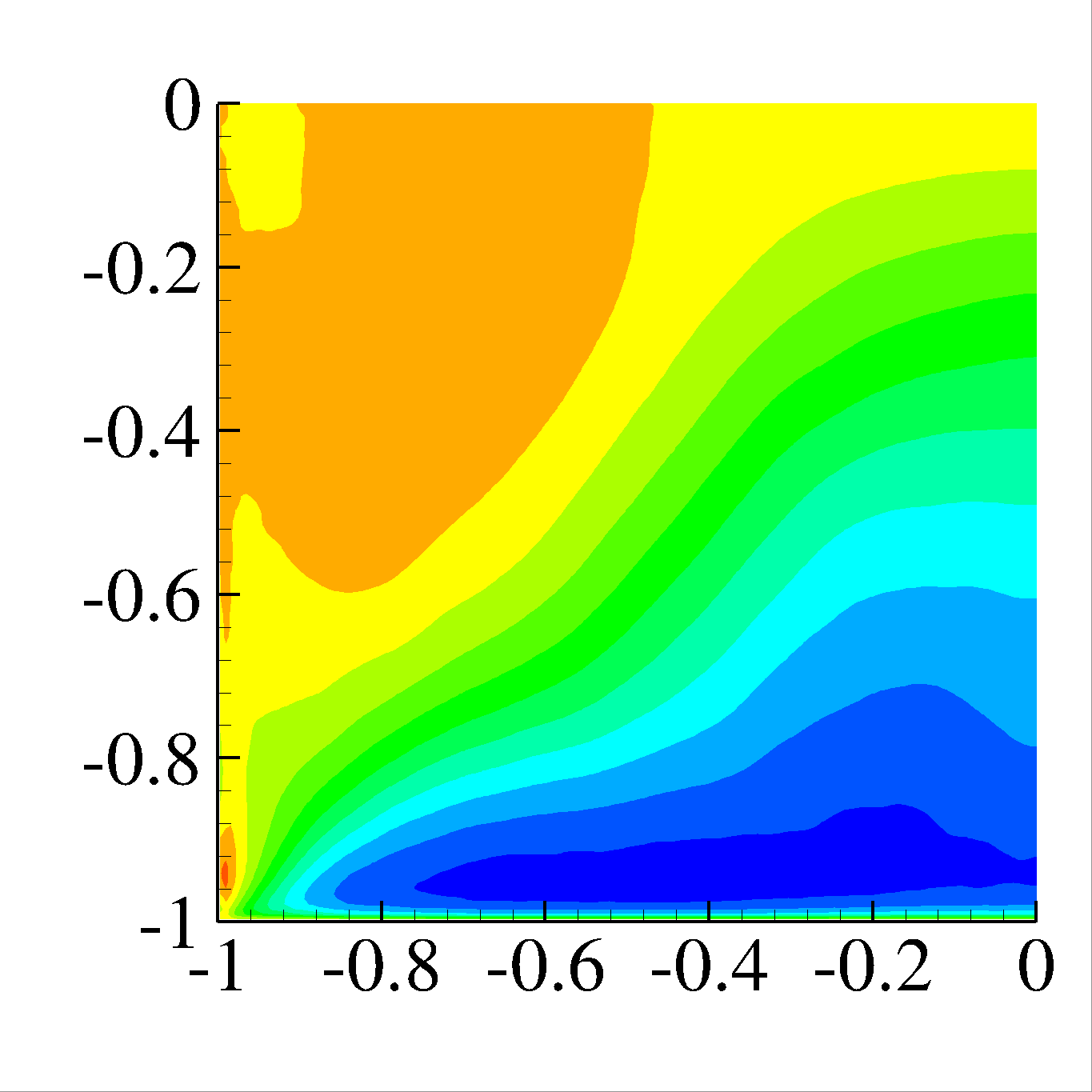}
   \label{fig:sub2}
   \end{subfigure}
   \hspace{-0.5em}
   \begin{subfigure}{0.24\textwidth}
   \centering
   \includegraphics[width=\linewidth]{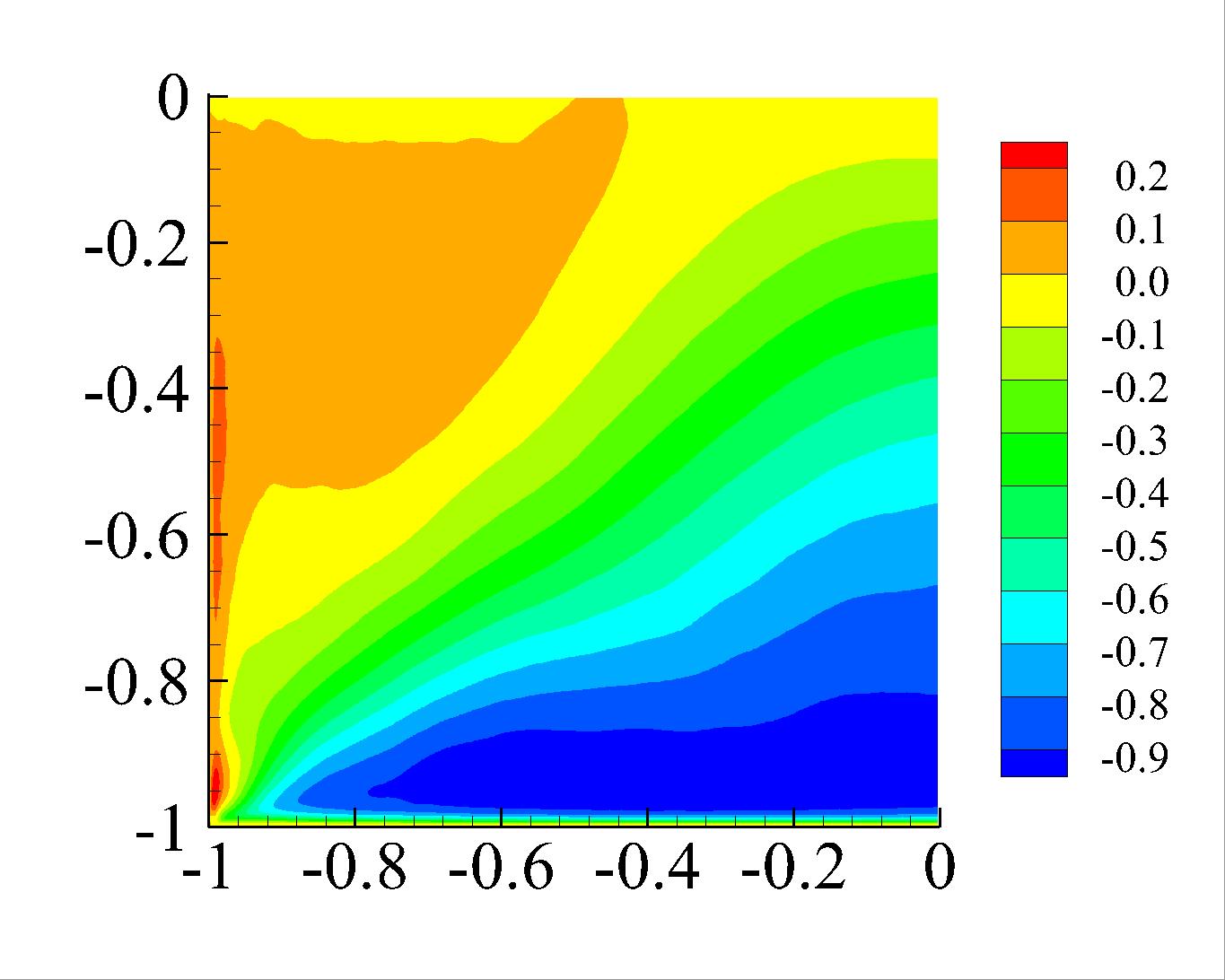}
   \label{fig:sub2}
   \end{subfigure}
   \caption{Turbulent square duct flow: cross-stream contours of mean velocity $U^+$ (row 1) and $V^+$ (row 2), turbulent normal stresses $\langle uu \rangle^+$ (row 3) and $\langle vv \rangle^+$ (row 4), and shear stress $\langle uv \rangle^+$ (row 5), computed on different grids using WMLES with the SM model. Normalization is based on the DNS wall units. Only a quarter of the full domain is presented. The horizontal axis is $z$ and the vertical axis is $y$. $\Delta_z/h=0.025$ (column 1), $\Delta_z/h=0.017$ (column 2), $\Delta_z/h=0.013$ (column 3), $\Delta_z/h=0.006$ (column 4) and DNS~\cite{pirozzoli2018turbulence} (column 5).}
   \label{fig:contour_smag}
\end{figure}

Figure~\ref{fig:contour_smag} presents contours obtained with the SM model for the mean velocity components ($U^+$, $V^+$), turbulent normal stresses ($\langle uu \rangle^+$, $\langle vv \rangle^+$), and turbulent shear stress ($\langle uv \rangle^+$). Only one quarter of the full domain is presented. The contours of $W^+$, $\langle ww \rangle^+$, and $\langle uw \rangle^+$ are not displayed, as they are symmetric with respect to the diagonal, to those of $V^+$, $\langle vv \rangle^+$, and $\langle uv \rangle^+$, respectively. The distributions obtained for various grids show little variations in the core region and agree reasonably well with the reference DNS data. The primary benefit of grid refinement is an increase in the overall accuracy in the near-wall region.

\begin{figure}[htbp]
   \centering
   \begin{subfigure}{0.49\textwidth}
       \centering
       \includegraphics[width=\linewidth]{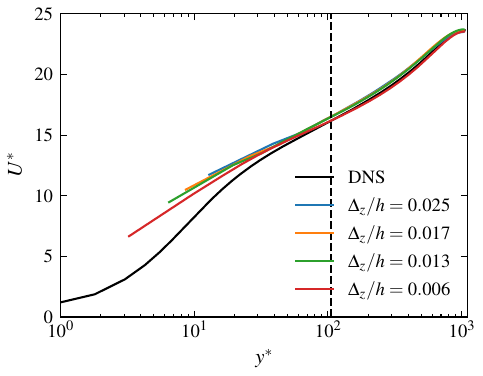}
       \caption{}
       \label{fig:sub1}
   \end{subfigure}
   \hfill
   \begin{subfigure}{0.49\textwidth}
       \centering
       \includegraphics[width=\linewidth]{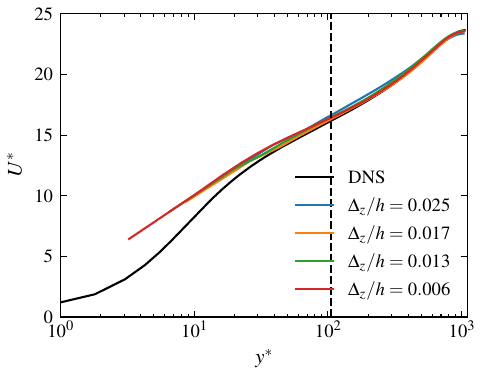}
       \caption{}
       \label{fig:sub2}
   \end{subfigure}
   \vfill
   \begin{subfigure}{0.49\textwidth}
       \centering
       \includegraphics[width=\linewidth]{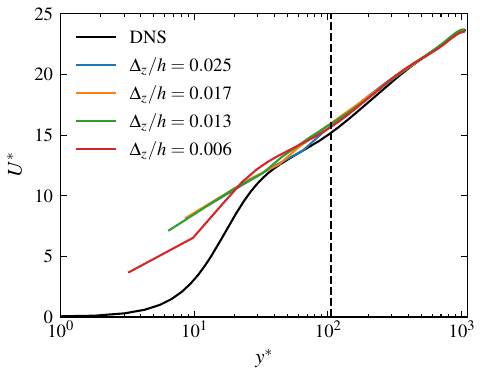}
       \caption{}
       \label{fig:sub1}
   \end{subfigure}
   \hfill
   \begin{subfigure}{0.49\textwidth}
       \centering
       \includegraphics[width=\linewidth]{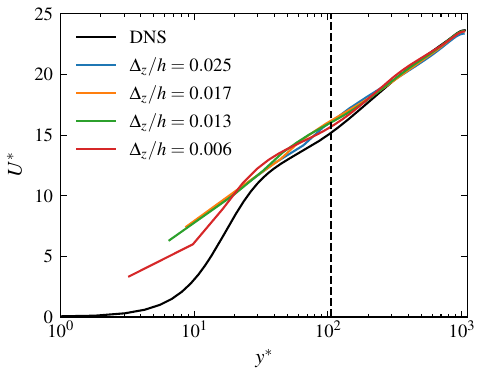}
       \caption{}
       \label{fig:sub2}
   \end{subfigure}
   \caption{Turbulent square duct flow: profiles of mean streamwise velocity along the wall bisector (a,b) and corner bisector (c,d) obtained from WMLES with the SM (a,c) and DSM (b,d) models. The dashed line denotes $y/h=0.1$. Normalization is based on the wall units of each simulation. The DNS data is from~\cite{pirozzoli2018turbulence}.}
   \label{fig:duct_vel}
\end{figure}

\begin{figure}[htbp]
   \centering
   \begin{subfigure}{0.49\textwidth}
   \centering
   \includegraphics[width=\linewidth]{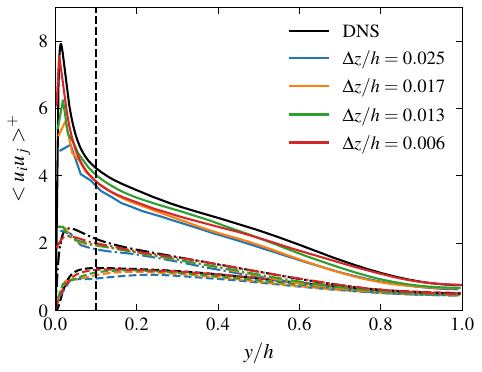}
   \caption{}
   \label{fig:sub2}
   \end{subfigure}
   \hfill
   \begin{subfigure}{0.49\textwidth}
   \centering
   \includegraphics[width=\linewidth]{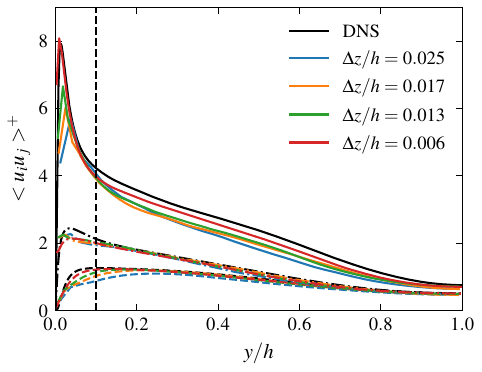}
   \caption{}
   \label{fig:sub2}
   \end{subfigure}
   \vfill
   \begin{subfigure}{0.49\textwidth}
   \centering
   \includegraphics[width=\linewidth]{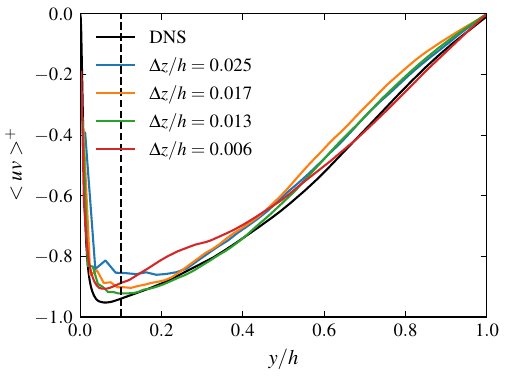}
   \caption{}
   \label{fig:sub2}
   \end{subfigure}
   \hfill
   \begin{subfigure}{0.49\textwidth}
   \centering
   \includegraphics[width=\linewidth]{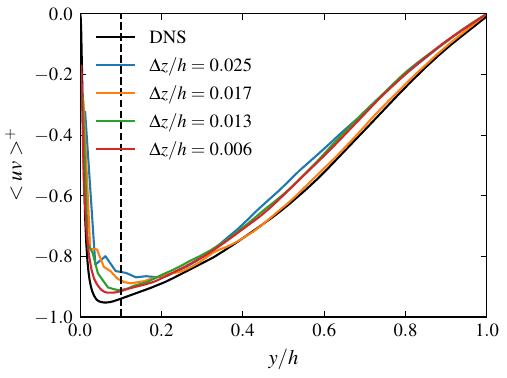}
   \caption{}
   \label{fig:sub2}
   \end{subfigure}
   \caption{Turbulent square duct flow: profiles of resolved turbulent normal stress (a,b) and shear stress (c,d) along the wall bisector as obtained from WMLES with the SM (a,c) and DSM (b,d) models. The dashed line denotes $y/h=0.1$. Normalization is based on the DNS wall units. The DNS data is from~\cite{pirozzoli2018turbulence}. Line codes: $\langle uu\rangle$ (solid), $\langle vv\rangle$ (dashed), $\langle ww\rangle$ (dash-dotted).}
   \label{fig:duct_rey}
\end{figure}

We further extract the profiles of these quantities along the wall and corner bisectors (see figure~\ref{fig:duct}(a)). Figure~\ref{fig:duct_vel} displays the profiles of mean streamwise velocity along both the wall bisector and the corner bisector. Consistent with the contours in figure~\ref{fig:contour_smag}, the mean streamwise velocity profiles show minor variations with grid refinement, indicating satisfactory grid convergence properties. In contrast, the most significant variations are observed in the turbulent normal stress $\langle uu \rangle$ (figure~\ref{fig:duct_rey}). Unlike the results for channel flow, the predicted peak in $\langle uu \rangle$ along the wall bisector increases with grid refinement, leading to improved accuracy. Small oscillations are observed near the turbulent stress peak for the coarser grids, which we attribute to the absence of grid clustering near the wall; these oscillations gradually disappear with grid refinement. Notably, the variations in shear stress exhibit a clearly non-monotonic pattern as the grid is refined.

\clearpage
\section{Conclusion}\label{sec:conclusion}
We have introduced CaLES, a GPU-accelerated solver designed for large-eddy simulation of incompressible wall-bounded flows. Based on the GPU-accelerated DNS solver CaNS, CaLES demonstrates significant computational efficiency improvements through GPU parallelization using a combination of CUDA Fortran and OpenACC directives, and good scalability on massively parallel architectures. The incorporation of SGS models, including static and dynamic Smagorinsky models, along with a wall model, extends the solver’s applicability to wall-resolved and wall-modeled LES.

Performance evaluation conducted on state-of-the-art high-performance computing clusters, shows that CaLES achieves substantial speed-ups using GPU acceleration compared to its CPU-only counterparts. The solver efficiently scales across multiple GPUs, achieving approximately $15\times$ speed-up on a single GPU compared to a full CPU node, making it highly effective for high-fidelity simulations on large computational grids. Validation cases—such as decaying isotropic turbulence, turbulent channel flow, and turbulent duct flow—confirm the solver's accuracy for LES, with the dynamic Smagorinsky model exhibiting somewhat superior performance in grid convergence and turbulence prediction.

CaLES has also been applied for wall-modeled LES of turbulent channel and duct flows. Its high computational efficiency enables WMLES for channels with $\Rey_\tau = 5200$ on progressively refined grids, meeting the grid requirements for wall-resolved LES. A key observation is the non-monotonic grid convergence in WMLES for channel and duct flows, particularly in wall friction, streamwise velocity fluctuations, and turbulent shear stresses. These results underscore the complexities inherent in WMLES and emphasize the need for continued research into SGS and wall models.

In summary, CaLES offers the computational efficiency and flexibility needed to investigate turbulent flows with moderate complexity at high Reynolds numbers. Its open-source availability makes it a valuable tool for fast simulations, which is particularly important in the application of machine learning to develop robust SGS and wall models, including data-driven approaches and reinforcement learning.

Finally, we should emphasize that the purpose of this work is not to carry out a physical investigation of the performance of SGS models and wall models. Instead, our aim is to incorporate well-established models into a baseline solver that can easily incorporate other models, such as the Vreman subgrid-scale model~\cite{vreman2004eddy} and the ordinary differential equation-based wall model~\cite{kawai2012wall}. 

\section*{Acknowledgements}
The simulations were performed using the EuroHPC Research Infrastructure resource LEONARDO, based at CINECA, Casalecchio di Reno, Italy, under a LEAP grant. Maochao Xiao would like to thank Di Zhou for carefully reviewing the manuscript. Additional thanks are extended to Jane H. Bae, Yufei Zhang and Xiao He for their helpful comments.

\bibliography{main_bibfile}
\end{document}